\documentclass[12pt,preprint]{aastex}
\usepackage{amssymb,amsmath,hyperref,rotating,natbib,morefloats,multirow,relsize}
\shorttitle{Ammonia and CCS in Infrared Dark Clouds}
\shortauthors{Dirienzo et al.}

\begin{document}

\title{Physical Conditions of the Earliest Phases of Massive Star Formation: Single-Dish and Interferometric Observations of Ammonia and CCS in Infrared Dark Clouds}

\author{William J. Dirienzo\altaffilmark{1,2}}
\email{william.dirienzo@uwc.edu}
\author{Crystal Brogan\altaffilmark{3}}
\author{R\'emy Indebetouw\altaffilmark{1,3}}
\author{Claire J. Chandler\altaffilmark{4}}
\author{Rachel K. Friesen\altaffilmark{5}}
\author{Kathryn E. Devine\altaffilmark{6}}

\altaffiltext{1}{Department of Astronomy, University of Virginia, P.O. Box 3818, Charlottesville, VA 22903, USA}
\altaffiltext{2}{CSEPA Department, University of Wisconsin-Sheboygan, One University Dr, Sheboygan, WI 53081, USA}
\altaffiltext{3}{National Radio Astronomy Observatory, 520 Edgemont Road, Charlottesville, VA 22903, USA}
\altaffiltext{4}{National Radio Astronomy Observatory, P.O. Box O, Socorro, NM 87801, USA}
\altaffiltext{5}{Dunlap Institute for Astronomy and Astrophysics, University of Toronto, 50 St. George Street, Toronto M5S 3H4, Ontario, Canada}
\altaffiltext{6}{Department of Mathematics and Physical Sciences, College of Idaho, 2112 Cleveland Blvd, Caldwell, ID 83605, USA}

\begin{abstract}

Infrared Dark Clouds (IRDCs) harbor the earliest phases of massive star formation, and many of the compact cores in IRDCs, traced by millimeter continuum or by molecular emission in high critical density lines, host massive young stellar objects (YSOs). We used the Robert C. Byrd Green Bank Telescope (GBT) and the Karl G. Jansky Very Large Array (VLA) to map NH$_{3}$ and CCS in nine IRDCs to reveal the temperature, density, and velocity structures and explore chemical evolution in the dense ($>10^{22}$ cm$^{-2}$) gas. Ammonia is an excellent molecular tracer for these cold, dense environments. The internal structure and kinematics of the IRDCs include velocity gradients, filaments, and possibly colliding clumps that elucidate the formation process of these structures and their YSOs. We find a wide variety of substructure including filaments and globules at distinct velocities, sometimes overlapping at sites of ongoing star formation. It appears that these IRDCs are still being assembled from molecular gas clumps even as star formation has already begun, and at least three of them appear consistent with the morphology of ``hub-filament structures'' discussed in the literature. Furthermore, we find that these clumps are typically near equipartition between gravitational and kinetic energies, so these structures may survive for multiple free-fall times.


\end{abstract}

\section{Introduction\label{sec-intro}}

Infrared Dark Clouds (IRDCs) are dense ($> 10^{5}$ cm$^{-3}$) and cold ($< 20$ K) collections of dust and molecular gas, typically arranged in filamentary and/or globular structures with compact cores. IRDCs were first observed as regions of high extinction in silhouette against the galactic background infrared (IR) emission by the \emph{Infrared Astronomical Satellite} (\emph{IRAS}) \citep{1994ApJS...95..457W}, the \emph{Midcourse Space Experiment} (\emph{MSX}) \citep{1998ApJ...508..721C,1998ApJ...494L.199E,2006ApJ...639..227S}, and the \emph{Infrared Space Observatory} (\emph{ISO}) \citep{2001A&A...365..598H}. The ability to identify and study these objects has been greatly improved by the \emph{Spitzer Space Telescope}, and in particular surveys with two of its instruments: the Infrared Array Camera (IRAC) \citep{2004ApJS..154...10F} and Multiband Imaging Photometer for \emph{Spitzer} (MIPS) \citep{2004ApJS..154...25R}.

The Galactic Legacy Infrared Mid-Plane Survey Extraordinaire (GLIMPSE) \citep{2003PASP..115..953B, 2009PASP..121..213C} covered the region $|b| \le 1^{\circ}$ and $ 10^{\circ} \le |\ell| \le 65^{\circ}$ in all the IRAC bands (3.6, 4.5, 5.8, and 8.0 $\micron$) with 1$\farcs$5 to 1$\farcs$9 resolution, while the MIPS Galactic Plane Survey (MIPSGAL) \citep{2009PASP..121...76C} was a complementary survey in the 24 and 70 $\micron$ MIPS wavebands with 6$\arcsec$ and 18$\arcsec$ resolution, respectively. These surveys observed much of the galactic plane, imaged the structure of IRDCs at higher resolution than before, and revealed embedded young stellar objects (YSOs), typically 1 to 10 per IRDC. (Note that in this paper we use the general term``YSO,'' encompassing both protostars and H {\smaller II} regions.) An extensive catalog of \emph{Spitzer} IRDCs is given by \cite{2009A&A...505..405P}, of which 80\% were previously not identified from \emph{MSX} images. Furthermore, \cite{2006ApJ...653.1325S} matched several hundred IRDCs to molecular clouds seen in $^{13}$CO $J$=1-0 by the Boston University Galactic Ring Survey (BU-GRS) \citep{2006ApJS..163..145J}, and thus determined velocities, kinematic distances, and physical properties of the population of clouds. The darkest clouds have very high column densities, as high as approximately $10^{24}$-$10^{25}$ cm$^{-2}$.

The terms ``core'' and ``clump'' are frequently used in the literature, but the meanings are not standardized. We use the term ``core'' to refer to an unresolved or marginally resolved overdense structure $<$0.1 pc across and tens of solar masses. We then use the term ``clump'' to refer to a resolved structure (e.g. projected area greater than three times the area of the beam) within an IRDC and is assigned by a clump deconvolution algorithm (see \S \ref{sec-clump} for description of algorithms used). ``Average'' results of spectral line fitting presented in this paper are averaged over clumps. We also occasionally refer to the velocity components within IRDCs when the IRDC's emission primarily occurs at two or more distinct velocities with little emission at the intermediate velocities; a velocity component may have one or more clump associated with it.

IRDCs are an active subject of study because they are nurseries of massive star formation and contain complex chemistry. It is now widely speculated that these objects contain the earliest stages of the formation of massive stars. Recent studies have revealed massive ($\gtrsim$10 $M_{\sun}$) YSOs \citep{2005ApJ...630L.181R, 2006A&A...447..929P, 2007ApJ...656L..85B,2008ApJ...672L..33W} and massive (10-1000 $M_{\sun}$) cores in IRDCs \citep{2006ApJ...641..389R, 2007ApJ...662.1082R,2009ApJ...705.1456S,2010A&A...518L..95H,2011ApJ...741..120R}. High mass YSOs are identified in these studies by the presence of masers (indicating accretion disks or outflows), radio continuum emission (from ionized gas), or fitting spectral energy distributions (SEDs) consistent with massive YSOs across mid-infrared (MIR) and submillimeter wavelengths. \cite{2010JKAS...43....9K} found that about 13\% of IRDC cores have YSOs identified by their IR or maser emission. Not all dense cores in IRDCs are obviously star-forming, which raises the question: ``Are these cores going to form stars and are just too young, or is there something fundamentally different about these cores preventing star formation?'' To answer this question, we must know about the physical properties of both starless and actively star-forming clumps in IRDCs.

\cite{2006ApJ...641..389R} used the Institut de Radioastronomic Millim\'etrique (IRAM) 30 m telescope to make 11$\arcsec$ resolution 1.2 mm dust continuum maps of 38 IRDCs to investigate the structure and clumps of IRDCs traced by dust emission. They found that these clumps had similar masses, sizes, and densities as hot ($>$ 50 K) clumps with massive YSOs, but were much colder (15-30 K), implying that the clumps were indeed an early evolutionary phase of massive star formation if the clumps were to subsequently collapse. The authors further suggested IRDCs may be precursors to star clusters, as they have masses comparable to young clusters and contain several compact clumps. Finally, they also asserted that better resolution was needed to distinguish better individual cores and investigate fragmentation.

Disentangling the substructure of IRDCs is important to determining their formation and how it relates to subsequent star formation. Early studies of filamentary clouds with low spatial and velocity resolution appear to show that such structures have radial density profiles of the form $\rho \propto r^{-2}$ \citep[see for example][]{1998ApJ...506..292A,1999ApJ...512..250L}. This density profile is consistent with the presence of a toroidal magnetic field \citep{2000MNRAS.311...85F}, which can prevent the filaments from expanding \citep{2013MNRAS.433..251C}. Higher resolution studies show that IRDCs typically have complex substructure, spatially and kinematically. Some studies observe distinct velocity components within IRDCs, seen in shock tracers such as SiO, that may be due to protostellar outflows or cloud-cloud collisions \citep{2013ApJ...773..123S}. Other studies observe collections of filaments with coherent velocity gradients consistent with gas flowing toward the most massive cores, perhaps guided by magnetic fields \citep{2014A&A...561A..83P}. \cite{2009ApJ...700.1609M} and \cite{2013MNRAS.436.3707L} describe a ``hub-filament structure,'' in which gas flows along filaments to a central hub where star formation is ongoing. Many studies have already described more generally how the collision and accumulation of smaller clouds and gas flows can assemble molecular cloud complexes \citep[see for example][]{1980ApJ...238..148B,2008MNRAS.391..844D,2009ApJ...696L.115F,2011ApJ...738...46T,2013ApJ...774L..31I,2014ApJ...780...36F,2014A&A...571A..32M,2015arXiv150405391F}.

Ammonia is an excellent probe of the molecular gas in IRDCs. It is a relatively abundant species, typically about $10^{-9}$ to $10^{-7}$ fractional abundance relative to molecular hydrogen \citep{2011ApJ...736..163R,2013A&A...552A..40C}. Unlike carbon-bearing species, NH$_{3}$ suffers minimal freeze out in these cold environments \citep{2007ARA&A..45..339B}. The ratio of the (1,1) to (2,2) line strengths leads to a rotation temperature that has been shown to be a good tracer of the kinetic temperature of the gas over the typical range for IRDCs \citep{1983ARA&A..21..239H}. Furthermore, the critical density of NH$_{3}$ is about $10^{4}$ cm$^{-3}$, well matched to the gas being studied \citep{2002ApJ...569..815T}. (\cite{2006ApJ...653.1325S} found typical average number densities for the majority of IRDCs to be about $10^{3}$ cm$^{-3}$ to $10^{4}$ cm$^{-3}$ by observing $^{13}$CO $J$=1-0 with a beam a few times larger than most of the IR extinction features, so when accounting for beam dilution we expect the dense gas associated with these features to be near or above the critical density for ammonia.) Ammonia has hyperfine structure, allowing determination of the optical depth in the NH$_{3}$ lines by comparing the observed hyperfine component strengths to those set by molecular physics \citep{1983ARA&A..21..239H}. The direct measurement of the temperature and optical depth allows for an unambiguous determination of the column density, perfect for environments in which the column can vary significantly from optically thin to optically thick in the same IRDC.

\cite{2011ApJ...733...44D} observed the NH$_3$ (1,1) and (2,2) and the CCS (2$_{1}$-1$_0{}$) transitions toward G19.30+0.07, and measured temperatures around 15-20 K, NH$_{3}$ column densities around 10$^{15}$-10$^{16}$ cm$^{-2}$, and linewidths of about 2 km s$^{-1}$. They also observed that NH$_{3}$ and CCS generally had spatial distributions that were anticorrelated. Furthermore, they found that G19.30+0.07 was composed of four distinct clumps in three distinct velocity components. These clumps had masses of tens to hundreds of solar masses and were virially unstable against self-gravitating collapse. This complex substructure has also been investigated in other star-forming regions with physical properties similar to IRDCs. \cite{2013A&A...554A..55H} additionally found that molecular filaments in Taurus could be deconvolved into a hierarchy of several sub-filaments at distinct velocities leading to core formation. (The smallest scale of the hierarchy, the ``filament bundles,'' have spatial widths and linewidths comparable to the separation of the bundles, about 0.1 pc and 0.5 km s$^{-1}$, so our study does not have the resolution to identify similar bundles in these IRDCs if they occur on the same scales as in Taurus.) \cite{2008ApJS..175..509R} also saw multiple velocity components in NH$_{3}$ toward dense cores in Perseus at 0.04 pc, 0.024 km s$^{-1}$ resolution. These individual components often had such narrow linewidths ($\sim$0.1-0.3 km s$^{-1}$) that turbulence was unlikely to contribute significantly to slowing collapse. \cite{2014ApJ...790...84L} studied 62 high-mass star-forming regions with Very Large Array (VLA) observations of the NH$_{3}$ (1,1) and (2,2) transitions. They found that parsec scale filaments were ubiquitous and often contained regularly spaced dense cores, though we note this regular spacing could be a result of using interferrometric observations. Furthermore, they suggested that the filaments could be supported by turbulence and found that the dense cores were near virial equilibrium.

\cite{2011ApJ...736..163R,2012ApJ...746..174R} performed a study of NH$_3$ (1,1) and (2,2) with the Robert C. Byrd Green Bank Telescope (GBT) and the Very Large Array (VLA) in six IRDCs at 3$\farcs$7-8$\farcs$3 and 0.6 km s$^{-1}$ resolution. They found that the majority of the gas had kinetic temperatures 8-13 K, indicating that protostellar heating was not significant for most of the IRDC. Furthermore, they found that velocity fields were generally coherent across the IRDCs, with local ($\sim$0.1 pc) disruptions of a few km s$^{-1}$ coincident with sites of local star formation. They argued that neither turbulence nor thermal pressure were sufficient to support the IRDCs and that the observed velocity structures were a result of ongoing collapse, fragmentation, and protostellar feedback.

\cite{2013A&A...559A..79R} further investigated the differences in filamentary and globular IRDCs and their hierarchical structure. They studied 11 IRDCs, covering a range of morphology and star formation activity, with \emph{Herschel} and the Submillimetre APEX Bolometer Camera (SABOCA) instrument on the Atacama Pathfinder Experiment (APEX) 12 m telescope at 350 $\micron$, resolving structure down to $\sim$0.1 pc. They performed a dendrogram analysis on the APEX data found that filamentary IRDCs tended to be more massive and have more hierarchical structure than clumpy IRDCs. This suggests that IRDCs may be divided into two relatively distinct morphological families. \cite{2014A&A...568A..73R} identified 7 Giant Molecular Filaments (GMFs) in which the molecular gas extend for $\sim$100 pc including IRDCs, IR bright structures, and more diffuse gas presumably enveloping these denser regions. The existence and structure of GMFs suggests that hierarchical structure may extend to even these large size scales.

Single-dish data are necessary for determining whether sharp transitions at the edges of the dense structures are real and not a consequence of interferometers filtering the extended emission. Fortunately, the NH$_3$ (1,1) and (2,2) and the CCS (2$_{1}$-1$_0{}$) transitions may be observed simultaneously with the GBT. It is also important to remember that we likely do not resolve the smallest substructure in these IRDCs and that interferometric data are even more important. The combination of interferometric and single-dish radio data from the VLA and the GBT gives us the high resolution we need to separate better individual cores and other structures compared with previous work, while also not resolving out emission. The limited number of studies in this regime suggest that the substructure is present because IRDCs represent a phase in which star formation proceeds even as the IRDC itself is still being constructed from colliding clumps and/or gas flowing into the IRDC along filaments \citep{2009ApJ...700.1609M,2013MNRAS.436.3707L,2013ApJ...773..123S}. A homogeneous study of several representative IRDCs combining total power and high resolution is thus necessary. For the current study, we mapped the NH$_3$ (1,1) and (2,2) and the CCS (2$_{1}$-1$_0{}$) transitions across nine IRDCs containing both ongoing star formation and more quiescent environments.

We describe the IRDC sample, with distance determination and previous studies of these IRDCs, in \S \ref{sec-sample}. Observations and data reduction are described in \S \ref{sec-data}. Our methodology is presented in \S \ref{sec-methods}, including NH$_{3}$ spectral line fitting, clump deconvolution, and IR extinction. Results of investigating the kinematics, the spectral line fitting, and clump properties are in \S \ref{sec-results}. A discussion of the results relating to kinematics, structure, gravitational stability, and chemical evolution is presented in \S \ref{sec-discussion}, and conclusions are in \S \ref{sec-conclusions}.

\section{Sample Selection and Distance Determination \label{sec-sample}}

This sample of nine IRDCs was originally selected in 2005 from extinction features in the 8 $\micron$ GLIMPSE images and chosen to cover a wide range of physical parameters. The IRDCs subsequently appeared in a catalog compiled by \cite{2006ApJ...639..227S}, identified by their high contrast against background emission in the 8.3 $\micron$ \emph{MSX} images. The sample size is large enough to contain tens of cores identified by dust thermal continuum, including about half that are coincident with IR point sources indicating ongoing star formation, but is also small enough for each object to be studied in detail. The larger IRDCs show both quiescent, starless cores and YSOs in close proximity to each other.

\begin{figure}
\begin{center}
\includegraphics[width=1\textwidth]{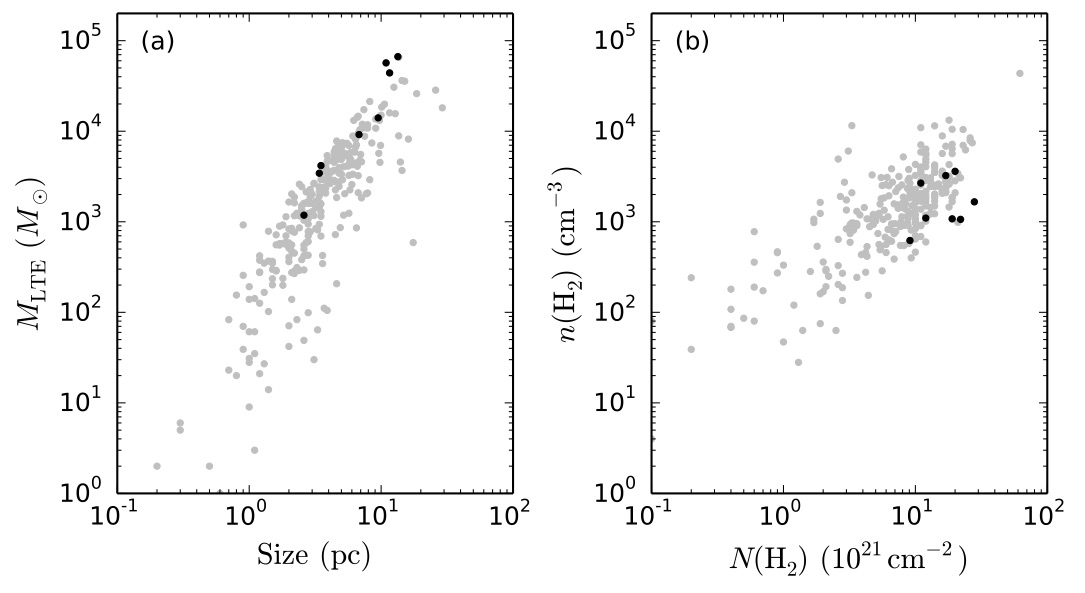}
\caption{A comparison of physical parameters of the IRDCs in this sample to all of those in the BU-GRS \citep{2006ApJS..163..145J} as analyzed by \cite{2006ApJ...653.1325S}. The black points represent IRDCs from this sample (except G010.74-00.13, which was not covered by the BU-GRS). \label{f1}}
\end{center}
\end{figure}

We include both filamentary and globular morphologies, apparently starless and protostellar cores, and ranges of physical sizes, masses, linewidths, distances, and IR contrast. A comparison of this sample with all the IRDCs analyzed by \cite{2006ApJ...653.1325S} in the BU-GRS data is shown in Figure \ref{f1} (except G010.74-00.13, which is not covered by the BU-GRS). The BU-GRS is a large scale survey of the 110.2 GHz $^{13}$CO $J$=1-0 transition in the disk of the Milky Way using the Five College Radio Astronomy Observatory (FCRAO) 14 m single dish telescope. The publicly available data cubes have velocity resolution of 0.2 km s$^{-1}$, angular resolution of 46$\arcsec$, and typical antenna temperature RMS sensitivity of 0.13 K \citep{2006ApJS..163..145J}. This sample tends toward higher mass and column density the general population of IRDCs in the BU-GRS.

We calculated the kinematic distances using the galactic rotation curve of \cite{2009ApJ...700..137R}. They adopted a galactocentric radius $R_{\circ}=8.4 \pm 0.6$ kpc and a circular rotation speed $\Theta_{\circ}=254 \pm 16$ km s$^{-1}$ kpc$^{-1}$, based on the results of their measured trigonometric parallaxes of massive star-forming regions. All of the regions in our sample were in the $|\ell| < 90^{\circ}$ regime, so there was naturally a near-far distance ambiguity. Since IRDCs are seen in extinction, we assumed for the rest of this work that all of the regions were at the near kinematic distance, except for G034.43+00.24. \cite{2011PASJ...63..513K} performed parallax measurements of an H$_2$O maser associated with G034.43+00.24 using Very Long Baseline Interferometry (VLBI) as part of the VLBI Exploration of Radio Astrometry (VERA) project and determined the distance to be $1.56^{+0.12}_{-0.11}$ kpc, less than half of the near kinematic distance of $3.55^{+0.36}_{-0.36}$. It should be noted that the distance to this IRDC is still under debate. \cite{2012ApJ...751..157F} used two near-infrared (NIR) extinction methods on two different NIR data sets to derive four extinction distances, all of them consistent with the near kinematic distance and three of them inconsistent with the maser distance (the distance estimate with the largest uncertainty was consistent with the maser distance). They further raise questions about the size of the uncertainties on the maser distance given the quality of the model fit to the data and the low declination of the source. We adopt the maser distance in this work, noting that sizes and masses reported for this IRDC may be factors of several times larger if the kinematic and extinction distances more accurate. A summary of the sample is presented in Table \ref{t1}.

\begin{deluxetable}{ccccc}
\tablecolumns{7}
\tablewidth{0pt}
\tablecaption{IRDC Sample \label{t1}}
\tablehead{
 \colhead{IRDC} &
 \colhead{R.A. (J2000)} &
 \colhead{Decl. (J2000)} &
 \colhead{$v_{\mathrm{LSR}}$} &
 \colhead{$D$} \\
 &
 \colhead{hh:mm:ss.s} & 
 \colhead{$\phantom{-}$dd:mm:ss} & 
 \colhead{(km s$^{-1}$)} & 
 \colhead{(kpc)}
 }
\startdata
G010.74-00.13 & 18:09:45.9 & -19:42:04 & 29.0 & $3.46^{+0.46}_{-0.55}$ \\
G022.56-00.20 & 18:32:59.6 & -09:20:08 & 76.8 & $4.55^{+0.26}_{-0.27}$ \\
G024.60+00.08 & 18:35:39.7 & -07:18:52 & 53.4 & $3.45^{+0.32}_{-0.34}$ \\
G028.23-00.19 & 18:43:30.2 & -04:12:56 & 80.0 & $4.50^{+0.30}_{-0.30}$ \\
G031.97+00.07 & 18:49:30.6 & -00:48:18 & 96.0 & $5.60^{+0.42}_{-0.42}$ \\
G032.70-00.30 & 18:52:06.2 & -00:20:26 & 90.2 & $5.23^{+0.45}_{-0.39}$ \\
G034.43+00.24 & 18:53:18.7 & \phantom{-}01:25:51 & 58.0 & $1.56^{+0.12}_{-0.11}$ \\
G035.39-00.33 & 18:57:09.4 & \phantom{-}02:07:48 & 44.5 & $2.98^{+0.37}_{-0.38}$ \\
G038.95-00.47 & 19:04:08.4 & \phantom{-}05:09:12 & 42.2 & $2.80^{+0.40}_{-0.40}$ \\
\enddata
\end{deluxetable}

\section{Observations and Data \label{sec-data}}

\subsection{GBT Observations \label{sec-gbtdata}}

Data from the GBT provide total flux on large spatial scales, which is not measured by our VLA observations. Dates of GBT observations for Project IDs AGBT05C\_014 and AGBT12B\_283 are listed for each source in Table \ref{t2}. The data are dual polarization taken in frequency switching mode using a 5 MHz shift. Observations in 2005 are in PointMap mode with the single pixel K-band receiver, while data in 2012 are in on-the-fly (OTF) mapping mode using beams 1 and 4 of the 7 possible beams in the K-band Focal Plane Array (KFPA). All observing configurations include simultaneous observation of the NH$_{3}$ (1,1) and (2,2) inversions lines and the CCS (2$_{1}$-1$_{0}$) line, with rest frequencies of 23.6945 GHz, 23.72263 GHz, and 22.34403 GHz, respectively. The expected beam FWHM at these frequencies is approximately 32$\farcs$ Hourly pointing scans indicated the actual FWHM of the beam during observing varied between 32$\arcsec$ and 34$\arcsec$, and the pointing correction was typically 6$\farcs$

\begin{deluxetable}{cccc}
\tablecolumns{4}
\tablewidth{0pt}
\tablecaption{GBT Observations \label{t2}}
\tablehead{
 \colhead{Date} &
 \colhead{Session \#} &
 \colhead{IRDC} &
 \colhead{Zenith Opacity}
 }
\startdata
\multicolumn{4}{c}{Project ID AGBT05C\_014} \\
\tableline
2005 Oct 19 & 1 & G010.74-00.13 & 0.123 \\
2005 Oct 19 & 2 & G038.95-00.47 & 0.115 \\
2005 Oct 23 & 3 & G010.74-00.13 & 0.128 \\
2005 Oct 30 & 6 & G022.56-00.20 & 0.057 \\
2005 Oct 31 & 7 & G022.56-00.206 & 0.083 \\
2005 Oct 31 & 7 & G024.60+00.08 & 0.084 \\
2005 Oct 31 & 7 & G032.70-00.30 & 0.084 \\
2005 Nov 02 & 8 & G032.70-00.30 & 0.067 \\
2005 Nov 03 & 8 & G038.95-00.47 & 0.067 \\
2005 Nov 17 & 9 & G032.70-00.30 & 0.039 \\
2005 Nov 17 & 9 & G038.95-00.47 & 0.04 \\
2005 Nov 18 & 11 & G028.23-00.19 & 0.058 \\
2005 Nov 19 & 11 & G038.95-00.47 & 0.056 \\
2005 Dec 24 & 12 & G028.23-00.19 & 0.066 \\
\tableline
\multicolumn{4}{c}{Project ID AGBT12B\_283} \\
\tableline
2012 Nov 05 & 2 & G034.43+00.24 & 0.048 \\
2012 Nov 14 & 3 & G034.43+00.24 & 0.043 \\
2012 Dec 05 & 4 & G031.97+00.07 & 0.041 \\
2012 Dec 05 & 4 & G035.39-00.33 & 0.040 \\
\enddata
\end{deluxetable}

The spectral resolution of the raw data varied between observations with different spectral setups, so all data were smoothed to the limiting velocity resolution of approximately 0.15 km s$^{-1}$. System temperatures typically varied between 40 and 60 K. Absolute flux density calibration of the 2012 observations was tied to beam nodding scans of Venus on 2012 December 05. We utilized the Green Bank Telescope Interactive Data Language (GBTIDL) procedures \texttt{venusmodel}, \texttt{venuscal}, and \texttt{venuscalget}, documented in GBT memo \#275,\footnote{\url{https://safe.nrao.edu/wiki/bin/view/GB/Knowledge/GBTMemos}} to model the antenna temperature of Venus at a given date, time, and elevation, and to calculate the aperture efficiency. This model corrected for the distance to Venus and any beam dilution if it was unresolved, but did not account for variations across the surface of Venus. The determined aperture efficiency, $ \eta_{A} $, of 0.64 was in good agreement with the value of 0.67 from the Ruze equation and the GBT sensitivity calculator for the NH$_{3}$ (1,1) line, although we note both of these are approximations for extended sources and there will be some variation at different frequencies. From this value, we also calculated the main beam efficiency to be $\eta_{\mathrm{mb}} = 1.37 \eta_{A} = 0.88$ as noted in ``Calibration of GBT Spectral Line Data in GBTIDL v2.1.''\footnote{\url{http://www.gb.nrao.edu/GBT/DA/gbtidl/gbtidl_calibration.pdf}}

The primary uncertainty in the efficiency estimates come from the possibility of pointing errors in the observations of Venus, which will tend to cause us to underestimate the aperture efficiency. Given typical pointing scan corrections, we do not expect this effect to be bigger than about 30\%, which would result in the flux densities in our final maps being 30\% higher. The good agreement between the measured aperture efficiency and the prediction form the Ruze equation implies that the errors are not this large.

Inspection of the observations taken in 2012 shows that while the relative flux density scales of dates 2012 November 14 and 2012 December 5 agree, observations on day 2012 November 5 have a lower flux density scale. This effect is seen in the maps of the same regions on different days, the peak antenna temperature in the Focus scans of our pointing source 1851+0035 (observed approximately every hour), and calibration observations of DR21. To determine the relative flux density scale, we averaged the results of the fractional difference in the maps and the Focus scans. We excluded DR21 in this calculation because the measured ratio may be affected by pointing errors since we neither mapped DR21 nor performed a pointing correction on it. We measured the flux density scale on 2012 November 5 to be 0.7 relative to that of the other days, so we adopted values of the aperture efficiency and main beam efficiency of 0.46, and 0.63, respectively, to bring data from that day into agreement with the others. This discrepancy may be a result of atmospheric effects, dish surface effects, or pointing or focus errors not accurately corrected by the standard calibration procedure.

No flux calibrators were observed during the 2005 observations, so we adopted 0.58 and 0.8 as the aperture efficiency and main beam efficiency, respectively, because the aperture efficiency at these frequencies in 2005 was typically 91\% of what is is now (T. Hunter 2015, private communication). We inspected the peak antenna temperature of the focus scans on the pointing sources across multiple epochs to look for variations in the relative flux density scales, assuming the flux densities of these sources to be stable over the approximately two month timescale of these observations. The most frequently observed pointing source was 1733-1304, being observed every day of the observations in 2005 except 2005-12-24, when only 1743-0350 was used. Additionally, 1751+0939 was observed sporadically on 2005 October 19, 2005 November 17, and 2005 November 18. Assuming these calibrators were constant in flux density, we found that the aperture efficiency only changed significantly over the course of 2005 December 24, possibly affecting the flux density scale by about 20\% across the maps of G028.23-00.19, though this may also be due to the intrinsic variability of the calibrators.

Initial calibration of the frequency switching data and conversion to main beam temperature was performed with the GBT pipeline for the KFPA provided by the NRAO. The pipeline queried the weather data from the GBT weather forecast database to determine the zenith opacity for each observing block. The pipeline also accepted values for the aperture efficiency and the main beam efficiency as inputs, determined as above.

After the initial calibration, we used GBTIDL to fit and subtract baselines from the spectra. Many of the baselines contained complex shapes with large amplitudes. We averaged the spectra for each individual combination of scan number, IF, polarization, and beam, then fit the line-free regions with a high order polynomial. The baseline shapes varied considerably between different IFs, beams, and polarizations, and showed moderate variation with time among different scans. The baselines were typically approximately linear in the vicinity of the line, though not always. We determined that the variation between individual integrations was within the noise of an individual integration, so that we could average over whole scans. For the 2012 observations taken in OTF mode, a single scan consisted of one row or column in the map. Since the 2005 observations were taken in PointMap mode, each scan was a single pointing. Baselines in the 2005 observations were generally less complex, so we used lower order polynomials at this stage.

Imaging was performed also using the GBT pipeline for the KFPA, which uses ParselTongue to call AIPS tasks from Python. Data were gridded using the \texttt{SDGRD} task. Since lower amplitude baseline effects were noticed on shorter timescales, we additionally performed lower order polynomial fitting and subtraction on an individual line of sight basis in the final imaged data cubes using the \texttt{imcontsub} task in the Common Astronomy Software Applications (CASA) package (\url{http://casa.nrao.edu}). By combining two baseline subtraction methods, we could address both the time variability in neighboring scans before the imaging step while also taking care to get the flattest baselines possible in the final data cubes. The data were converted from main beam temperature to Jy beam$^{-1}$ by multiplying by $ (2k_{\mathrm{B}}\eta_{\mathrm{mb}})/(A_{\mathrm{p}}\eta_{A}) = 0.483 $, where $ k_{\mathrm{B}} $ is the Boltzmann constant and $ A_{\mathrm{p}} $ is the physical collecting area of the GBT. Typical RMS noise in the final data cubes was 25-40 mJy beam$^{-1}$ per 0.15 km s$^{-1}$ channel. We estimate our absolute flux density uncertainty to be approximately 30\%.

\subsection{VLA Observations \label{sec-vladata}}

NH$_{3}$ (1,1) and (2,2) and CCS (2$_{1}$-1$_{0}$) emission in our sample of IRDCs was observed with the VLA in D configuration, primarily in 2005 and 2006 (NRAO Proposal ID AD0516), with observations of CCS in G028.83-00.19, G031.97+00.07, and G034.43+00.24 in 2007 (NRAO Proposal ID AD0556). The 3.125 MHz bandwidth covered only the main and innermost satellite hyperfine lines of the NH$_{3}$ (1,1) and (2,2) transitions. The spectral setup had 24.414 kHz resolution. Different numbers of pointings per IRDC were used to fully cover the highest opacity regions with the approximately 1\farcm 9 (FWHM) primary beam, as determined from the 8 $\micron$ images. G031.97+00.07 and G034.43+00.24 each were observed with five pointings (though only four were observed in CCS), G028.83-00.19 was observed with two pointings, and the remaining IRDCs were observed with one pointing each. J1820-2528 was used as the phase and amplitude calibrator for G010.74-00.13, while J1832-1035 was used for G022.56-00.20 and G024.60+00.08, and J1851+0035 was used for the remaining IRDCs. 3C286 (J1331+3030) and 3C48 (J0137+3309) were used as flux density calibrators. Table \ref{t3} summarizes the parameters of the VLA observations, including bandpass calibrators, beam sizes, and flux-density-to-temperature conversion factors.

\begin{deluxetable}{cccr@{$\times$}lr@{$\times$}lcc}
\tabletypesize{\footnotesize}
\tablecolumns{9}
\tablewidth{0pt}
\tablecaption{VLA Observations \label{t3}}
\tablehead{
 \colhead{IRDC} & 
 \colhead{Project} & 
 \colhead{Bandpass} & 
 \multicolumn{5}{c}{Synthesized Beam} & 
 \colhead{Flux to $T_{\mathrm{B}}$} \\
 & 
 \colhead{ID} & 
 \colhead{Calibrator\tablenotemark{a}} & 
 \multicolumn{4}{c}{Major Axis $\times$ Minor Axis} & 
 \colhead{PA} & 
 \colhead{(K Jy$^{-1}$)} \\
 & 
 & 
 & 
 \multicolumn{2}{c}{$(\arcsec)\times(\arcsec)$} & 
 \multicolumn{2}{c}{(pc)$\times$(pc)} & 
 \colhead{(deg)} & 
 }
\startdata
\multicolumn{9}{c}{NH$_{3}$ (1,1) 23.6945 GHz} \\
\tableline
G010.74-00.13 & AD0516 & J1851+0035 & 4.8 & 3.4 & 0.081 & 0.057 & -4.6 & 134.6 \\
G022.56-00.20 & AD0516 & J1832-1035 & 4.3 & 3.3 & 0.095 & 0.073 & -2.3 & 155.3 \\
G024.60+00.08 & AD0516 & J1832-1035 & 4.2 & 3.3 & 0.070 & 0.055 & 7.3 & 158.2 \\
G028.23-00.19 & AD0516 & J1851+0035 & 4.0 & 3.6 & 0.087 & 0.079 & -9.6 & 154.2 \\
G031.97+00.07 & AD0516 & J1851+0035 & 3.9 & 3.6 & 0.106 & 0.010 & 30.3 & 153.6 \\
G032.70-00.30 & AD0516 & 3C273, J1851+0035 & 3.8 & 3.8 & 0.096 & 0.096 & 64.9 & 149.3 \\
G034.43+00.24 & AD0516 & 3C273, J1851+0035 & 3.8 & 3.7 & 0.029 & 0.028 & 79.2 & 156.3 \\
G035.39-00.33 & AD0516 & 3C273, J1851+0035 & 4.2 & 3.6 & 0.061 & 0.052 & 62.6 & 143.8 \\
G038.95-00.47 & AD0516 & 3C273, J1851+0035 & 3.9 & 3.6 & 0.053 & 0.049 & 87.0 & 154.2 \\
\tableline
\multicolumn{9}{c}{NH$_{3}$ (2,2) 23.72263 GHz} \\
\tableline
G010.74-00.13 & AD0516 & J1832-1035 & 4.7 & 3.4 & 0.079 & 0.057 & -7.6 & 134.0 \\
G022.56-00.20 & AD0516 & 3C273, J1832-1035 & 4.4 & 3.3 & 0.097 & 0.073 & -6.3 & 149.3 \\
G024.60+00.08 & AD0516 & 3C273, J1832-1035 & 4.2 & 3.3 & 0.070 & 0.055 & 1.7 & 154.8 \\
G028.23-00.19 & AD0516 & 3C273, J1851+0035 & 4.2 & 3.5 & 0.092 & 0.076 & -23.1 & 149.3 \\
G031.97+00.07 & AD0516 & 3C273, J1851+0035 & 3.9 & 3.6 & 0.106 & 0.098 & -9.1 & 154.3 \\
G032.70-00.30 & AD0516 & 3C273, J1851+0035 & 4.1 & 3.7 & 0.104 & 0.094 & -45.2 & 144.3 \\
G034.43+00.24 & AD0516 & 3C273, J1851+0035 & 3.8 & 3.7 & 0.029 & 0.028 & -69.3 & 153.0 \\
G035.39-00.33 & AD0516 & 3C273, J1851+0035 & 4.2 & 3.7 & 0.061 & 0.053 & 64.9 & 141.0 \\
G038.95-00.47 & AD0516 & 3C273, J1851+0035 & 4.0 & 3.7 & 0.054 & 0.050 & -78.4 & 148.9 \\
\tableline
\multicolumn{9}{c}{CCS (2$_{1}$-1$_0{}$) 22.34403 GHz} \\
\tableline
G010.74-00.13 & AD0516 & 3C454.3 & 5.4 & 3.2 & 0.091 & 0.054 & 15.8 & 140.3 \\
G022.56-00.20 & AD0516 &3C454.3 & 4.4 & 3.5 & 0.097 & 0.077 & 8.0 & 160.9 \\
G024.60+00.08 & AD0516 & 3C273, 3C345, 3C454.3 & 4.3 & 3.6 & 0.072 & 0.060 & -0.5 & 158.5 \\
G028.23-00.19 & AD0556 & 3C345 & 4.8 & 3.5 & 0.105 & 0.076 & -5.3 & 147.5 \\
G031.97+00.07 & AD0556 & 3C345 & 5.0 & 3.4 & 0.136 & 0.092 & -9.2 & 142.5 \\
G032.70-00.30 & AD0516 & 3C273, 3C454.3 & 4.3 & 3.6 & 0.109 & 0.091 & -46.2 & 159.8 \\
G034.43+00.24 & AD0556 & 3C345 & 4.8 & 3.4 & 0.036 & 0.026 & -10.8 & 147.1 \\
G035.39-00.33 & AD0516 & 3C84, 3C345, 3C454.3 & 4.1 & 3.6 & 0.059 & 0.052 & -49.2 & 162.5 \\
G038.95-00.47 & AD0516 & 3C84, 3C345, 3C454.3 & 4.1 & 3.6 & 0.056 & 0.049 & -55.1 & 166.7 \\
\enddata
\tablenotetext{a}{Bandpass calibrators were selected by the best bandpass solution for each observing date. CCS observations often also used the phase and amplitude and/or flux density calibrators to improve the bandpass solution. 3C273 is also known as J1229+020, 3C454.3 is also known as J2253+161, 3C345 is also known as J1642+398, and 3C84 is also known as J0319+041.}
\end{deluxetable}

All data were calibrated using AIPS. The 2007 CCS observations were conducted during the EVLA upgrade, and eight of the twenty-seven antennas had been converted to EVLA antennas. Data were obtained using both the VLA and EVLA antennas. Doppler tracking could not be used during the upgrade period, so the AIPS task \texttt{CVEL} was applied during calibration to correct for motion of the Earth relative to the LSR during the observations. In the 22 GHz data sets, the uncertainty of the absolute flux densities was estimated to be 10\%. We estimated the absolute positional accuracy to be better than 1$\farcs$

Before imaging, data were Hanning smoothed, bringing the effective velocity resolution to approximately 0.6 km s$^{-1}$. Continuum subtraction was performed with the CASA task \texttt{uvcontsub} on all the data for G034.43+00.24 and the CCS observations of G031.97+00.07, as there were significant continuum point sources seen in these four dirty images. The data were imaged in CASA using the \texttt{clean} task in mosaic mode with natural weighting of the visibilities, deconvolved with 0.33 km s$^{-1}$ channels and 0$\farcs$7 pixels. We also employed the multiscale capability of \texttt{clean} to include clean components approximately one and three times the size of the synthesized beam, as well as the standard point-like components. The CASA task \texttt{pbcor} was used to apply a primary beam correction. The mosaics were imaged to the 35\% power level relative to the peak of the mosaic primary beam response.

The synthesized beam in the cubes is 3$\arcsec$-5$\arcsec$. The RMS noise is 1-4.5 mJy beam$^{-1}$ per 0.33 km s$^{-1}$ channel. The conversion between from flux density in Jy beam$^{-1}$ to brightness temperature, $T_{\mathrm{B}}$ (K), for the various data cubes are listed in Table \ref{t3}.

\subsection{Combining Single-dish and Interferometric Data \label{sec-combination}}

Successful combination of single dish and interferometric data relies upon good astrometric alignment and calibration between the two data sets. The estimated positional accuracy of our GBT observations, 6$\arcsec$, while small compared to the GBT beam, were larger than the VLA beam. This allowed for positional errors that might introduce significant artifacts in our combined maps. For example, using the GBT as a clean model for the VLA with a positional offset between the two can cause negative features neighboring the emission in the combined image. To address this issue, we computed the amount of positional shift required to align the GBT and VLA data by smoothing our VLA-only (1,1) cubes to the GBT angular resolution and smoothed the GBT-only (1,1) cubes to the VLA velocity resolution. We then determined the positional shift required to align the maxima in the emission. Shifts were restricted to integer numbers of pixels (6$\arcsec$ each) in the native GBT cubes. The largest total shift was 18$\arcsec$, or approximately half a GBT beam, for G035.39-00.33. All other IRDCs required shifts less than 10$\arcsec$, and three IRDCs (G032.70-00.30, G034.43+00.24, and G038.95-00.47) did not require any shift.

The GBT data were combined with the VLA data in a two-stage process. First, the GBT cube was used as a starting model for the VLA data in the CASA task \texttt{clean}, and then cleaning proceeded as for the VLA only, described above in \S \ref{sec-vladata}. The resulting cube was then combined with the GBT cube again, using the CASA task \texttt{feather}. Before feathering, the VLA primary beam response was applied to GBT cube with the same cutoff as \texttt{clean} to ensure that emission in the GBT cube that was outside the VLA footprint was not included in the Fourier transform. Primary beam correction was then applied to the resulting feathered cubes. No relative calibration factor was applied during feathering, implicitly assuming that there was no systematic offset in the absolute calibration of the datasets.

Feathering Fourier transforms the two images, adds the two filtered Fourier cubes, and trans- forms back. Feathering two images is a simple way to combine the two data sets, but can be sensitive to the shape of the tapering function and to relative calibration uncertainties. Using a cleaned image that has the single dish data as a starting model, rather than the cleaned VLA image alone, mitigates these effects because the cleaning process can correct any potential issues with the feathered image that might be inconsistent with the interferometric visibilities. Combining the data sets by using a clean model alone, however, is also sensitive to calibration errors and can result in a cube with more total flux than the single dish alone, which is not physical. Feathering these resulting cubes with the single-dish brings the total flux in the final combined images back into agreement with the single-dish data.

Using both methods produced a result that is as consistent as possible with the both the GBT and VLA data taken individually. The RMS noise was approximately the same as the VLA-only cubes. The synthesized beams, and thus the conversion between brightness temperature and flux density, were exactly the same as for the VLA data alone (Table \ref{t3}).

\begin{figure}
\begin{center}
\includegraphics[width=1\textwidth]{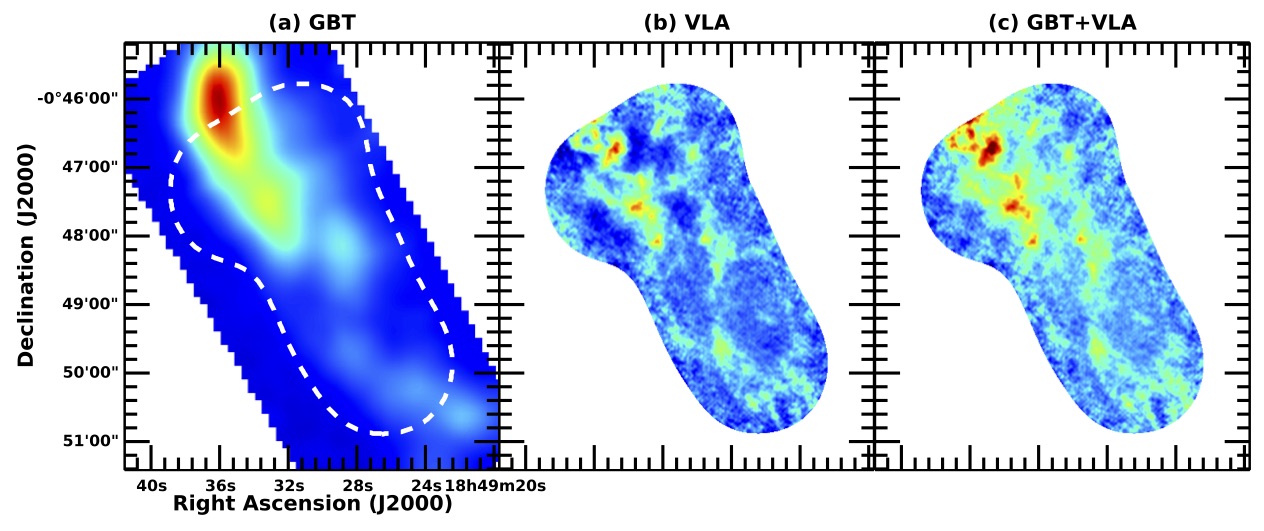}
\caption{A comparison of the NH$_{3}$ (1,1) data for the GBT-only, VLA-only, and combined images. Images of G031.97+00.07 are integrated from 94 km s$^{-1}$ to 97 km s$^{-1}$. The VLA footprint is plotted on the GBT image as a reference. The VLA and combined images have both been corrected for the response of the primary beam and are shown on the same flux scale (all three flux scales are linear). Aside from the improved resolution over the GBT along, the combined image also recovers more extended emission than the VLA alone. The combined image recovers the total flux of the GBT image, and recovers more flux than the VLA for even relatively compact, marginally resolved sources. \label{f2}}
\end{center}
\end{figure}

As a check on the flux in our combined cubes, we smoothed them to the resolution of the GBT and divided by the GBT cubes. This should have produced a cube of values close to 1, as the GBT is sensitive to total flux. We found that within regions of significant emission, our smoothed, combined cubes were almost always within the 20\% flux uncertainty of the GBT. Our final cubes were therefore within the uncertainty of recovering the correct total flux. Comparing to the VLA data cubes alone, the morphology of the combined data cubes is similar on size scales approximately the size of the synthesized beam, however using the VLA alone misses most the extended, diffuse emission. The fluxes of clump-sized sources seen in the VLA are tens of percent lower than what is seen in the combined images because of the contribution of this diffuse gas across the IRDCs. The total flux in the VLA cubes is typically less than half of the total flux measured by the GBT, and thus the combined images. An image showing the comparison between GBT-only, VLA-only, and GBT+VLA images is shown in Figure \ref{f2}.

\section{Methods \label{sec-methods}}

\subsection{Clump Deconvolution \label{sec-clump}}

Detailed knowledge of the kinematic and spatial structure of these IRDCs provides constraints on the NH$_{3}$ spectral line fitting (see \S \ref{sec-line}). It is readily apparent in the data that regions containing at least two strong, distinct velocity components, if not three or four, are common. Attempts to fit a single velocity component to the NH$_{3}$ spectra in these regions result in poor fits with unphysically large linewidths and optical depths, so multiple components must be included. By performing clump deconvolution on the data before fitting, it is trivial to determine the number of components to include at each line of sight. Sophisticated clump deconvolution algorithms make use of the data cube and its noise properties as a whole, and so are more robust determinations of the number of components to fit than any method making this determination on a line of sight (pixel-by-pixel) basis. Additionally, identifying significant, coherent emission in the data allows us to make very good initial estimates of the central velocities and velocity widths at each line of sight for each component by making first- and second-order moment maps (velocity field and velocity dispersion, respectively) restricted to emission within the identified clump. Finally, using these clumps provides a straightforward and physically motivated way to analyze the physical parameter results from the fitting for different substructures within the IRDCs.

We perform clump deconvolution on the NH$_{3}$ (1,1) data using the \texttt{cprops} package described by \cite{2006PASP..118..590R}. We use the main hyperfine component of the (1,1) line because it typically has the highest signal-to-noise ratio and the (2,2) line does not trace the coldest, and often lowest column density, gas. The velocity offsets between cospatial velocity components are sometimes comparable to the NH$_{3}$ hyperfine splitting, so the deconvolution must be performed carefully. We use the GBT and VLA combined data cubes before primary beam correction so the noise is roughly constant across the images to generate the clump assignment cubes. The VLA primary beam response was applied to the GBT data before combination with the uncorrected VLA data specifically to allow the cubes to be used for clump deconvolution in this way; applying the VLA primary beam response to the final combined data cubes would be equivalent. The initial mask only includes voxels (single elements in the position-position-velocity cubes, akin to pixels in position-position images) with values greater than 7$\sigma$, and then the mask is expanded to include voxels above 5$\sigma$ that are connected to the initial mask via a path only passing through significant emission. The \texttt{cprops} algorithm then identifies ``kernels'' in the data, consisting of local maxima significantly above a ``merge level'' -- the contour value at which multiple kernels are connected in the data and the significant emission cannot be uniquely assigned to one kernel over another. The \texttt{cprops} documentation refers to this unassigned emission as the ``watershed.'' We further restrict the list of kernels to those that lead to clumps with projected area on the sky greater than three times the synthesized beam area, i.e. only well-resolved clumps are assigned as independent structures so that we can accurately probe their physical properties. Kernels must also result in clumps that extend over at least 3 velocity channels (recall channels are 0.33 km s$^{-1}$). Kernels that lead to these unresolved clumps are rejected before the final \texttt{cprops} assignments are determined, so the voxels, and thus emission, from these clumps is free to be reassigned to another clump or the watershed. This rejection ensures that all clumps are resolved, effectively rejecting cores. This algorithm produces clump assignments from our data that map excellently to clumps discerned by eye.

An alternate method in the \texttt{cprops} package uses the \texttt{clumpfind} algorithm \citep{1994ApJ...428..693W} to assign emission to the same kernels list as the standard \texttt{cprops} algorithm, but proceeds by assigning voxels to clumps in discrete contours continuing all the way to the noise floor. Assignment degeneracies in this method are broken by evaluating the proximity of voxels to the clump peaks in position-position-velocity space. This method has the advantage of assigning all of the significant emission into clumps, however the assignment cubes tend to have a ``patchwork'' appearance that likely does not represent physically accurate clump boundaries. The \texttt{clumpfind} algorithm is also fairly sensitive to the input parameters that determine the contouring scheme used to make the clump assignments. This will in turn affect the clump parameters we calculate in \S \ref{sec-discussion}. The \texttt{cprops} algorithm is generally less sensitive than \texttt{clumpfind} to the inputs. The specific assignments can be altered by varying the parameters, but the typical clump sizes, aspect ratios, etc. are not significantly affected.

For our study, we start with the standard \texttt{cprops} assignments, and then further used the \texttt{clumpfind} assignments to assign the watershed emission. The result is an assignment cube in which all of the significant emission is assigned to exactly one clump, and the patchwork assignments from the clumpfind algorithm are restricted to the weakest emission. The clump-averaged properties discussed in this work will be dominated by the strongest emission, and thus by the \texttt{cprops} assignments. Results of the deconvolution are presented in \S \ref{sec-cdresults} and discussed in \S \ref{sec-clumpstab}.

\subsection{IR Extinction \label{sec-extinction}}

IRDCs are initially identified by their apparent extinction in MIR images, so it is natural to use their contrast with the background emission as a measure of physical properties in the IRDC. In the simplest terms, a greater contrast, i.e. darker cloud compared to the background, indicates a greater column of dust. It is rather straightforward to generate extinction maps if one has an estimate of the background, and with a few assumptions the dust mass surface density may also be calculated across a map.

Any method of estimating this optical depth must account for the following complications: (1) we do not have a direct measurement of the background IR emission at the location of the IRDC, so it must be estimated from an irregular, varying background measured off the IRDC while attempting to avoid contamination from foreground objects, such as other dark clouds and bright nebulae; (2) there exists foreground emission from the dust between the IRDC and the observer; and (3) IRDCs often contain embedded point sources that do not probe the full column of the IRDCs and contaminate our contrast measurements.

We adopt a modified version of the Large Median Filter (LMF) method presented by \cite{2009ApJ...696..484B} to map the 24 $\micron$ optical depth, $\tau_{24\mu\mathrm{m}}$, using the \emph{Spitzer} MIPSGAL data. This method is summarized in detail in Appendix \ref{sec-irextmaps}. In our optical depth maps, the IRDCs typically peak at optical depth of about 0.25-0.5. G028.23-00.19 has noticeably higher optical depth than the other IRDCs by about a factor of two, and peaks at an approximate optical depth of 1. Further results of this analysis are presented in \S \ref{sec-irresults}.

\subsection{Ammonia Spectral Line Fitting \label{sec-line}}

We fit the spectra along individual lines of sight in our combined GBT and VLA data. The (1,1) and (2,2) lines were fit simultaneously. The fitting routine was written in Python using the \texttt{nmpfit} package, which performed a least-squares (Levenberg-Marquardt algorithm) fit and returned the fit parameters and the full covariance matrix. We include the full hyperfine structure of NH$_{3}$ from the components and intrinsic strengths documented by \cite{1967PhRv..156...83K}. The limited bandwidth of the VLA data restricted our fit in practice to the main and inner satellites components of the spectral lines. We simultaneously fit the central velocity ($v_{c,\mathrm{LSR}}$), velocity FWHM ($\Delta v$), total optical depth in the (1,1) component ($\tau_{0}(1,1)$, abbreviated henceforth as $\tau_{0}$), excitation temperature ($T_{\mathrm{ex}}$), and rotation temperature ($T_{\mathrm{R}}$). Note that here we use $\tau_{0}$ for the total opacity in the entire (1,1) line, which is precisely a factor of two larger than the total opacity in the main component of the (1,1) line.

There is a natural degeneracy between the excitation temperature and optical depth if they are low. However, the main hyperfine component is typically optically thick in this study (see \S \ref{sec-slresults}), and so the degeneracy is broken. Since we observe lines that are so optically thick ($\tau_{0} > 5$), our fitted values of $\tau_{0}$ may only be lower limits. The fitting routine is capable of handling multiple velocity components simultaneously and independently along the same line of sight (see Section \ref{sec-clump}).

\begin{figure}
\begin{center}
\includegraphics[width=1\textwidth]{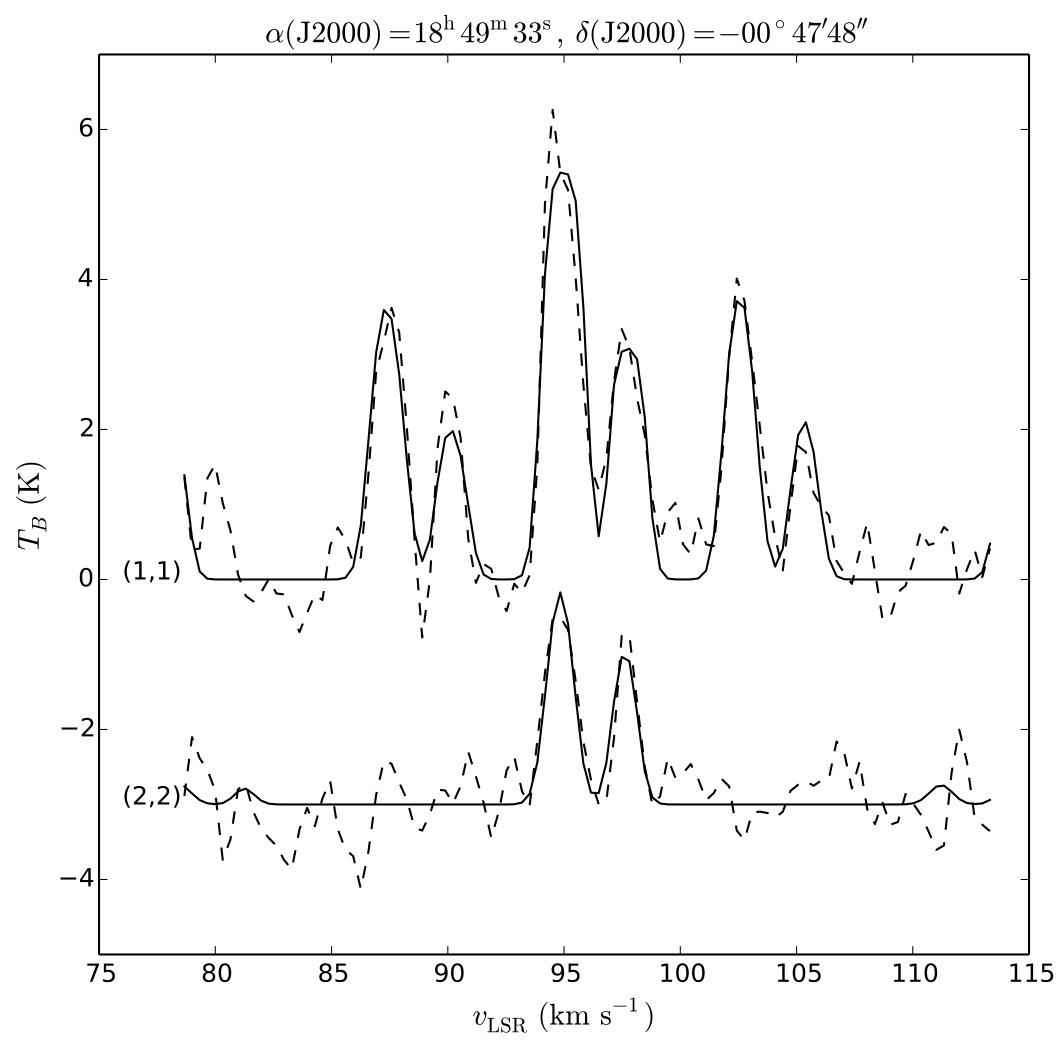}
\caption{An example spectrum of the NH$_{3}$ (1,1) and (2,2) lines with fit to the spectrum. This spectrum is from one line of sight toward G031.97+00.07 with two distinct velocity components. The data are the dashed lines and the fit is the solid line. The (2,2) spectrum and fit have been vertically offset by -3 K for clarity. The brightness temperature scale assumes a beam filling factor of unity. \label{f3}}
\end{center}
\end{figure}

The details of the spectral line fitting routine and the determination of physical parameters, including the kinetic temperature, $T_{K}$, and the column density of ammonia are presented in Appendix \ref{sec-linefit}. A list of key parameters for the analysis, including the parameters used in the spectral line fitting, is given in Table \ref{t4}. Results of the spectral line fitting are presented in \S \ref{sec-slresults}. An example spectrum and fit of a single line of sight with two distinct velocity components is shown in Figure \ref{f3}.

\begin{deluxetable}{rl}
\tablecolumns{2}
\tablewidth{0pt}
\tablecaption{Selected Physical Parameters of Interest \label{t4}}
\tablehead{
 \colhead{Parameter} &
 \colhead{Description}
}
\startdata
\multicolumn{2}{c}{Line of Sight Parameters} \\
\tableline
\sidehead{Calculated Directly from the Radio Data:}
$I_{(1,1)}$ & Moment 0 (integrated intensity) of the NH$_{3}$ (1,1) line \\
$\langle v_{(1,1)} \rangle$ & Moment 1 (intensity-weighted velocity field) of the NH$_{3}$ (1,1) line \\
$\langle \sigma_{v,(1,1)} \rangle$ & Moment 2 (intensity-weighted velocity dispersion) of the NH$_{3}$ (1,1) line \\
$I_{\mathrm{CCS}}$ & Moment 0 (integrated intensity) of the CCS line \\
\sidehead{Calculated Directly from the IR Data:}
$\tau_{24\mu\mathrm{m}}$ & Optical depth at 24 $\micron$ from IR extinction \\
\sidehead{From NH$_{3}$ (1,1) and (2,2) Spectral Line Fitting\tablenotemark{a}:}
$\nu_{c}$ & Central Doppler shifted NH$_{3}$ frequency \\
$\Delta \nu$ & NH$_{3}$ frequency FWHM \\
$\tau_{0}$ & Total optical depth in the NH$_{3}$ (1,1) line \\
$T_{\mathrm{ex}}$ & NH$_{3}$ excitation temperature \\
$T_{\mathrm{R}}$ & NH$_{3}$ rotation temperature \\
$\chi^{2}$ & Goodness of fit statistic \\
\sidehead{Calculated from Spectral Line Fitting Results\tablenotemark{a}:}
$T_{K}$ & Kinetic Temperature \\
$v_{c,\mathrm{LSR}}$ & Central NH$_{3}$ LSR velocity \\
$\Delta v$ & NH$_{3}$ velocity FWHM \\
$\sigma_{\mathrm{obs}}$ & Observed line of sight velocity dispersion \\
$\sigma_{\mathrm{line}}$ & Line of sight velocity dispersion corrected for spectral resolution \\
$\sigma_{T}$ & NH$_{3}$ Thermal velocity dispersion \\
$\sigma_{\mathrm{NT}}$ & NH$_{3}$ Nonthermal velocity dispersion \\
$\mathcal{M}$ & Nonthermal (turbulent) Mach number \\
$N(\mathrm{NH}_{3})$ & Total column density of NH$_{3}$ \\
$N(\mathrm{H}_{2})$ & Column density of molecular Hydrogen \\
\tablebreak
\multicolumn{2}{c}{Clump-by-clump Parameters} \\
\tableline
\sidehead{Directly from \texttt{cprops} for each clump:}
R.A. (J2000) & R. A. of the peak NH$_{3}$ (1,1) main beam temperature \\
Decl. (J2000) & Decl. of the peak NH$_{3}$ (1,1) main beam temperature \\
$R_{\mathrm{eff}}$ & Effective radius of a circle with the same projected area as the clump \\
$A_{0}$ & (Initial) Aspect ratio projected onto the sky \\
\sidehead{Calculated clump-by-clump:}
$M_{\mathrm{cl}}$ & Clump mass from total NH$_{3}$ and $X(\mathrm{NH}_{3})$ \\
$t_{\mathrm{ff,sph}}$ & Spherical free-fall time \\
$t_{\mathrm{ff,cyl}}$ & Cylindrical free-fall time along the axis \\
$M_{\mathrm{vir,sph}}$ & Spherical virial mass \\
$M_{\mathrm{vir,cyl}}$ & Cylindrical virial mass \\
$\alpha_{\mathrm{vir,sph}}$ & Spherical virial parameter \\
$\alpha_{\mathrm{vir,cyl}}$ & Cylindrical virial parameter \\
$B_{\mathrm{cr}}$ & Minimum magnetic field strength to support clump against collapse \\
\tableline
\multicolumn{2}{c}{Whole IRDC Parameters} \\
\tableline
$D$ & Distance to IRDC \\
$f_{\mathrm{fore}}$ & Fraction of IR galactic emission that is foreground to the IRDC \\
$M_{\mathrm{IR}}$ & Mass lower limit from 24 $\micron$ extinction \\
$M_{\mathrm{13CO}}$ & Mass lower limit from BU-GRS $^{13}$CO ($J$=1-0) \\
$M_{\mathrm{mm}}$ & Mass from BGPS 1.12 mm emission \\
\enddata
\tablenotetext{a}{These parameters may also be evaluated as clump-averaged values, weighted by the $(\chi^{2}_{\mathrm{red}})^{1/2}$ statistic over the clump. Both line of sight and clump-averaged values separate different velocity components that overlap spatially.}
\end{deluxetable}

\section{Results \label{sec-results}}

\begin{figure}
\begin{center}
\includegraphics[width=0.7\textwidth]{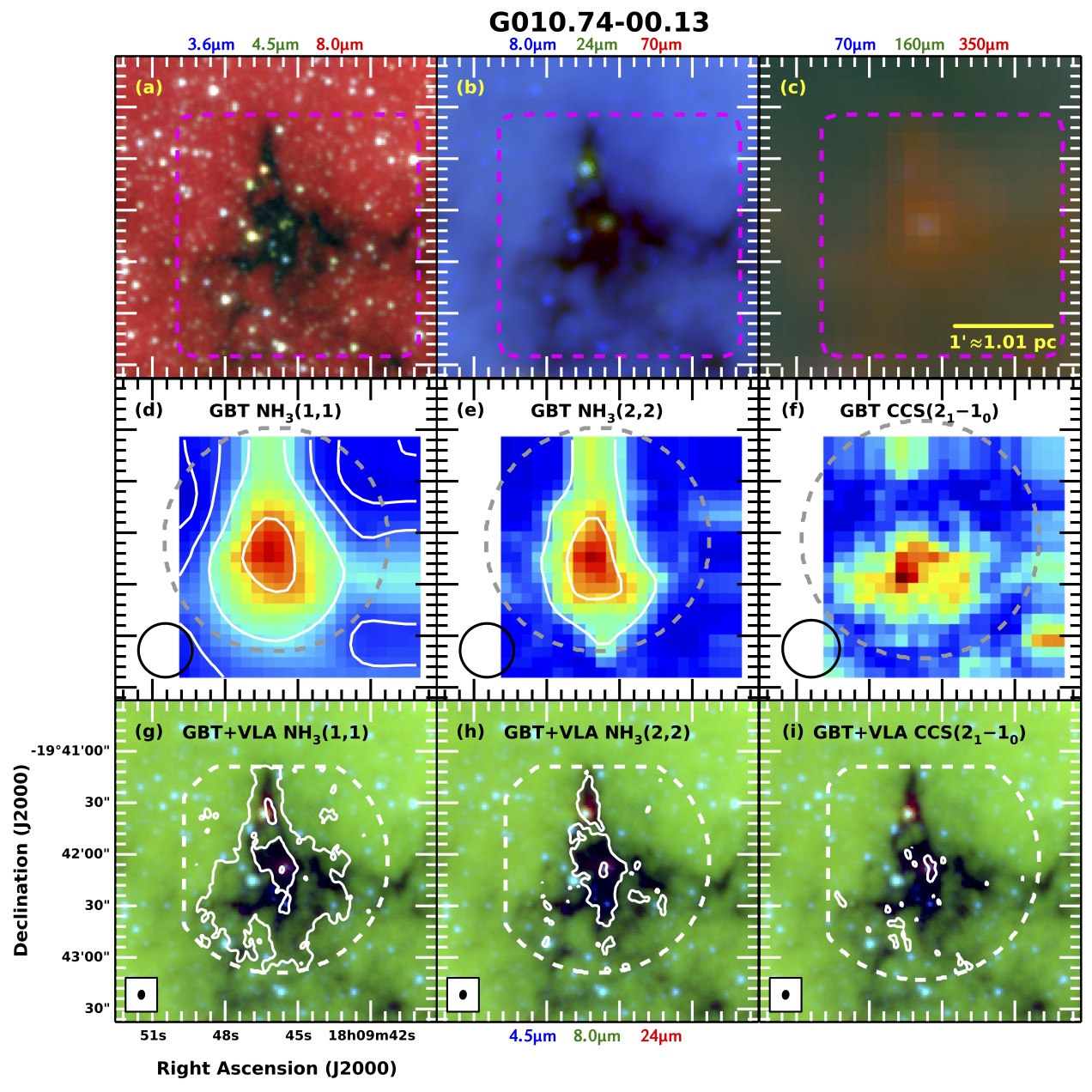}
\caption{\emph{Spitzer Space Telescope} and \emph{Herschel Space Observatory} images of G010.74-00.13, along with our GBT and VLA data. The IRDC is seen against the diffuse mid-infrared emission of the galaxy toward the shorter wavelengths (panels (a) and (b)), though embedded infrared point sources are visible. At longer wavelengths (panel (c)), the IRDC becomes visible in the thermal emission of the dust. The highest 24 $\micron$ optical depth, $\tau_{24 \mu\mathrm{m}}$, that we calculate in this IRDC is 0.42. A yellow scalebar representing 1$\arcmin$ with the physical size at our adopted distance is shown at the lower right of (c). The GBT footprint is overplotted in dashed magenta in panels (a)-(c). Panels (d)-(f) show the NH$_{3}$ (1,1), NH$_{3}$ (2,2), and CCS emission, respectively, as seen by the GBT. An arbitrary linear color stretch is shown with white contours showing the signal-to-noise ratio at 5$\sigma$, 10$\sigma$, 20$\sigma$, and 40$\sigma$. The size of the GBT beam is shown in the lower left of each panel, and the VLA footprint is overplotted in dashed gray. Panels (g)-(i) show \emph{Spitzer} images of G010.74-00.13 with contours of the combined GBT and VLA NH$_{3}$ (1,1), NH$_{3}$ (2,2), and CCS emission, respectively. The contours show the signal-to-noise ratio at 5$\sigma$, 25$\sigma$, and 50$\sigma$. The NH$_{3}$ generally traces the MIR extinction and peaks near IR point sources. The NH$_{3}$ (2,2) is generally less extended than the (1,1), as it traces warmer gas. The CCS, though weak, typically follows the highest extinction parts of the IRDCs when it is detected. The combined footprint of the GBT and VLA observations (their intersection) is shown in dashed white. The synthesized beam is shown at the lower left of each panel. \label{f4}}
\end{center}
\end{figure}

\begin{figure}
\begin{center}
\includegraphics[width=0.7\textwidth]{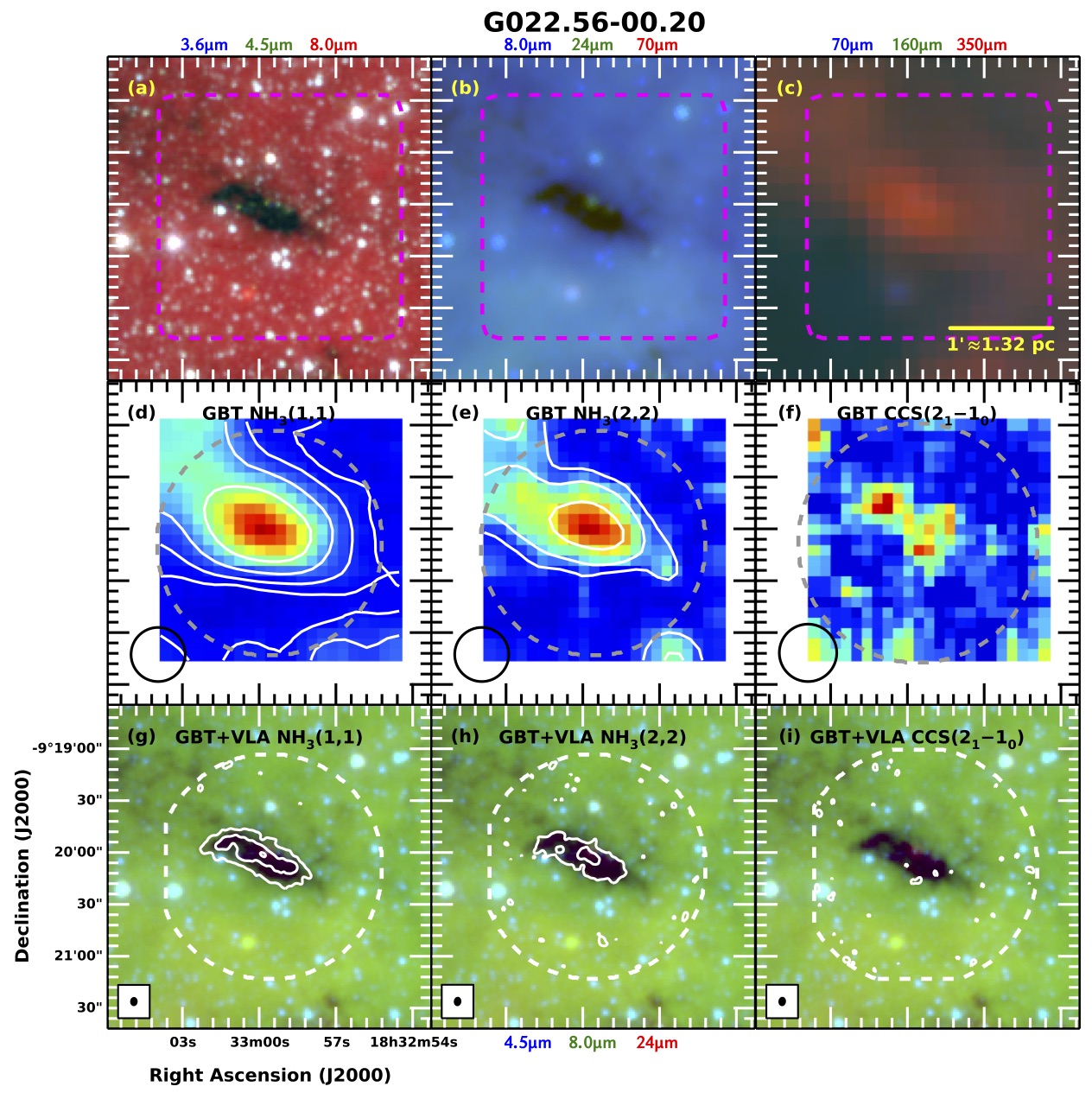}
\caption{Same as Figure \ref{f4} but for G022.56-00.20. The highest value of $\tau_{24 \mu\mathrm{m}}$ is 0.27. \label{f5}}
\end{center}
\end{figure}

\begin{figure}
\begin{center}
\includegraphics[width=0.7\textwidth]{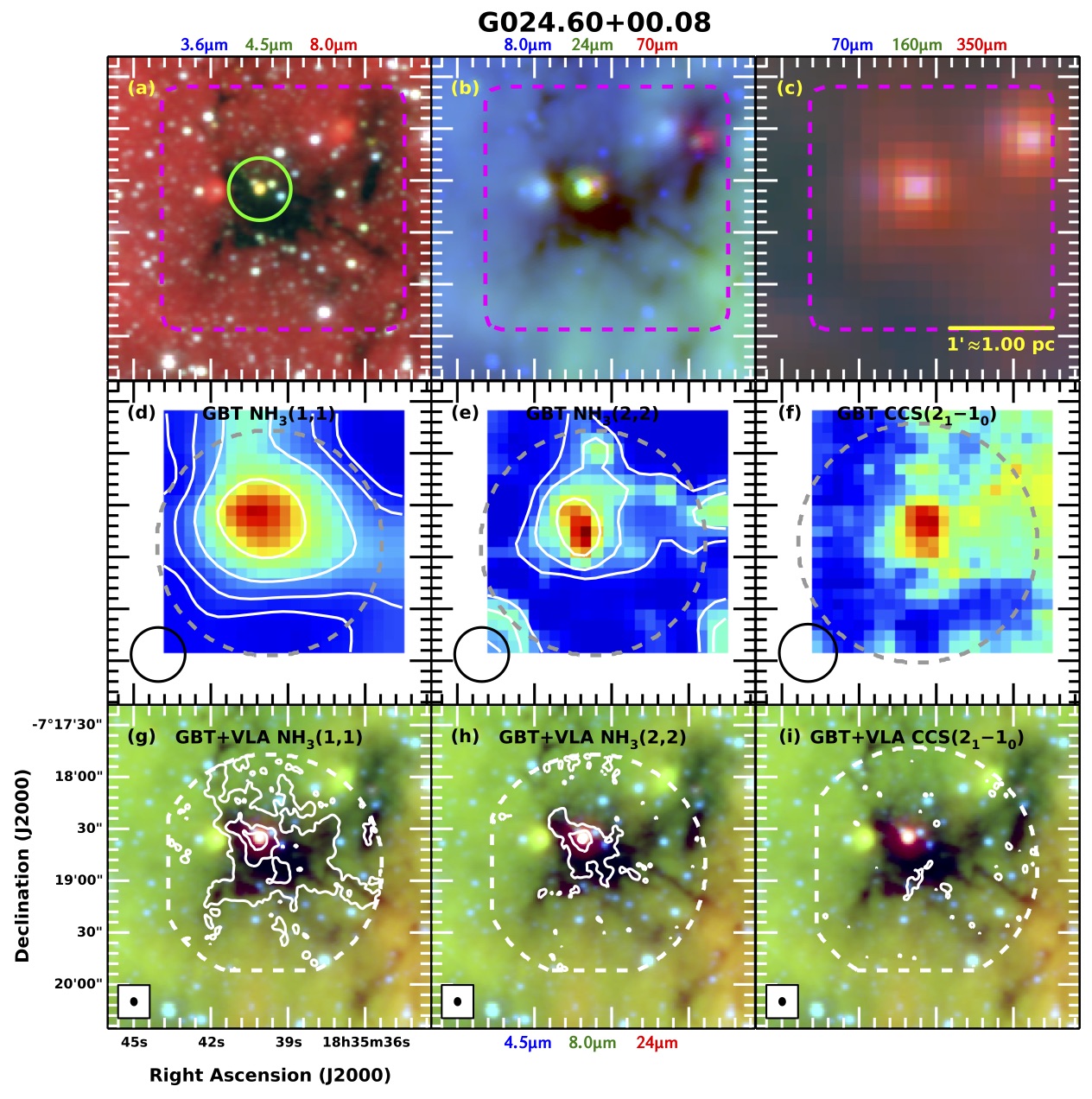}
\caption{Same as Figure \ref{f4} but for G024.60+00.08. The highest value of $\tau_{24 \mu\mathrm{m}}$ is 0.47. The green circle in panel (a) shows the location of the extended green object (EGO) identified by \cite{2008AJ....136.2391C}. \label{f6}}
\end{center}
\end{figure}

\begin{figure}
\begin{center}
\includegraphics[width=0.7\textwidth]{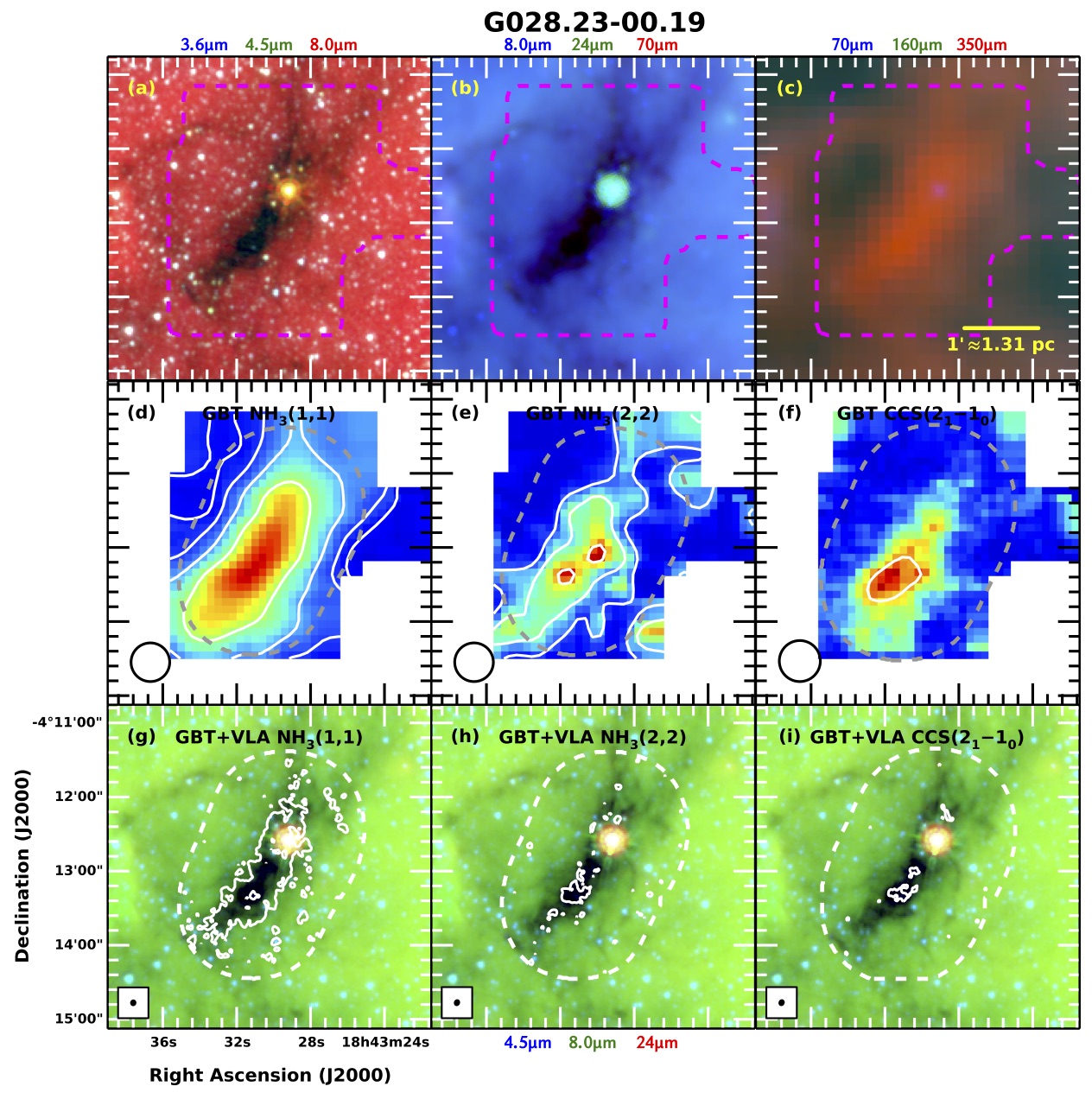}
\caption{Same as Figure \ref{f4} but for G028.23-00.19 with GBT+VLA contours at 5$\sigma$ and 25$\sigma$. The bright point source is an unrelated, foreground late-type star \citep{1989ApJ...347..325B}. The highest value of $\tau_{24 \mu\mathrm{m}}$ is 0.89. \label{f7}}
\end{center}
\end{figure}

\begin{figure}
\begin{center}
\includegraphics[width=0.7\textwidth]{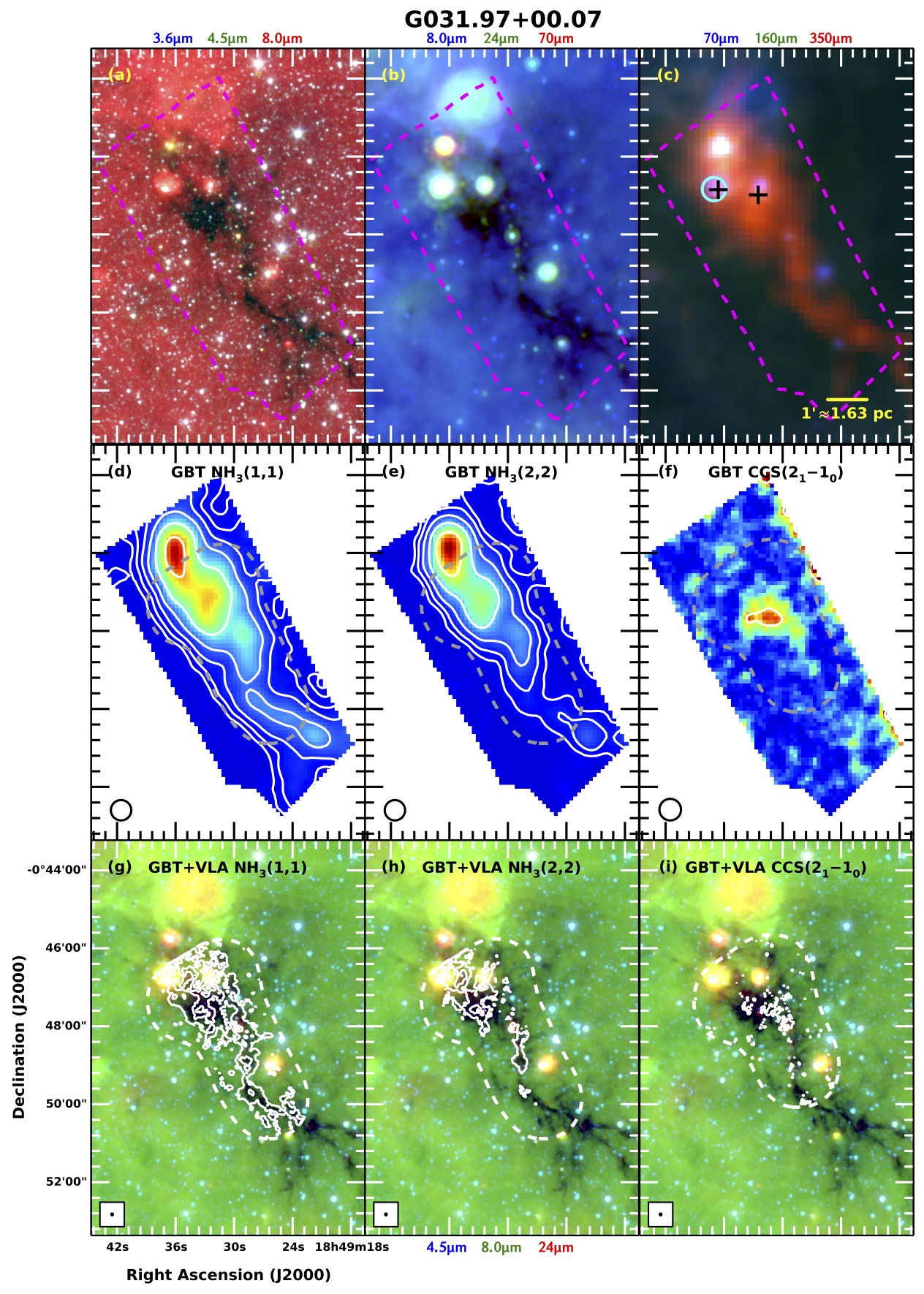}
\caption{Same as Figure \ref{f4} but for G031.97+00.07 with GBT contours at 5$\sigma$, 10$\sigma$, 20$\sigma$, 40$\sigma$, 80$\sigma$, and 160$\sigma$, and with GBT+VLA contours at 5$\sigma$, 25$\sigma$, 50$\sigma$, and 100$\sigma$. The highest value of $\tau_{24 \mu\mathrm{m}}$ is 0.43. The black crosses in panel (c) are H$_{2}$O masers reported by \cite{2006ApJ...651L.125W}. The cyan circle shows the location of the H {\smaller II} region identified by \cite{2009A&A...501..539U}. \label{f8}}
\end{center}
\end{figure}

\begin{figure}
\begin{center}
\includegraphics[width=0.7\textwidth]{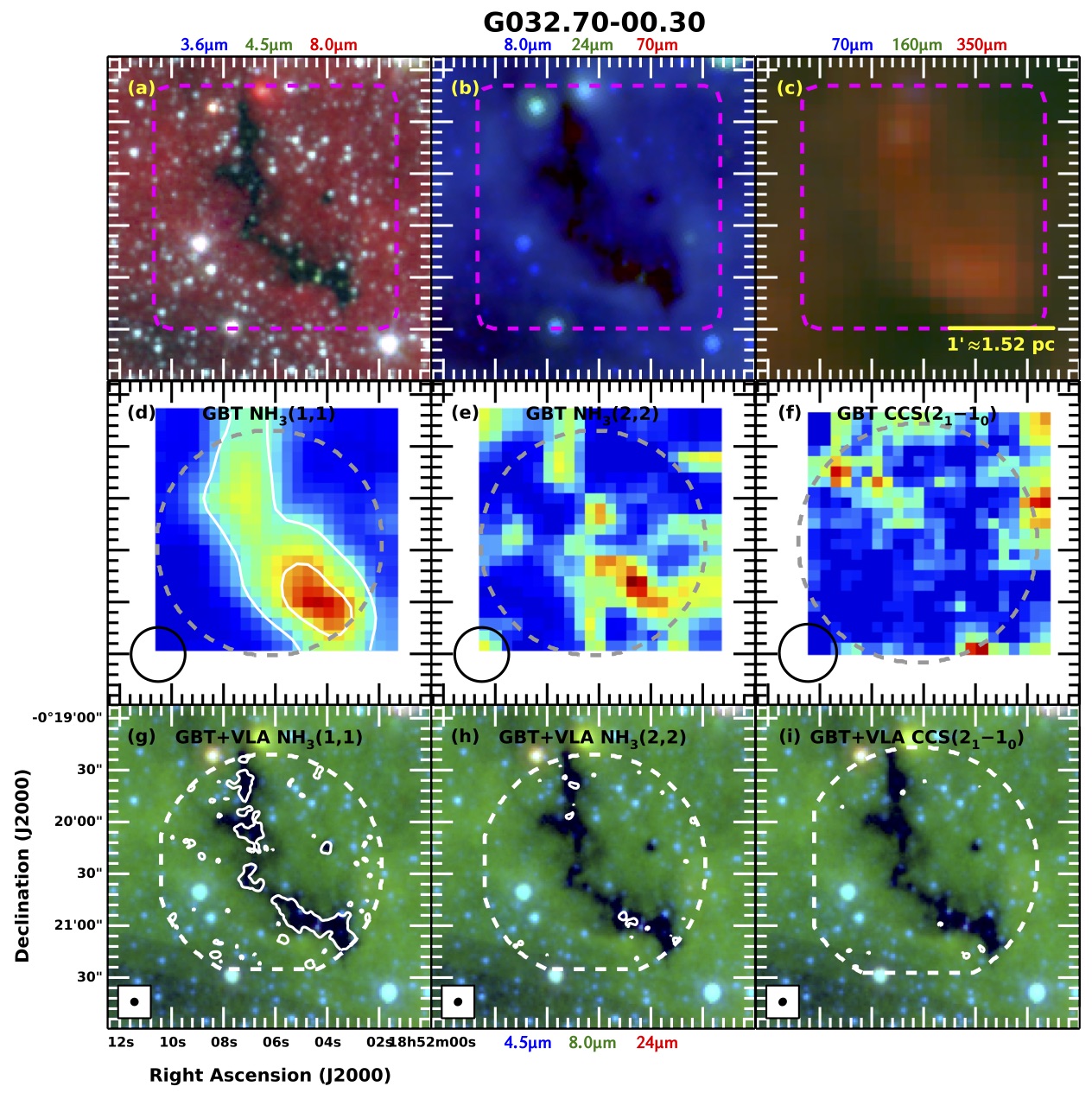}
\caption{Same as Figure \ref{f4} but for G032.70-00.30 with GBT contours at 5$\sigma$ and 10$\sigma$, and with GBT+VLA contours at 5$\sigma$. The highest value of $\tau_{24 \mu\mathrm{m}}$ is 0.36. \label{f9}}
\end{center}
\end{figure}

\begin{figure}
\begin{center}
\includegraphics[width=0.7\textwidth]{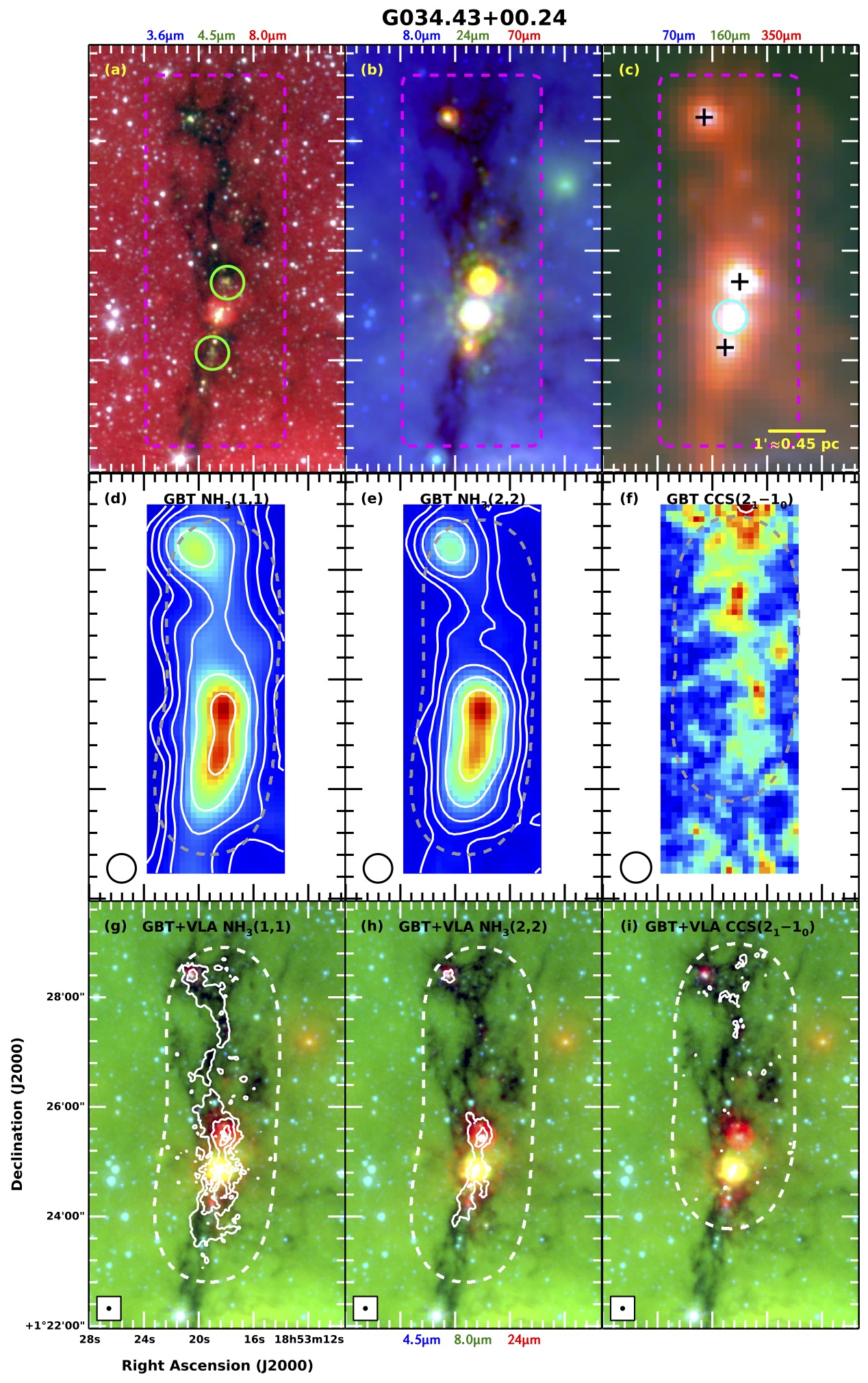}
\caption{Same as Figure \ref{f4} but for G034.43+00.24 with GBT contours at 5$\sigma$, 10$\sigma$, 20$\sigma$, 40$\sigma$, 80$\sigma$, and 160$\sigma$, and GBT+VLA contours at 5$\sigma$, 25$\sigma$, 50$\sigma$, and 100$\sigma$. The highest value of $\tau_{24 \mu\mathrm{m}}$ is 0.14. The green circles in panel (a) show the location of the extended green objects (EGOs) identified by \cite{2008AJ....136.2391C}. The black crosses in panel (c) are H$_{2}$O masers reported by \cite{2006ApJ...651L.125W}. The cyan circle shows the location of the H {\smaller II} region identified by \cite{2009A&A...501..539U}. \label{f10}}
\end{center}
\end{figure}

\begin{figure}
\begin{center}
\includegraphics[width=0.7\textwidth]{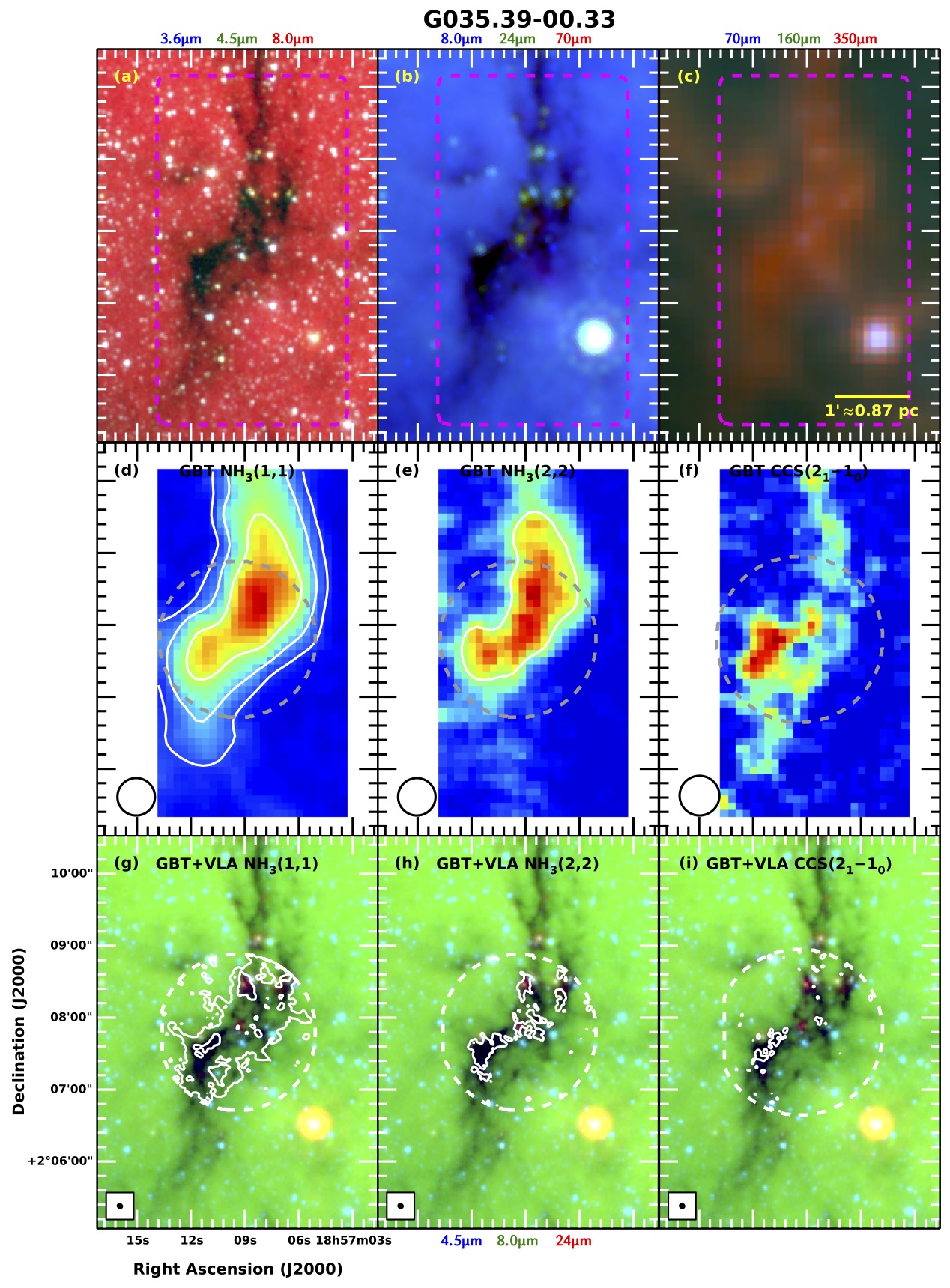}
\caption{Same as Figure \ref{f4} but for G035.39-00.33 with GBT contours at 5$\sigma$, 10$\sigma$, and 20$\sigma$. The highest value of $\tau_{24 \mu\mathrm{m}}$ is 0.44. \label{f11}}
\end{center}
\end{figure}

\begin{figure}
\begin{center}
\includegraphics[width=0.7\textwidth]{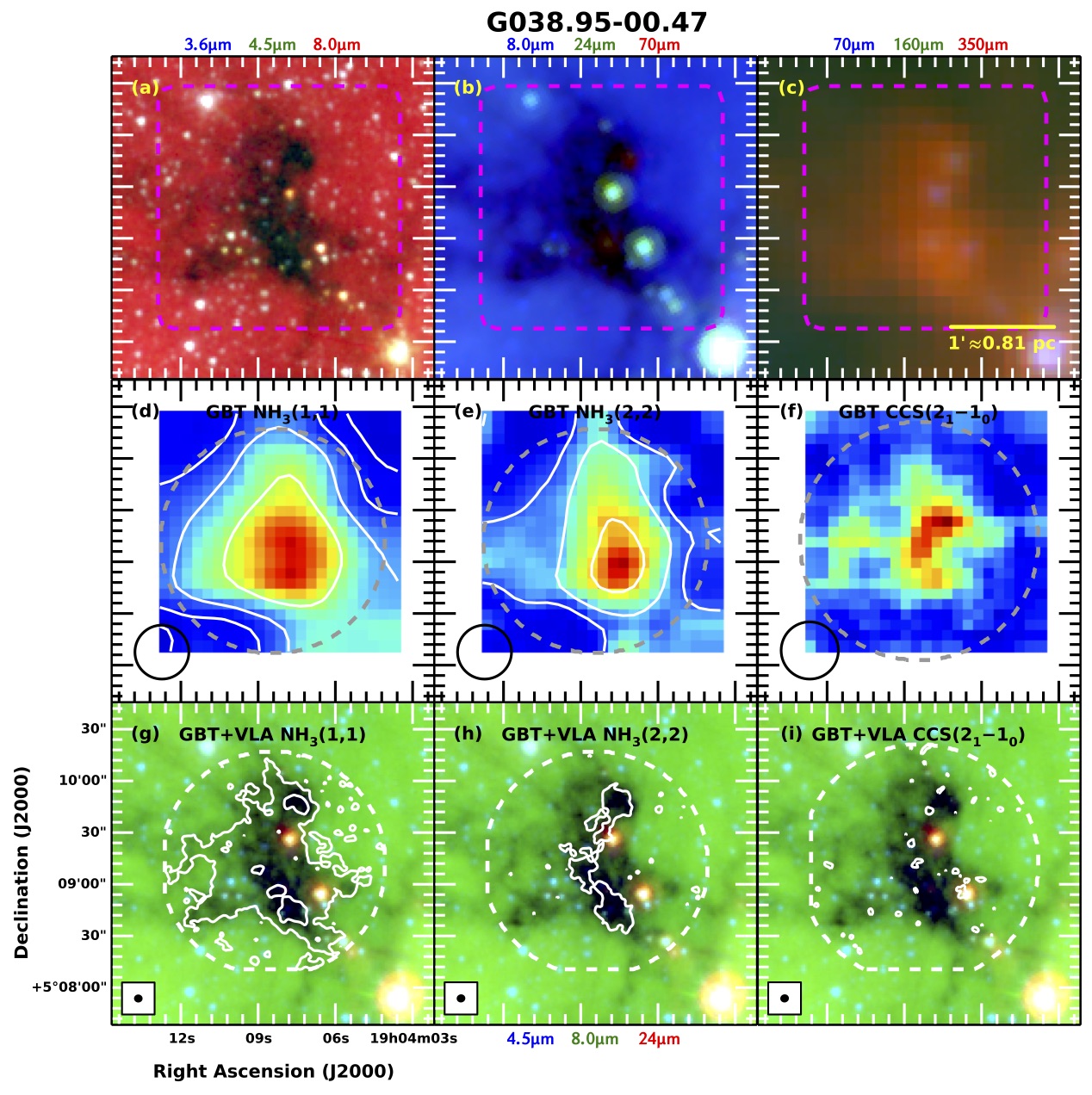}
\caption{Same as Figure \ref{f4} but for G038.95-00.47 with GBT+VLA contours at 5$\sigma$ and 25$\sigma$. The highest value of $\tau_{24 \mu\mathrm{m}}$ is 0.34. \label{f12}}
\end{center}
\end{figure}

Images of the sample from the \emph{Spitzer Space Telescope} and the \emph{Herschel Space Observatory} with the GBT and VLA data are shown in Figures \ref{f4}, \ref{f5}, \ref{f6}, \ref{f7}, \ref{f8}, \ref{f9}, \ref{f10}, \ref{f11}, and \ref{f12}. The NH$_{3}$ (1,1) distribution generally traces the IR extinction very closely and peaks around infrared point sources and, to a lesser extent, the IR extinction peaks. The NH$_{3}$ (2,2) distribution is more compact and generally correlates with stronger (1,1) emission. The CCS emission had systematically lower signal-to-noise ratio ($<5\sigma$) than the NH$_{3}$ and is typically only marginally detected. The CCS that is observed, however, does not typically cover the full spatial extent of the NH$_{3}$. Extended CCS emission may be resolved out by the VLA, however the GBT images in Figures \ref{f4}-\ref{f12} show that frequently the strongest CCS emission is not cospatial with the strongest NH$_{3}$ emission.

\subsection{Clump Deconvolution Results \label{sec-cdresults}}

A summary of the clump properties is presented in Table \ref{t5}. Coordinates and velocities are given at the peaks of the emission. The effective radius, $R_{\mathrm{eff}}$, is the radius of a circle on the sky that has the same projected area as the full extent of the clump using the clump assignment from \texttt{cprops}, which ranges from approximately 0.02 pc to 0.28 pc in our sample, with a median of about 0.16 pc. The physical resolution (i.e. synthesized beam) of the (1,1) cubes ranges from 0.02 pc to 0.1 pc, so clumps are typically well resolved.

\begin{deluxetable}{cccccccccccc}
\tabletypesize{\scriptsize}
\tablecolumns{12}
\tablewidth{0pt}
\tablecaption{Clump Properties \label{t5}}
\tablehead{
 \colhead{Parent} &
 \multicolumn{3}{c}{Peak Coordinates} &
 \multicolumn{2}{c}{\texttt{cprops}} &
 \colhead{\emph{Herschel}} &
 \multicolumn{4}{c}{Spectral Line Fitting} &
 \colhead{\emph{Spitzer}}  \\
 \colhead{IRDC\tablenotemark{a}} &
 \colhead{R.A. (J2000)} &
 \colhead{Decl. (J2000)} &
 \colhead{$v_{\mathrm{LSR}}$} &
 \colhead{$R_{\mathrm{eff}}$} &
 \colhead{$A_{0}$} &
 \colhead{70 $\micron$} &
 \colhead{$\langle \sigma_{\mathrm{line}} \rangle$} &
 \colhead{$\langle \tau_{0} \rangle$} &
 \colhead{$\langle T_{K} \rangle$} &
 \colhead{$\langle N(\mathrm{NH}_{3}) \rangle$} &
 \colhead{$\langle \tau_{24 \mu\mathrm{m}} \rangle$\tablenotemark{b}} \\
 &
 \colhead{hh:mm:ss} & 
 \colhead{$\phantom{-}$dd:mm:ss} & 
 \colhead{(km s$^{-1}$)} &
 \colhead{(pc)} &
 &
 \colhead{source?} &
 \colhead{(km s$^{-1}$)} &
 &
 \colhead{(K)} &
 \colhead{($10^{23}$ cm$^{-2}$)} &
 }
\startdata
G10 & 18:09:45 & -19:42:29 & 28.50 & 0.17 & 2.1 & N & 1.0 & 7.3 & 14.1 & 0.8 & 0.35 \\
G10 & 18:09:45 & -19:42:07 & 28.50 & 0.16 & 1.2 & Y & 1.0 & 8.3 & 15.1 & 1.1 & 0.34 \\
G10 & 18:09:44 & -19:42:06 & 29.16 & 0.11 & 2.1 & N & 0.7 & 7.9 & 14.0 & 0.6 & 0.37 \\
G10 & 18:09:46 & -19:41:47 & 29.82 & 0.16 & 2.5 & Y & 1.0 & 8.1 & 14.7 & 0.9 & 0.27 \\
G22 & 18:33:00 & -09:19:59 & 74.84 & 0.15 & 3.0 & Y & 1.3 & 6.7 & 16.5 & 1.0 & 0.20 \\
G22 & 18:33:01 & -09:19:56 & 76.82 & 0.09 & 1.6 & N & 1.6 & 6.0 & 15.7 & 0.9 & 0.19 \\
G22 & 18:33:00 & -09:20:04 & 78.14 & 0.09 & 2.8 & N & 1.6 & 5.1 & 16.3 & 0.7 & 0.22 \\
G24 & 18:35:40 & -07:18:36 & 52.49 & 0.12 & 1.4 & Y & 1.4 & 4.5 & 19.8 & 0.8 & 0.19 \\
G24 & 18:35:39 & -07:18:55 & 53.15 & 0.17 & 1.5 & N & 0.8 & 6.1 & 14.0 & 0.6 & 0.28 \\
G24 & 18:35:40 & -07:19:06 & 53.81 & 0.09 & 1.2 & N & 0.6 & 6.1 & 14.2 & 0.4 & 0.23 \\
G24 & 18:35:40 & -07:18:31 & 53.81 & 0.10 & 1.2 & N & 1.0 & 4.5 & 18.1 & 0.6 & 0.15 \\
G24 & 18:35:41 & -07:18:03 & 53.81 & 0.03 & 3.4 & N & 0.3 & 8.4 & 11.7 & 0.6 & 0.06 \\
G24 & 18:35:41 & -07:19:08 & 54.14 & 0.05 & 1.6 & N & 0.4 & 5.6 & 16.0 & 0.4 & 0.19 \\
G28 & 18:43:31 & -04:13:23 & 79.15 & 0.09 & 1.5 & N & 1.2 & 12.2 & 13.9 & 1.5 & 0.84 \\
G28 & 18:43:30 & -04:13:09 & 79.81 & 0.08 & 1.3 & N & 1.0 & 9.5 & 13.6 & 0.8 & 0.77 \\
G28 & 18:43:30 & -04:12:59 & 80.14 & 0.08 & 1.8 & N & 1.5 & 6.2 & 12.3 & 0.8 & 0.70 \\
G28 & 18:43:29 & -04:12:34 & 80.14 & 0.05 & 1.8 & N & 0.9 & 4.6 & 13.8 & 0.4 & \nodata \\
G28 & 18:43:29 & -04:12:27 & 80.14 & 0.06 & 2.9 & N & 1.1 & 3.5 & 14.6 & 0.4 & 0.44 \\
G28 & 18:43:30 & -04:12:30 & 80.80 & 0.06 & 4.4 & N & 1.1 & 9.9 & 13.1 & 0.9 & 0.63 \\
G28 & 18:43:30 & -04:13:32 & 81.13 & 0.05 & 1.7 & N & 0.8 & 8.5 & 14.0 & 0.7 & 0.67 \\
G28 & 18:43:29 & -04:12:51 & 81.13 & 0.04 & 1.8 & N & 1.2 & 13.0 & 11.4 & 1.2 & 0.56 \\
G31 & 18:49:29 & -00:48:53 & 93.85 & 0.18 & 1.3 & N & 1.0 & 7.6 & 15.3 & 0.9 & 0.15 \\
G31 & 18:49:26 & -00:49:52 & 94.18 & 0.09 & 2.0 & N & 0.6 & 7.3 & 12.9 & 0.5 & 0.29 \\
G31 & 18:49:31 & -00:47:09 & 94.18 & 0.10 & 1.2 & N & 2.3 & 4.3 & 18.9 & 0.9 & 0.12 \\
G31 & 18:49:28 & -00:48:30 & 94.51 & 0.17 & 1.5 & N & 1.2 & 4.7 & 18.1 & 0.6 & 0.18 \\
G31 & 18:49:29 & -00:48:03 & 94.51 & 0.17 & 1.9 & Y & 1.3 & 5.5 & 18.1 & 0.7 & 0.18 \\
G31 & 18:49:30 & -00:47:23 & 94.51 & 0.07 & 1.7 & N & 0.7 & 6.6 & 16.4 & 0.6 & 0.17 \\
G31 & 18:49:30 & -00:48:18 & 94.84 & 0.07 & 2.9 & N & 1.0 & 6.1 & 16.5 & 0.5 & 0.28 \\
G31 & 18:49:32 & -00:48:04 & 94.84 & 0.17 & 1.5 & N & 1.7 & 5.4 & 17.1 & 0.9 & 0.29 \\
G31 & 18:49:30 & -00:46:46 & 94.84 & 0.08 & 1.4 & N & 1.6 & 4.4 & 16.9 & 0.7 & 0.14 \\
G31 & 18:49:32 & -00:46:29 & 94.84 & 0.08 & 2.3 & N & 1.8 & 2.8 & 19.5 & 0.6 & \nodata \\
G31 & 18:49:26 & -00:50:28 & 95.50 & 0.06 & 3.4 & N & 1.1 & 6.8 & 12.7 & 0.8 & 0.28 \\
G31 & 18:49:31 & -00:47:17 & 95.50 & 0.10 & 1.3 & N & 1.8 & 3.6 & 17.3 & 0.6 & 0.25 \\
G31 & 18:49:33 & -00:47:15 & 95.50 & 0.16 & 1.7 & N & 2.2 & 5.4 & 18.4 & 1.1 & 0.14 \\
G31 & 18:49:24 & -00:50:07 & 95.83 & 0.09 & 3.3 & N & 1.0 & 5.6 & 13.8 & 0.5 & 0.29 \\
G31 & 18:49:34 & -00:46:44 & 95.83 & 0.28 & 1.1 & Y & 1.8 & 5.2 & 18.9 & 1.2 & 0.01 \\
G31 & 18:49:28 & -00:49:39 & 96.49 & 0.18 & 2.0 & N & 0.9 & 7.2 & 13.2 & 0.8 & 0.28 \\
G31 & 18:49:33 & -00:47:33 & 96.49 & 0.25 & 2.1 & N & 1.7 & 6.6 & 16.7 & 1.2 & 0.25 \\
G31 & 18:49:25 & -00:50:09 & 96.82 & 0.08 & 4.7 & N & 1.5 & 7.4 & 12.2 & 1.0 & 0.25 \\
G31 & 18:49:31 & -00:46:46 & 98.47 & 0.06 & 2.1 & N & 1.3 & 4.1 & 18.0 & 0.5 & \nodata \\
G31 & 18:49:32 & -00:46:59 & 99.46 & 0.18 & 1.3 & Y & 2.0 & 6.3 & 18.0 & 1.2 & 0.12 \\
G31 & 18:49:31 &  00:46:41 & 99.46 & 0.12 & 2.5 & N & 0.9 & 5.4 & 17.6 & 0.6 & 0.05 \\
G32 & 18:52:03 &  00:21:09 & 89.83 & 0.04 & 1.3 & N & 0.5 & 4.9 & 10.5 & 0.5 & 0.31 \\
G32 & 18:52:05 &  00:20:54 & 90.16 & 0.08 & 3.7 & Y & 0.8 & 5.2 & 14.3 & 0.5 & 0.25 \\
G32 & 18:52:07 &  00:20:06 & 90.82 & 0.05 & 2.0 & N & 0.4 & 3.6 & 12.6 & 0.3 & 0.32 \\
G34 & 18:53:17 &  01:23:45 & 55.86 & 0.10 & 2.1 & N & 1.4 & 5.7 & 19.7 & 0.8 & 0.01 \\
G34 & 18:53:17 &  01:24:32 & 56.19 & 0.08 & 2.0 & N & 1.6 & 5.3 & 18.9 & 1.3 & \nodata \\
G34 & 18:53:17 &  01:26:52 & 56.19 & 0.05 & 1.8 & N & 0.6 & 6.5 & 12.8 & 0.5 & 0.01 \\
G34 & 18:53:18 &  01:24:51 & 56.52 & 0.08 & 1.1 & Y & 2.6 & 4.4 & 24.6 & 2.0 & \nodata \\
G34 & 18:53:18 &  01:26:37 & 56.52 & 0.04 & 2.8 & N & 0.4 & 6.6 & 11.9 & 0.5 & 0.01 \\
G34 & 18:53:16 &  01:26:30 & 56.85 & 0.04 & 2.6 & Y & 1.3 & 6.5 & 13.6 & 0.9 & \nodata \\
G34 & 18:53:15 &  01:26:58 & 56.85 & 0.02 & 2.2 & N & 0.4 & 9.4 & 13.1 & 0.6 & \nodata \\
G34 & 18:53:17 &  01:23:59 & 57.18 & 0.02 & 1.9 & N & 1.2 & 4.7 & 13.8 & 0.7 & \nodata \\
G34 & 18:53:21 &  01:23:08 & 57.51 & 0.02 & 2.5 & N & 0.4 & 8.6 & 11.2 & 0.6 & 0.00 \\
G34 & 18:53:18 &  01:25:25 & 57.51 & 0.08 & 1.4 & Y & 2.3 & 4.3 & 27.9 & 1.8 & \nodata \\
G34 & 18:53:19 &  01:26:27 & 57.51 & 0.09 & 2.5 & N & 1.0 & 5.6 & 14.1 & 0.7 & 0.04 \\
G34 & 18:53:19 &  01:24:29 & 57.84 & 0.14 & 2.7 & Y & 1.7 & 6.1 & 20.6 & 1.6 & 0.01 \\
G34 & 18:53:18 &  01:25:37 & 58.17 & 0.11 & 1.2 & N & 1.5 & 5.5 & 19.8 & 1.3 & \nodata \\
G34 & 18:53:19 &  01:23:13 & 58.50 & 0.05 & 2.3 & N & 1.0 & 6.7 & 14.7 & 0.8 & \nodata \\
G34 & 18:53:18 &  01:25:13 & 58.50 & 0.06 & 1.6 & N & 1.5 & 6.0 & 21.3 & 1.7 & \nodata \\
G34 & 18:53:19 &  01:27:49 & 58.50 & 0.13 & 2.9 & Y & 1.1 & 6.3 & 13.9 & 0.9 & 0.06 \\
G34 & 18:53:20 &  01:28:21 & 59.49 & 0.09 & 1.4 & Y & 1.2 & 5.2 & 20.3 & 1.1 & 0.05 \\
G34 & 18:53:21 &  01:26:57 & 59.82 & 0.02 & 1.8 & N & 0.9 & 5.4 & 13.8 & 0.7 & 0.01 \\
G35 & 18:57:08 &  02:08:25 & 42.53 & 0.09 & 2.1 & N & 0.8 & 7.1 & 12.8 & 0.7 & 0.20 \\
G35 & 18:57:10 &  02:07:36 & 44.84 & 0.14 & 1.6 & N & 0.8 & 7.9 & 14.2 & 0.8 & 0.34 \\
G35 & 18:57:09 &  02:07:08 & 45.17 & 0.04 & 1.9 & N & 0.4 & 6.7 & 11.8 & 0.5 & 0.27 \\
G35 & 18:57:09 &  02:08:18 & 45.17 & 0.09 & 1.4 & Y & 1.0 & 5.2 & 14.4 & 0.6 & 0.23 \\
G35 & 18:57:07 &  02:08:41 & 45.17 & 0.03 & 2.0 & N & 0.7 & 5.2 & 14.0 & 0.7 & 0.25 \\
G35 & 18:57:08 &  02:07:52 & 45.50 & 0.10 & 1.3 & Y & 0.9 & 7.4 & 13.5 & 0.8 & 0.32 \\
G35 & 18:57:09 &  02:07:52 & 45.50 & 0.11 & 1.8 & Y & 1.0 & 5.3 & 14.5 & 0.6 & 0.26 \\
G35 & 18:57:08 &  02:08:03 & 45.50 & 0.13 & 1.4 & Y & 0.7 & 8.1 & 13.9 & 0.7 & 0.29 \\
G38 & 19:04:07 &  05:08:47 & 42.16 & 0.10 & 1.3 & N & 1.2 & 5.2 & 17.6 & 0.9 & 0.23 \\
G38 & 19:04:10 &  05:09:14 & 42.16 & 0.04 & 2.0 & N & 0.6 & 4.9 & 22.9 & 0.3 & 0.22 \\
G38 & 19:04:06 &  05:08:50 & 42.49 & 0.07 & 1.9 & Y & 1.3 & 4.0 & 13.2 & 0.6 & 0.16 \\
G38 & 19:04:08 &  05:09:30 & 42.49 & 0.10 & 1.7 & Y & 1.3 & 3.8 & 15.1 & 0.5 & 0.16 \\
G38 & 19:04:10 &  05:08:43 & 42.82 & 0.04 & 1.1 & N & 0.6 & 4.5 & 13.6 & 0.5 & 0.24 \\
G38 & 19:04:10 &  05:08:54 & 42.82 & 0.05 & 1.6 & N & 1.1 & 3.9 & 15.0 & 0.5 & 0.10 \\
G38 & 19:04:08 &  05:08:56 & 42.82 & 0.10 & 2.0 & Y & 1.3 & 3.2 & 15.3 & 0.5 & 0.26 \\
G38 & 19:04:08 &  05:09:09 & 42.82 & 0.11 & 1.1 & N & 1.0 & 3.4 & 17.1 & 0.4 & 0.27 \\
G38 & 19:04:07 &  05:09:46 & 42.82 & 0.08 & 1.4 & Y & 1.1 & 4.9 & 18.3 & 0.8 & 0.27 \\
G38 & 19:04:10 &  05:09:05 & 43.15 & 0.04 & 1.3 & N & 0.8 & 4.2 & 15.0 & 0.4 & 0.17 \\
\enddata
\tablenotetext{a}{Parent IRDC (values of $f_{\mathrm{fore}}$ given in parentheses): G10 = G010.74-00.13 (0.095), G22 = G022.56-00.20 (0.173), G24 = G024.60+00.08 (0.127), G28 = G028.23-00.19 (0.202), G31 = G031.97+00.07 (0.328), G32 = G032.70-00.30 (0.289), G34 = G034.43+00.24 (0.058), G35 = G035.39-00.33 (0.131), G38 = G038.95-00.47 (0.131).}
\tablenotetext{b}{We cannot compute $\tau_{24 \mu\mathrm{m}}$ for some clumps because IR emission covers their entire angular extent or the contrast with the background is too low.}
\end{deluxetable}

Additionally, \texttt{cprops} performs an elliptical fit to each clump with a variable position angle to determine sizes along the major and minor axes. The aspect ratio, $A_{0}$, is calculated from the \texttt{cprops} results as the ratio of the major axis length to the minor axis length, which is confirmed visually to be an accurate method. The aspect ratios of most of the clumps are approximately 1-2, but extend to about 3 to 4 for the more filamentary clumps, while the highest aspect ratio in our sample is 6.

The effective radii and aspect ratios we measure will be sensitive to the orientations of elongated structures with respect to the line of sight. The values we report for $R_{\mathrm{eff}}$ and $A_{0}$ will be lower limits, and so clumps will be systematically more filamentary than discussed here. We further discuss the projection effects in Appendix \ref{sec-projection}.

\subsection{IR Extinction Results \label{sec-irresults}}

The IR optical depth, $\tau_{24 \mu\mathrm{m}}$, is calculated as described in \S\ref{sec-extinction} and then averaged over the full extent of the clump on the sky. Since the extinction is calculated from two-dimensional images, we cannot accurately separate the extinction from multiple components along the same line of sight, however the majority of individual pixels in the extinction maps are included in at most one velocity component. Values of $\tau_{24 \mu\mathrm{m}}$ range from approximately 0 to 0.85. Clumps at the lower end of this distribution are those dominated by IR point sources or other emission, such that we cannot reliably determine an IR optical depth. Since none of the optical depths averaged over a whole clump are greater than 1, we can conclude that while the centers of clumps may be very optically thick, there are at least significant outer portions of the clumps that are susceptible to heating by external or internal IR radiation fields. They are, however, still opaque to optical and ultraviolet radiation since the extinction at 550 nm is approximately 50 times greater than at 24 $\micron$ \citep{2003ARA&A..41..241D}.

\subsection{Spectral Line Fitting Results \label{sec-slresults}}

\begin{figure}[h!]
\begin{center}
\includegraphics[width=1\textwidth]{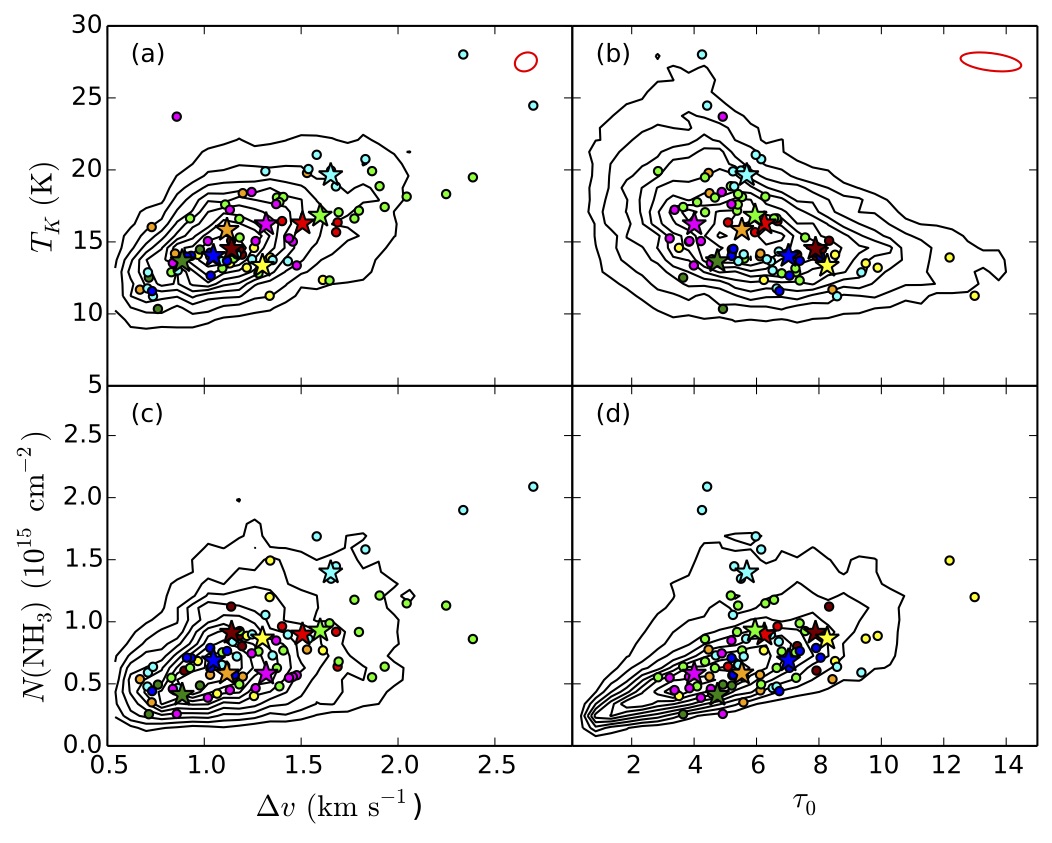}
\caption{Plots of the results of NH$_{3}$ spectral line fitting. The circles are $(\chi^{2}_{\mathrm{red}})^{1/2}$-weighted averages over clumps defined by \texttt{cprops}, and stars are $(\chi^{2}_{\mathrm{red}})^{1/2}$-weighted averages over whole IRDCs. Symbols are colored by IRDC: G010.74-00.13, G022.56-00.20, G024.60+00.08, G028.23-00.19, G031.97+00.07, G032.70-00.30, G034.43+00.24, G035.39-00.33, and G038.95-00.47 are maroon, red, orange, yellow, bright green, dark green, cyan, blue, and magenta, respectively. The background black contours show the $(\chi^{2}_{\mathrm{red}})^{1/2}$-weighted density of all lines of points in the sample (individual lines of sight and separated by distinct velocity components). Panels (a) and (b) also show red representative covariance ellipses from the spectral line fitting in the upper right. \label{f13}}
\end{center}
\end{figure}

Plots of the results of pixel-by-pixel NH$_{3}$ spectral line fitting are shown in Figure \ref{f13}, as well as averages over clumps (also in Table \ref{t5}) and whole IRDCs. The parameter averages are weighted by the square root of the reduced chi-squared statistic from the fit,
\begin{equation}
\left( \chi^{2}_{\mathrm{red}} \right)^{1/2} = \left( \dfrac{\chi^{2}}{n_{\mathrm{dof}}} \right)^{1/2} ,
\end{equation}
where $\chi^{2}$ is the goodness of fit statistic and $n_{\mathrm{dof}}$ is the number of degrees of freedom, i.e. the difference between the number of data points and the number of fit parameters (five per velocity component). This is a natural choice for the weighting because the relative uncertainties of the fit parameters scale with $(\chi^{2}_{\mathrm{red}})^{1/2}$. Experimentation shows that the averages are not very sensitive to the exact weighting scheme; for example, the results change by only a few percent if weighting by $\chi^{2}_{\mathrm{red}}$ is used instead of $(\chi^{2}_{\mathrm{red}})^{1/2}$.

The typical uncertainties for single lines of sight and single velocity components for $v_{c,\mathrm{LSR}}$, $\Delta v$, $\tau_{0}$, $T_{\mathrm{ex}}$, and $T_{\mathrm{R}}$ are 0.04 km s$^{-1}$, 0.09 km s$^{-1}$, 1.4, 0.58 K, and 1.1 K, respectively. The velocity FWHM, kinetic temperature, optical depth in the (1,1) line, and column density are generally correlated. G034.43+00.24 (cyan) is notable as having the highest $\Delta v$, $T_{K}$, and $N(\mathrm{NH}_{3})$ in the sample, dominated by the emission surrounding the bright YSOs with outflows as traced by masers and extended green objects (EGOs). EGOs are bright, resolved 4.5 $\micron$ sources often attributed to shocked molecular hydrogen emission in protostellar outflows \citep{2008AJ....136.2391C}. G028.23-00.19 is also notable for being apparently starless throughout, and is colder and has a higher $\tau_{0}$ than the rest of the sample.

The clump-averaged values quoted for the optical depth, $\tau_{0}$, and the kinetic temperature of the gas, $T_{K}$, are averages weighted by the reduced $(\chi^{2}_{\mathrm{red}})^{1/2}$ value from the spectral line fitting, and can be separated kinematically by individual clump even when they overlap spatially. Most of the clumps have an average NH$_{3}$ optical depth $\tau_{0}$ greater than 5 and all are greater than 1, indicating that these clumps are typically optically thick in NH$_{3}$ emission. We note that G028.83-00.19 shows significantly higher NH$_{3}$ optical depth than the rest of the sample, matching its high IR optical depth.

\subsubsection{Linewidths \label{sec-linewidths}}

In further analysis we adjust the values of the velocity dispersion to account for observational effects. We first convert the linewidth from the fit to a velocity dispersion, $\sigma_{\mathrm{obs}} = \Delta v (8 \ln 2)^{-1/2}$. The data are discretely sampled along the spectral axis and thus limited by the spectral resolution of the VLA observations. The effective velocity resolution is $\sigma_{\mathrm{res}} \approx 0.6$ km s$^{-1}$. Then
\begin{equation}
\sigma_{\mathrm{line}} \approx \sqrt{\sigma_{\mathrm{obs}}^{2} - \sigma_{\mathrm{res}}^{2}}
\end{equation}
is the ``true'' velocity dispersion along the line of sight. These values of the velocity dispersion averaged over the \texttt{cprops} clump boundaries are listed in Table \ref{t5}.

We further calculate the one-dimensional thermal linewidth calculated from the kinetic temperature,
\begin{equation}
\sigma_{T} = \sqrt{\dfrac{k_{\mathrm{B}} T_{K}}{\mu_{\mathrm{NH_{3}}} m_{\mathrm{H}}}} ,
\end{equation}
where $\mu_{\mathrm{NH_{3}}} = 17.03$ is the mean molecular weight of ammonia and $m_{\mathrm{H}}$ is the mass of a hydrogen atom. We can then calculate the nonthermal contribution to the velocity dispersion via
\begin{equation}
\sigma_{\mathrm{NT}} = \sqrt{\sigma_{\mathrm{line}}^{2} - \sigma_{T}^{2}} .
\end{equation}
We measure dispersions typically around 2 km s$^{-1}$, greater than the thermal linewidths, which are typically less than 0.1 km s$^{-1}$. The effect of this nonthermal dispersion is discussed in \S \ref{sec-clumpstab}.

\subsubsection{Comparison to Previous Studies \label{sec-previous}}

Taking the optical depth to be typically 5-10 for the majority of clumps, and the kinetic temperature to be 12-25 K for the majority of the clumps, we compare our results to similar studies. These values are in agreement with the dense clumps in G19.30+0.07 observed by \cite{2011ApJ...733...44D} with the VLA and the same spectral lines. Compared to the sample of \cite{2011ApJ...736..163R}, who also used the (1,1) and (2,2) lines observed by the GBT and the VLA, we find slightly higher kinetic temperatures in our study. They measured kinetic temperatures in the range of about 8 to 13 K, however their sample was selected to be devoid of star formation indicators and thus be in the earliest evolutionary phases. It is possible that our sample generally reflects a slightly later stage in IRDC evolution, in which star formation activity has increased and the gas is showing the affects of protostellar heating. Indeed, the clumps in our sample that are coincident with one or more 70 $\micron$ point sources in the \emph{Herschel} images have kinetic temperatures 2-3 K higher on average than clumps without 70 $\micron$ point sources.

Our clumps are slightly different than dense cores and core candidates in Perseus observed by \cite{2008ApJS..175..509R} with the GBT, also with the same spectral lines. They found colder kinetic temperatures around 10-12 K and optical depths usually less than 5, but as high as 15 in the densest cores. Furthermore, they observed column densities around $5 \times 10^{14}$ cm$^{-2}$ with only the densest cores exceeding $10^{15}$ cm$^{-2}$, while we observe column densities $10^{14}$-$10^{16}$ cm$^{-2}$ and nearly all of our clumps have peak column densities of about $3 \times 10^{15}$ cm$^{-2}$. We would expect higher column densities in our sample given that we selected IRDCs, which must have high enough dust column densities to be seen in extinction at MIR wavelengths. Their GBT observations of cores at a distance of 260 pc have comparable physical resolution (0.04 pc) to our study (0.02-0.1 pc), so the differences cannot be explained by beam dilution effects. The higher column densities in our study can be partially explained by using a slightly different method of calculation that includes a correction for both parity states of the (1,1) \citep{2009ApJ...697.1457F}. This correction typically increases the column densities by less than a factor of two for the measured excitation temperatures. We therefore are observing slightly warmer gas and slightly higher column densities in IRDCs than lower mass and apparently more quiescent star-forming environments like Perseus. This is not surprising if star formation activity is largely controlled by mass surface density; the average mass surface density in Perseus is approximately 90 $M_{\sun}$ pc$^{-2}$ \citep{2010ApJ...723.1019H}, but is approximately 150 $M_{\sun}$ pc$^{-2}$ in IRDCs (based on masses and sizes reported by \cite{2006ApJ...653.1325S}).

\section{Discussion \label{sec-discussion}}

\subsection{Kinematics and Previous Studies of Individual Sources \label{sec-kinematics}}

All nine IRDCs show evidence of clumps and velocity substructure. The IRDCs are generally composed of many distinct clumps that sometimes can be grouped into distinct velocity components. These clumps also overlap spatially and show apparent interactions in position-position-velocity space, with star formation tracers (IR point sources, masers, H {\smaller II} regions, etc.) coincident with these overlapping sites. We also observe smoother velocity gradients along the most filamentary (high aspect ratios, $A_{0} \gtrsim 3$) clumps and across IRDCs as a whole. It is clear that the internal structure of varies from IRDC to IRDC, both in terms of the complexity of velocity substructure and in the prevalence of filamentary structures compared to the prevalence of globule structures. We discuss the characteristics of the individual IRDCs below.

\subsubsection{G010.74-00.13 \label{sec-G10}}

\begin{figure}[h!]
\begin{center}
\includegraphics[width=1\textwidth]{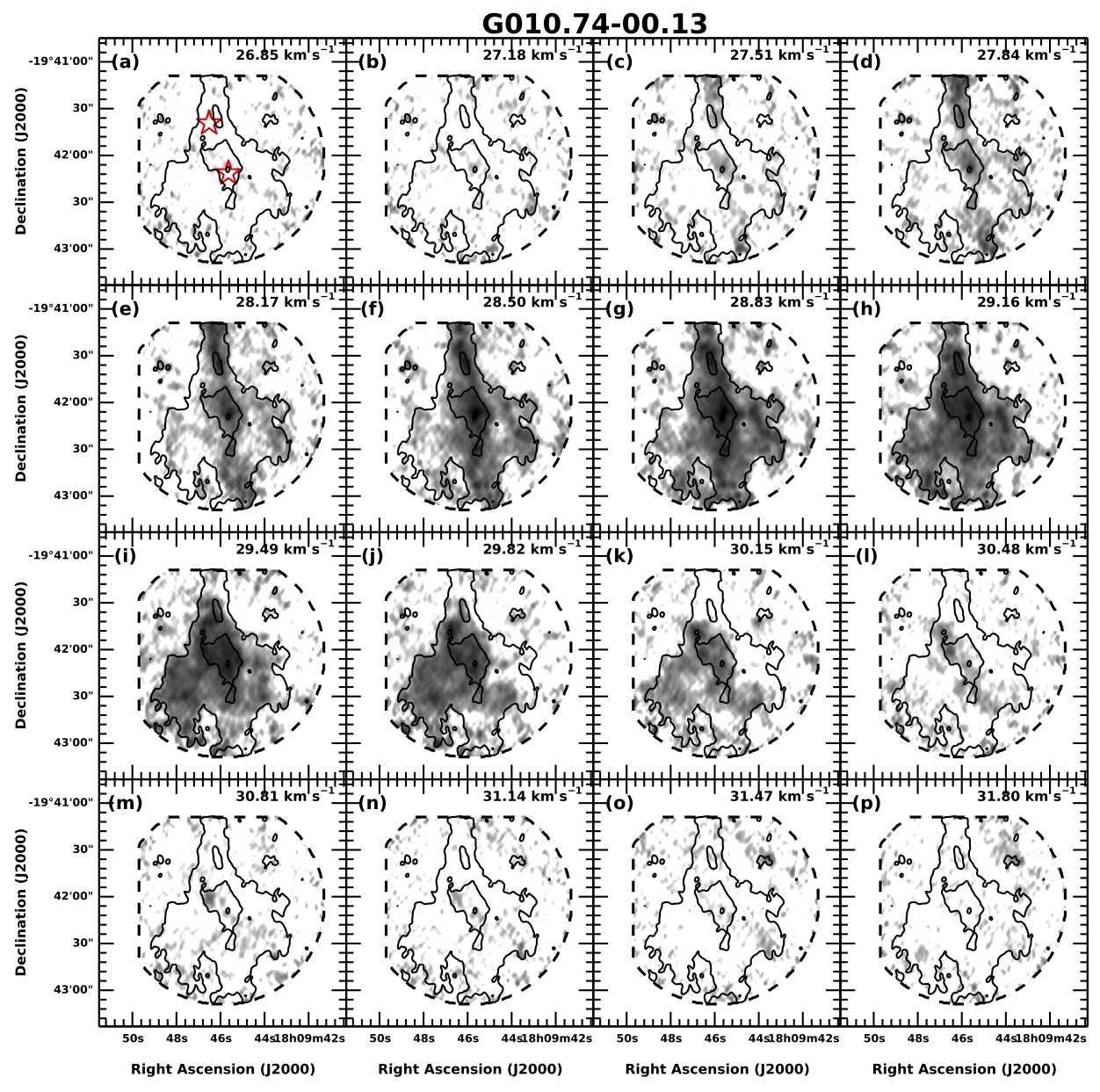}
\caption{Channel maps of the main hyperfine component of the NH$_{3}$ (1,1) emission in G010.74-00.13, shown in a logarithmic scale from 5 to 70 mJy beam$^{-1}$. Black contours show the integrated NH$_{3}$ (1,1) intensity and overlapping GBT and VLA footprint, same as the white contours in Figure \ref{f4}. Channel velocities, $v_{\mathrm{LSR}}$, are labeled in the upper right corner of each panel in km s$^{-1}$. The 70 $\micron$ infrared point sources are marked by stars in panel (a) (red in the online figure). There is a clear gradient from the filamentary structure running north to south into the globular structure extending to the southeast. The largest linewidth is at the center of the IRDC, coincident with both structures and with an embedded infrared point source. \label{f14}}
\end{center}
\end{figure}

G010.74-00.13 shows a clear velocity gradient west to east across the IRDC. Channel maps of the NH$_{3}$ (1,1) emission are presented in Figure \ref{f14}. At low velocity ($v_{\mathrm{LSR}} \approx$ 28 km s $^{-1}$), the NH$_{3}$ morphology appears filamentary, however it gradually transitions to a more globular morphology at higher velocity ($v_{\mathrm{LSR}} \approx$ 30 km s$^{-1}$). This structure appears to have two distinct velocity components with different morphologies. It is notable that both IR point sources (possible YSOs; see Figure \ref{f4}) are coincident with the filamentary component. There is an elevated velocity dispersion (peak in the second-moment maps of about 0.9 km s$^{-1}$) in the center of the IRDC where these two velocity components overlap and may be colliding in this scenario. This location is also coincident with one of the protostellar candidates. The majority of the NH$_{3}$ (2,2) emission is within the filamentary velocity component, whereas the CCS more closely follows the 30 km s $^{-1}$ globular component. In G010.74-00.13, it may be that this cloud-cloud collision has triggered both the formation of the YSO and the high velocity dispersion, or it may be that the YSO is responsible for the high velocity dispersion directly and its location is coincidental.

\subsubsection{G022.56-00.20 \label{sec-G22}}

\begin{figure}[h!]
\begin{center}
\includegraphics[width=1\textwidth]{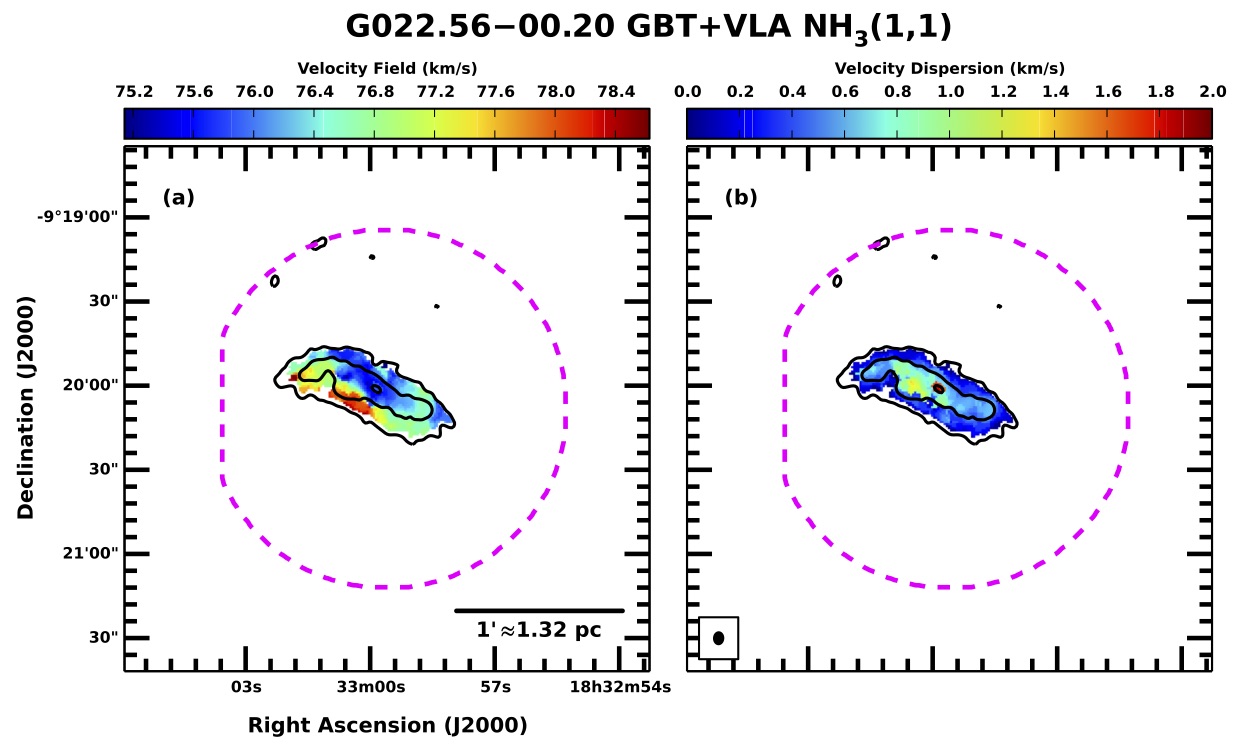}
\caption{Moment 1 and moment 2 maps showing the velocity field ($\langle v_{(1,1)} \rangle$) and velocity dispersion ($\langle v_{(1,1)}^{2} \rangle$), respectively, from the NH$_{3}$ (1,1) emission in G022.56-00.20. Clump deconvolution with \texttt{cprops} identifies three clumps: one filament running along the northwestern edge, one globule at the northeastern end, and one filament running along the southeastern edge. The gradient in the velocity velocity field is explained by the velocity offset between the two filaments, and the elevated velocity dispersion through the center of the IRDC is coincident with the overlap of the two filaments. The peak in the velocity dispersion, the peak in the integrated NH$_{3}$ emission, and an infrared point source are all coincident at the center of the IRDC. \label{f15}}
\end{center}
\end{figure}

\begin{figure}[h!]
\begin{center}
\includegraphics[width=1\textwidth]{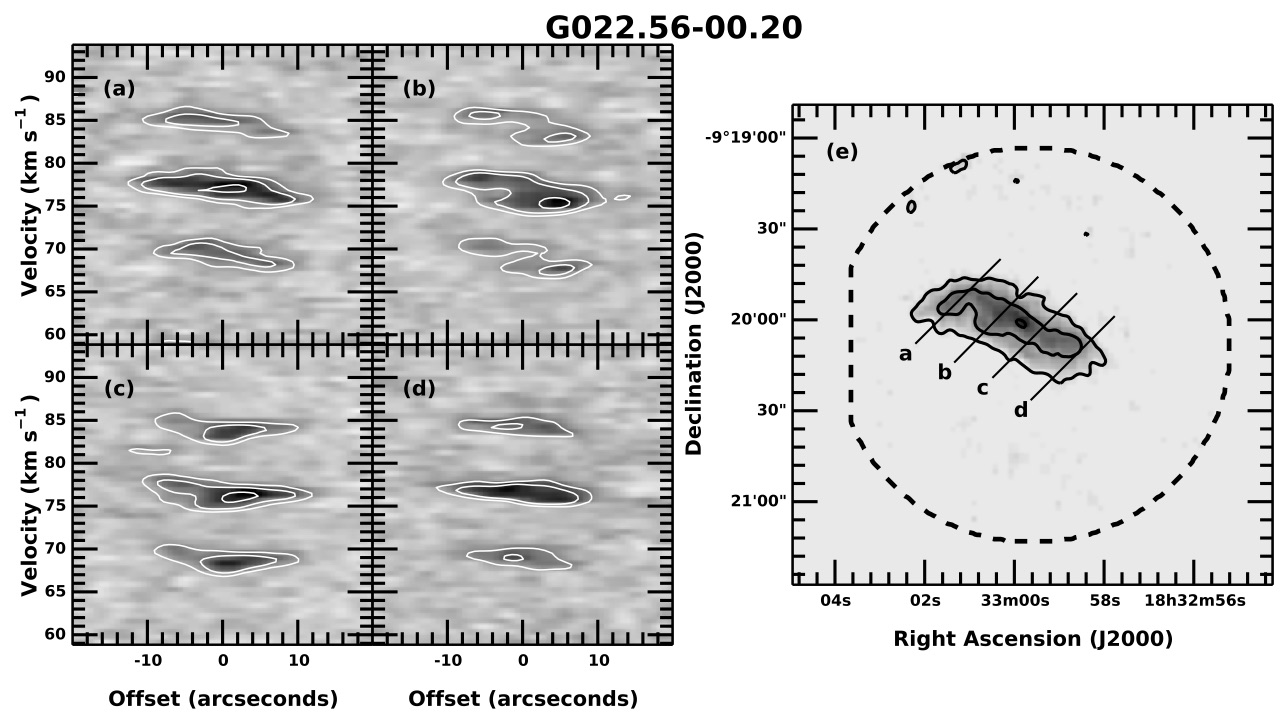}
\caption{Position-velocity diagrams in (a) through (d), with white contours at 2$\sigma$, 4$\sigma$, and 8$\sigma$. Slices for the diagrams are labeled and overplotted on the integrated intensity map of G022.56-00.20 in (e) in grayscale with a linear stretch. The black contours for the integrated intensity and combined GBT and VLA footprint are the same as the white contours in Figure \ref{f5}. The angular offsets in the diagrams increase from east to west. The three most distinct structures in each diagram are the main and two inner satellite hyperfine components of the NH$_{3}$ (1,1) line. The velocity offset between the two primary filaments is most apparent through the center of the IRDC. \label{f16}}
\end{center}
\end{figure}

G022.56-00.20 is deconvolved into three clumps with \texttt{cprops}. The central velocity component ($v_{\mathrm{LSR}} \approx$ 77 km s$^{-1}$) is located at the northeastern end of the IRDC, while the high ($v_{\mathrm{LSR}} \approx$ 78 km s$^{-1}$) and low ($v_{\mathrm{LSR}} \approx$ 75 km s$^{-1}$) velocity components have high aspect ratios ($A_{0} \approx $3) and run nearly parallel to each other along the IRDC's major axis. The 75 km s$^{-1}$ velocity component is located slightly to the north of the 78 km s$^{-1}$ velocity component, though they overlap spatially, and the IRDC shows a high velocity dispersion of about 2 km s$^{-1}$ in the overlap region. It is noteworthy that the peak in integrated intensity, the peak in velocity dispersion, and an infrared point source are all coincident in the center of this overlap region, as can be seen in Figures \ref{f5} and \ref{f15}. In Figure \ref{f16}, the position-velocity diagrams taken across the IRDC parallel to the minor axis clearly show these major velocity components and their interaction. We identify two possible scenarios for G022.56-00.20: (1) this IRDC represents a collision of multiple velocity components, apparently triggering the formation of the YSO; or (2) the YSO is driving expansion of the molecular gas around it. The separation between the two primary components is $\lesssim$ 5 km s$^{-1}$ and they extend $\lesssim$ 1 pc, making either scenario plausible.

\subsubsection{G024.60+00.08 \label{sec-G24}}

\begin{figure}[h!]
\begin{center}
\includegraphics[width=1\textwidth]{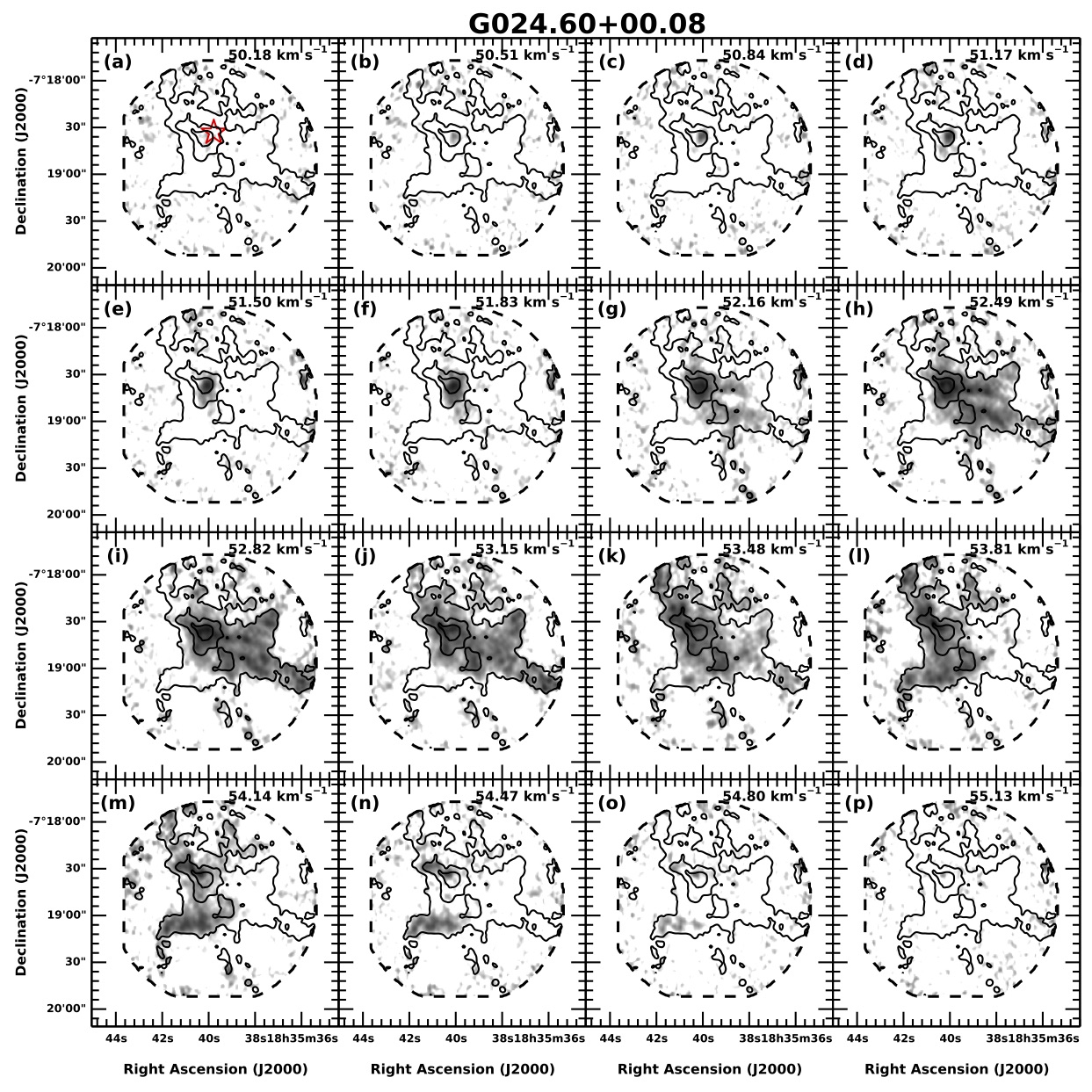}
\caption{Channel maps of the NH$_{3}$ (1,1) emission as in Figure \ref{f14} but for G024.60+00.08, with contours and field of view matching Figure \ref{f6}, and only one 70 $\micron$ source. There is a clear gradient from the northwest to southeast along the filamentary structures. These filaments extend further in the infrared extinction maps beyond the coverage of the NH$_{3}$ observations. Deconvolution with \texttt{cprops} identifies multiple clumps with offset velocities along these filaments. The largest linewidth is at the center of the IRDC, coincident with both structures and with an embedded infrared point source. \label{f17}}
\end{center}
\end{figure}

G024.60+00.08 shows a clear gradient in the velocity field from about $v_{\mathrm{LSR}} \approx$ 50.5 km s$^{-1}$ in the west to about $v_{\mathrm{LSR}} \approx$ 55 km s$^{-1}$ in the east-southeast. Channel maps of the IRDC in Figure \ref{f17} do not clearly distinguish between the possibility of two distinct components or a single gradient across one major component. There is a velocity dispersion peak of about 1.2 km s$^{-1}$, cospatial with a large NH$_{3}$ clump and an IR point source near the center of the IRDC (see Figure \ref{f6}). \cite{2007ApJ...662.1082R} identified two protostellar condensations (bright, compact, millimeter cores with IR emission indicative of star formation) in G024.60+00.08 with IRAM Plateau de Bure 1.3 and 3 mm continuum and \emph{Spitzer} images, one of which is the central IR point source. \cite{2008AJ....136.2391C} identified an EGO in G024.60+00.08. The position of the EGO is marked in Figure \ref{f6}, and is also coincident with the protostellar candidate.

\subsubsection{G028.23-00.19 \label{sec-G28}}

G028.23-00.19 is an apparent starless, dark cloud. The bright point source (Figure \ref{f7}) is an unrelated, foreground late-type star \citep{1989ApJ...347..325B}. \cite{2006ApJ...641..389R} made millimeter continuum maps of G028.23-00.19 with the IRAM 30 m single dish telescope and observed three dense clumps at 11$\arcsec$ resolution with masses ranging from 38 to 705 $M_{\sun}$. The primary IRDC has one filament, with at least 4 connecting filaments extending beyond our maps, but seen in IR extinction. \cite{2013ApJ...773..123S} found two distinct velocity components in CARMA observations of this IRDC, with 0.3 km s$^{-1}$, 10$\farcs$9 resolution. In our data, the position-velocity diagram shows that the velocity structure in NH$_{3}$ is a smoother gradient across the filament, from about $v_{\mathrm{LSR}} \approx$ 79 km s$^{-1}$ to $v_{\mathrm{LSR}} \approx$ 82 km s$^{-1}$. It is unlikely that we are seeing a resolution effect, given that we have comparable velocity resolution and higher angular resolution than the CARMA study. The discrete clumps may be explained by a combination of chemical differentiation across the IRDC and/or missing the largest spatial scales in the CARMA data. There is also an ammonia velocity dispersion peak of about 0.9 km s$^{-1}$ coincident with the optical depth peak. It is possible that the core in G028.83-00.19 may be a deeply embedded YSO with an outflow exciting the SiO and CH$_{3}$OH. However, this is unlikely given the lack of IR emission, the absence of a $T_{K}$ increase in the ammonia fitting results toward this source, and the detection of NH$_{2}$D by \cite{2013ApJ...773..123S}. These characteristics are most consistent with a cold core.

\subsubsection{G031.97+00.07 \label{sec-G31}}

G031.97+00.07 has the most complex substructure in our sample. \cite{2006ApJ...641..389R} made millimeter continuum maps of G031.97+00.07 with the IRAM 30 m single dish telescope at 11$\arcsec$ resolution and observed nine dense clumps with masses ranging from 151 to 1890 $M_{\sun}$. \cite{2006ApJ...651L.125W} reported H$_{2}$O masers in G031.97+00.07 and \cite{2009A&A...501..539U} identified an H {\smaller II} region in G031.97+00.07, all marked in Figure \ref{f8}. The IR morphology has a long, thin filamentary structure leading to a higher contrast globule neighbored by at least 3 YSOs with H {\smaller II} regions and masers. The IRDC itself is part of a much larger molecular complex seen in $^{13}$CO in the BU-GRS with IR dark filaments extending beyond our observations.

\begin{figure}[h!]
\begin{center}
\includegraphics[width=0.75\textwidth]{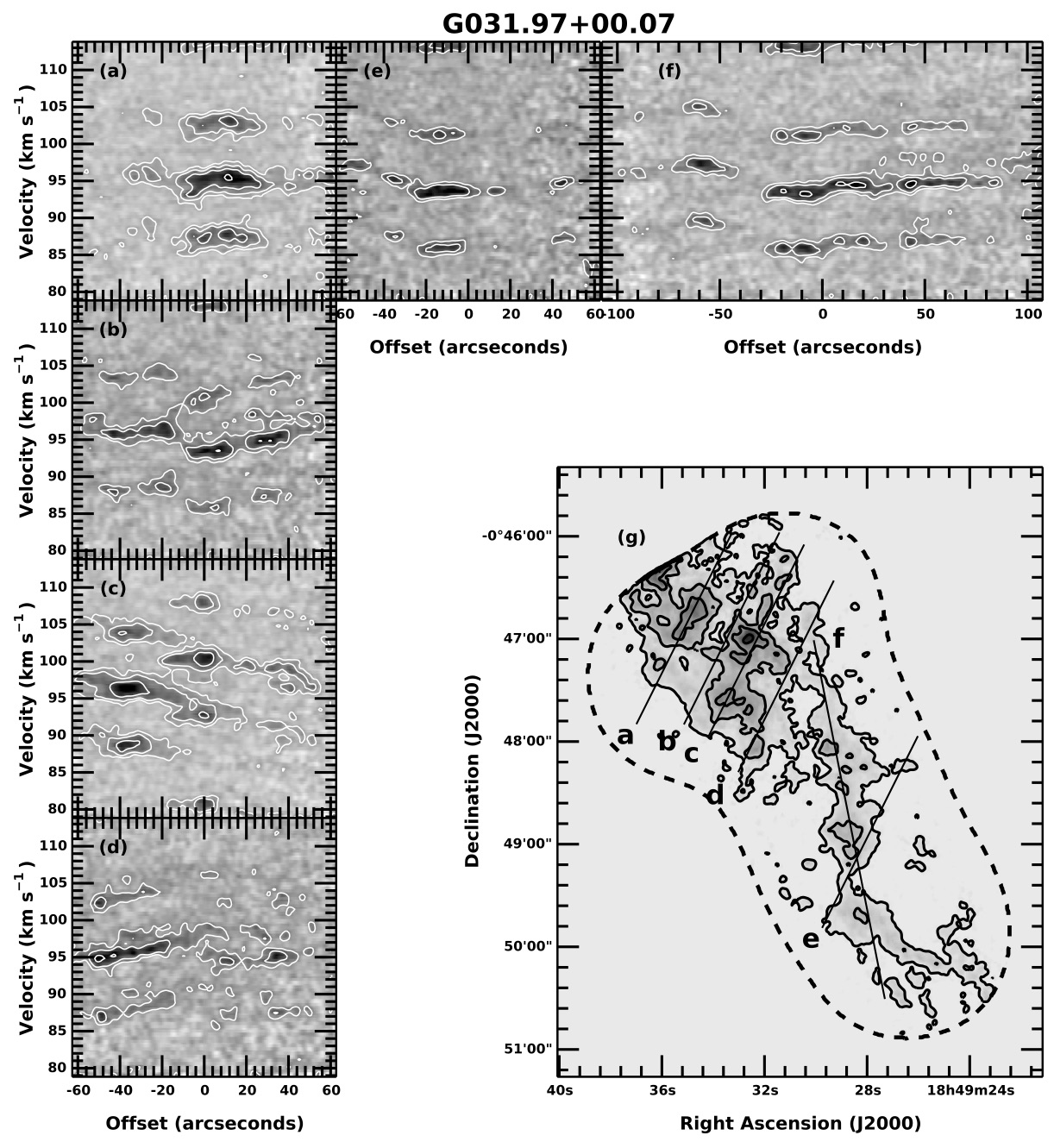}
\caption{Position-velocity diagrams in (a) through (f) as in Figure \ref{f16}, but for G031.97+00.07 with white contours at 2$\sigma$, 4$\sigma$, 8$\sigma$, and 16$\sigma$. Slices and the integrated intensity map are shown in (g) also as in Figure \ref{f16}, but with contours taken from Figure \ref{f8}. G031.97+00.07 contains a complex substructure of globules and filaments at offset velocities and with internal velocity gradients. \label{f18}}
\end{center}
\end{figure}

\begin{figure}
\begin{center}
\includegraphics[width=1\textwidth]{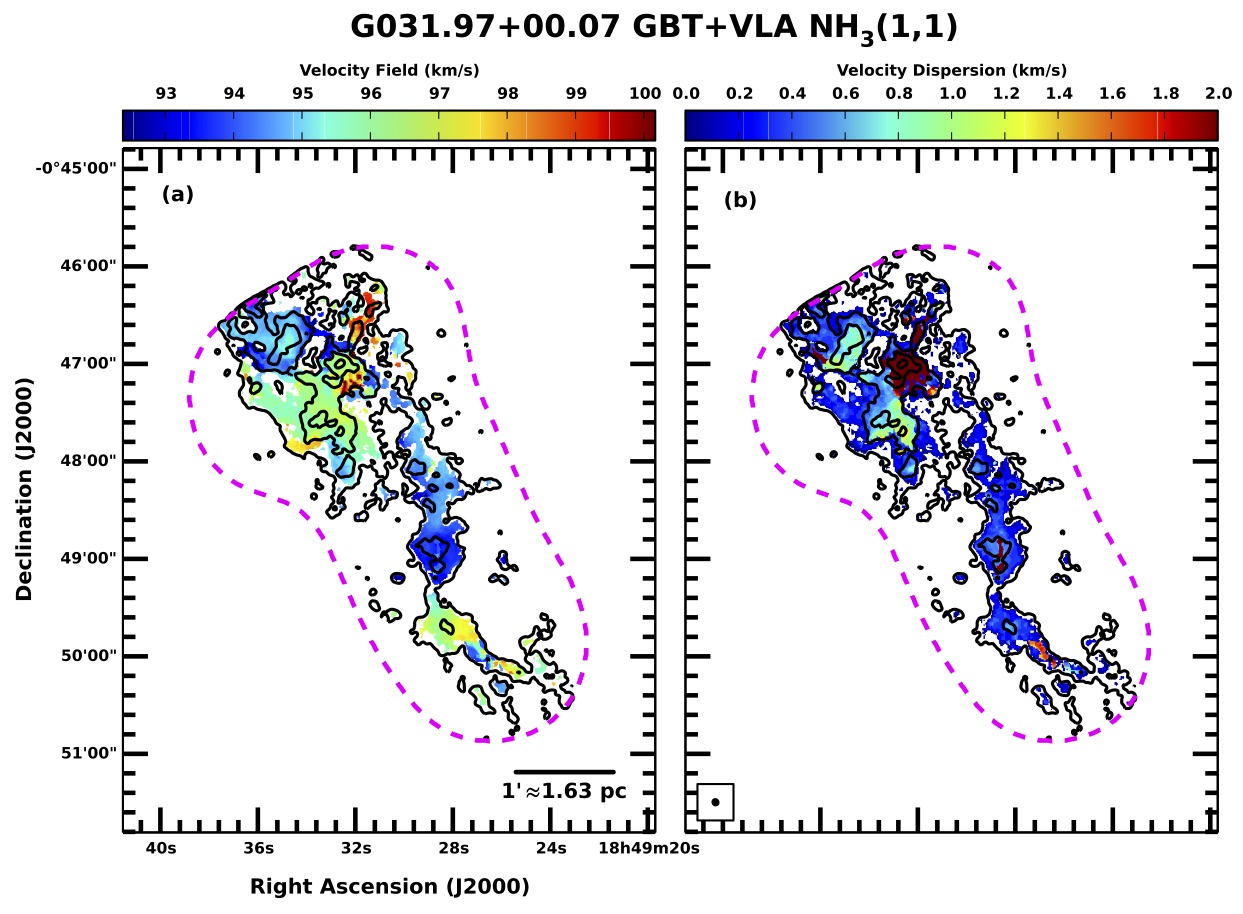}
\caption{Moment 1 and moment 2 maps showing the velocity field ($\langle v_{(1,1)} \rangle$) and velocity dispersion ($\langle v_{(1,1)}^{2} \rangle$), respectively, from the NH$_{3}$ (1,1) emission in G031.97+00.07. Deconvolution with \texttt{cprops} reveals a complex substructure with 21 globules and filaments. The majority of the emission is in a lower velocity component around 92-99 km s$^{-1}$, though a few clumps form a higher velocity component around 97-102 km s$^{-1}$. These velocity components are cospatial, as shown by the elevated values in the velocity dispersion map. There are additionally velocity gradients along the substructures with higher aspect ratios, such as the filament in the middle of the IRDC. \label{f19}}
\end{center}
\end{figure}

Seen in Figure \ref{f8}, the NH$_{3}$ emission closely matches the IR contrast, and it is deconvolved into 21 distinct clumps. As seen in the position-velocity slices in Figure \ref{f18}, the structures are a mix of filaments and globules, and tend to fall in one of two distinct velocity ranges: $v_{\mathrm{LSR}} \approx$ 92-99 km s$^{-1}$ and $v_{\mathrm{LSR}} \approx$ 97-102 km s$^{-1}$. The majority of the emission is in the 92-99 km s$^{-1}$ velocity range, however the 97-102 km s$^{-1}$ velocity range is coincident with at least one YSO. The filamentary structures also show velocity gradients spanning about 2 km s$^{-1}$ along their major axis, as seen in Figure \ref{f19}. This IRDC also has the strongest CCS emission in our sample, with a peak signal-to-noise ratio of about 22. G031.97+00.07 may be a region where several weaker filaments extending tens of parsecs are feeding molecular gas and star formation is progressing most quickly at the collision points. The morphology and velocity structure of this IRDC is consistent with the ``hub-filament structure'' described in \cite{2009ApJ...700.1609M} and \cite{2013MNRAS.436.3707L}, in which gas flows along the filaments to a central hub where it feeds star formation.

\cite{2014ApJ...787..113B} recently studied G031.97+00.07 (called G32.02+0.06 in their work) in the NH$_{3}$ (1,1), (2,2), and (4,4) transitions with the VLA and discussed the environment of the larger molecular complex. They observed two pointings, one toward the active region at the north end of the IRDC, and one quiescent region near the south end of the IRDC (their pointings partially overlap our combined maps, but do not cover the central portion of the IRDC). They found dense parsec scale filaments of 10-100 $M_{\sun}$ with dense cores less than 0.1 pc in size, in agreement with our findings. The authors reported that the dense cores were virially unstable to gravitational collapse, and that turbulence likely set the fragmentation length scale in the filaments. In the quiescent region, they observed the two distinct velocity components that continue along the IRDC toward the more complex active region in our data. Finally, they note the existence of at least three bubbles all seen in \emph{Spitzer}, \emph{Herschel}, and the BU-GRS that are likely older H {\smaller II} regions from previous generations of massive stars, and may have compressed the molecular gas to form and/or shape the IRDC and trigger more recent massive star formation.

\subsubsection{G032.70-00.30 \label{sec-G32}}

G032.70-00.30 has the weakest NH$_{3}$ emission in our sample. As shown in Figure \ref{f5}, the (2,2) line is only marginally detected, and there is no detected CCS. Weak IR point sources indicate protostellar candidates at both ends of the IRDC. There is a velocity gradient from $v_{\mathrm{LSR}} \approx$ 89 km s$^{-1}$ to $v_{\mathrm{LSR}} \approx$ 91 km s$^{-1}$ across the filament from southeast to northwest, with the highest velocity dispersion of 0.8 km s$^{-1}$ near the southwestern end.

\subsubsection{G034.43+00.24 \label{sec-G34}}

\cite{2006ApJ...651L.125W} reported H$_{2}$O masers in G034.43+00.24 and \cite{2009A&A...501..539U} identified an H {\smaller II} region in G034.43+00.24, all marked in Figure \ref{f10}. \cite{2005ApJ...630L.181R} observed the millimeter/submillimeter continuum in G034.43+00.24 with IRAM, the James Clerk Maxwell Telescope (JCMT), and the Caltech Submillimeter Observatory (CSO). They identified three compact clumps of several hundred solar masses each. The SEDs stretching from the millimeter to the IR indicated high luminosities consistent with YSOs of $\sim$10 $M_{\sun}$ each. Moreover, \cite{2005ApJ...630L.181R} also observed HCN, CS, and SiO in G034.43+00.24 with IRAM and CSO. The large line widths of $\sim$10 km s$^{-1}$ in HCN and CS along with the detection of SiO indicated outflows and shocked gas, further evidence of ongoing star formation. \cite{2008AJ....136.2391C} identified two EGOs, marked in Figure \ref{f10}. \cite{2010ApJ...715...18S} observed this region in multiple molecular gas tracers with the APEX 12 m telescope, the Nobeyama Radio Observatory (NRO) 45 m, and the Swedish-ESO 15 m Submillimeter Telescope (SEST). They found 4 molecular cores with velocity profiles indicative of outflows and large scale infall toward the most massive core, further strengthening the evidence for ongoing massive star formation in this IRDC.

\begin{figure}
\begin{center}
\includegraphics[width=1\textwidth]{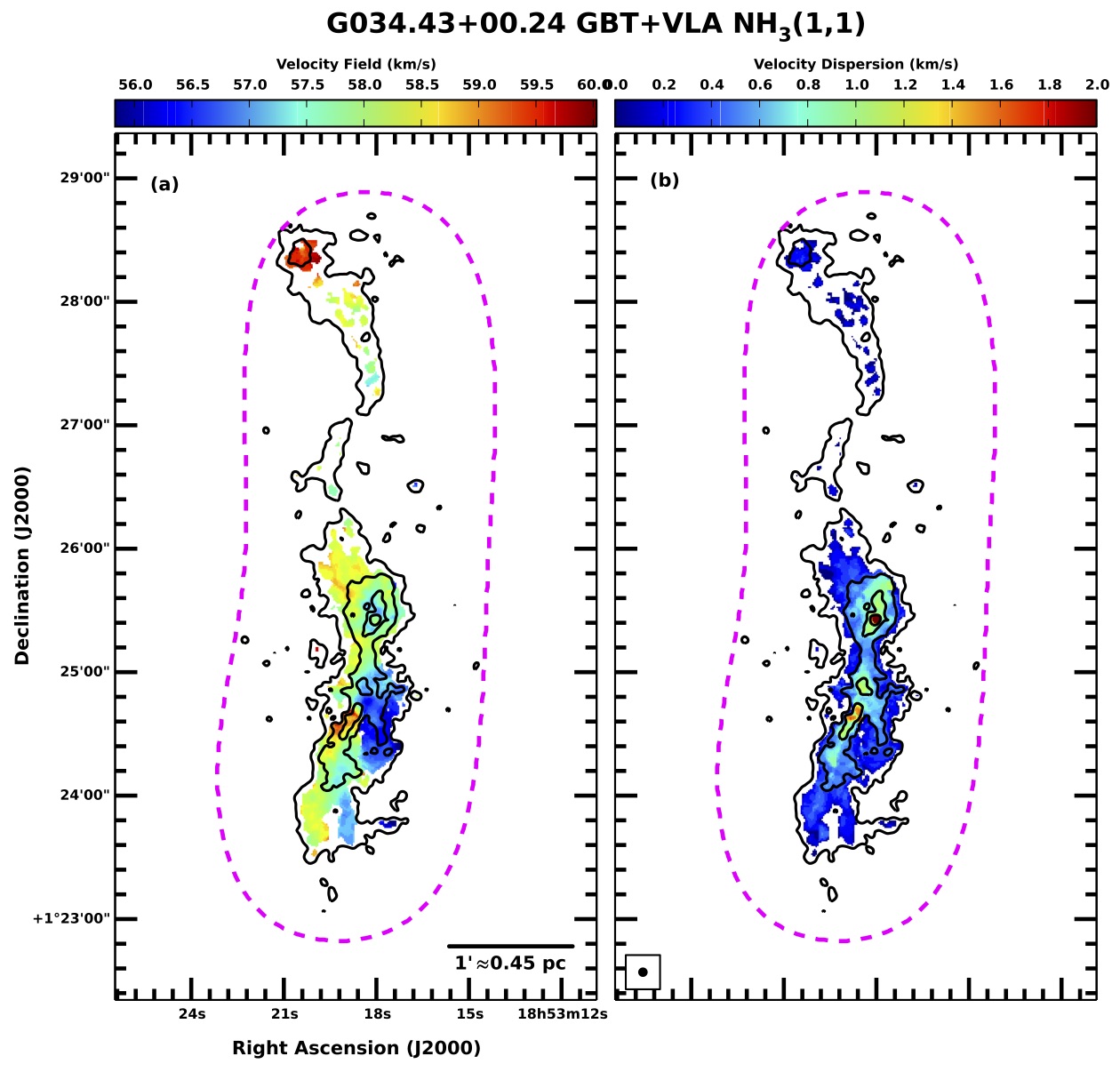}
\caption{Moment 1 and moment 2 maps showing the velocity field ($\langle v_{(1,1)} \rangle$) and velocity dispersion ($\langle v_{(1,1)}^{2} \rangle$), respectively, from the NH$_{3}$ (1,1) emission in G034.43+00.24. The IRDC as a whole is very filamentary, but deconvolution with \texttt{cprops} reveals clumps ranging from globule to filamentary within the larger filamentary structure. The southern half of the IRDC contains three strong protostellar candidates. The two with known EGOs, indicative of molecular outflows, are coincident with the peaks in the velocity dispersion. An overall velocity gradient in the IRDC is seen from the western edge to the eastern edge in this part of the IRDC, then continues along the fainter filament to the YSO in the north. \label{f20}}
\end{center}
\end{figure}

G034.43+00.24 has a complex substructure. The overall shape is very filamentary (see Figure \ref{f10}), and it is a portion of the Giant Molecular Filament GMF38.1-32.4a identified by \cite{2014A&A...568A..73R}. The IRDC is known to have three YSOs with masers, EGOs, and millimeter cores along the primary filament, with an additional YSO at the northern end of the IRDC. Shown in Figure \ref{f20}, The velocity dispersion is elevated to over 2 km s$^{-1}$ at two of the YSO positions. There is an overall gradient from the southwest to the northeast. The major filament has a velocity gradient from the west to the east, leading into a gradient from south ($v_{\mathrm{LSR}} \approx$ 56 km s$^{-1}$) to north ($v_{\mathrm{LSR}} \approx$ 60 km s$^{-1}$) along the weaker filament toward the northern YSO. This gradient may be the result of gas flowing along the larger GMF, or may be composed of multiple unresolved velocity components.

\subsubsection{G035.39-00.33 \label{sec-G35}}

G035.39-00.33 is a long, filamentary, high IR contrast IRDC with protostellar candidates (see Figure \ref{f11}). The velocity structure shows two distinct velocity components: a $v_{\mathrm{LSR}} \approx$ 43 km s$^{-1}$ velocity component at the northern end of our observations, and a $v_{\mathrm{LSR}} \approx$ 45 km s$^{-1}$ velocity filament with a gradient from the southeast to the northwest. The two components are well separated in velocity space, and position-velocity slices do not show any apparent interaction between the two. We observe high velocity dispersion around most of the YSOs, only one of which is coincident with the lower velocity component. Our GBT data, which cover much more of the IRDC and have higher velocity resolution, show that the velocity gradient in the 45 km s$^{-1}$ component continues along the northern extent of the filament.

\cite{2010MNRAS.406..187J} observed G035.39-00.33 in the SiO $J$=2-1 line (a shock tracer) with the IRAM 30 m, and found both bright, compact emission with broad linewidths and weaker, extended emission with narrow linewidths. The compact emission was consistent with protostellar outflows, and they proposed three explanations for the extended emission: (1) a collection of low mass outflows throughout the IRDC, (2) a recently processed, more massive outflow whose energy was distributed throughout the IRDC, or (3) the IRDC was recently assembled from collisions of smaller clumps. They detected three filaments with distinct velocities at approximately 44.1, 45.3, and 46.6 km s$^{-1}$. \cite{2014MNRAS.439.1996J} further found velocity gradients of 0.4 to 0.8 km s$^{-1}$ pc$^{-1}$ from north to south along the filaments, as seen in multiple transitions of $^{13}$CO and C$^{18}$O with the IRAM 30 m and the JCMT. They further found that dense cores were preferentially located at the intersection of the (possibly interacting) filaments. \cite{2014MNRAS.440.2860H} observed N$_{2}$H$^{+}$ (1-0) in G035.39-00.33 with the Plateau de Bure Interferometer (PdBI) array and noted that the local kinematics of the filaments were affected the dense cores, either through accretion or expanding envelopes. \cite{2012ApJ...756L..13H} found that this parsec-scale region was consistent with being in virial equilibrium without requiring support from magnetic fields, and that the timescale for achieving equilibrium was less that the estimated 1.4 Myr for formation from two converging flows.

While these studies focused primarily on the narrower portion of the IRDC north of our VLA maps, our larger GBT maps cover the region in these studies, and we also detect these three velocity components in NH$_{3}$ (though the 46.6 km s$^{-1}$ component is seen only weakly). The two velocity components we observed in our combined maps are the extensions of two of these filaments to the south. Most of the emission we detected here is part of the 45 km s$^{-1}$ velocity component, and this southern region is both wider and darker (in MIR extinction) than the northern portion of the IRDC. We suggest that this IRDC may also represent an example of the hub-filament structure, in which gas is flowing along the parsec-scale 45 km s$^{-1}$ filament to the region we studied, where it feeds the ongoing star formation traced by IR point sources.

\subsubsection{G038.95-00.47 \label{sec-G38}}

G038.95-00.47 is a globular IRDC with long, low IR contrast filaments extending from a few parsecs away from it to the east. Seen in Figure \ref{f12}, it also contains at least 2 protostellar candidates along the western edge, and neighbors two IR bubbles to the southeast and southwest. The velocity field shows a clear gradient from the southwest to the east ($v_{\mathrm{LSR}} \approx$ 40.5 km s$^{-1}$) and north ($v_{\mathrm{LSR}} \approx$ 43.5 km s$^{-1}$). This IRDC is also consistent with the ``hub-filament structure'' described in \cite{2009ApJ...700.1609M} and \cite{2013MNRAS.436.3707L}.

\cite{2013A&A...559A.113X} studied G038.95-00.47 in relationship to its environment, specifically the two adjacent H {\smaller II} regions G38.91-0.44 (N74) and G39.30-1.04 (N75). They used 1.4 GHz radio continuum and single dish $^{12}$CO, $^{13}$CO, and C$^{18}$O $J$=1-0 observations to probe the physical characteristics of the bubbles and used GLIMPSE data to identify YSOs in the region. They found a molecular clump associated with G038.95-00.47 that showed a sharp integrated intensity gradient toward the H {\smaller II} regions consistent with it being compressed by their expansion. They further identify an overabundance of YSOs in the IRDC and propose that star formation was triggered via radiatively driven implosion.

\subsection{Mass Estimates \label{sec-mass}}

\begin{figure}
\begin{center}
\includegraphics[width=0.5\textwidth]{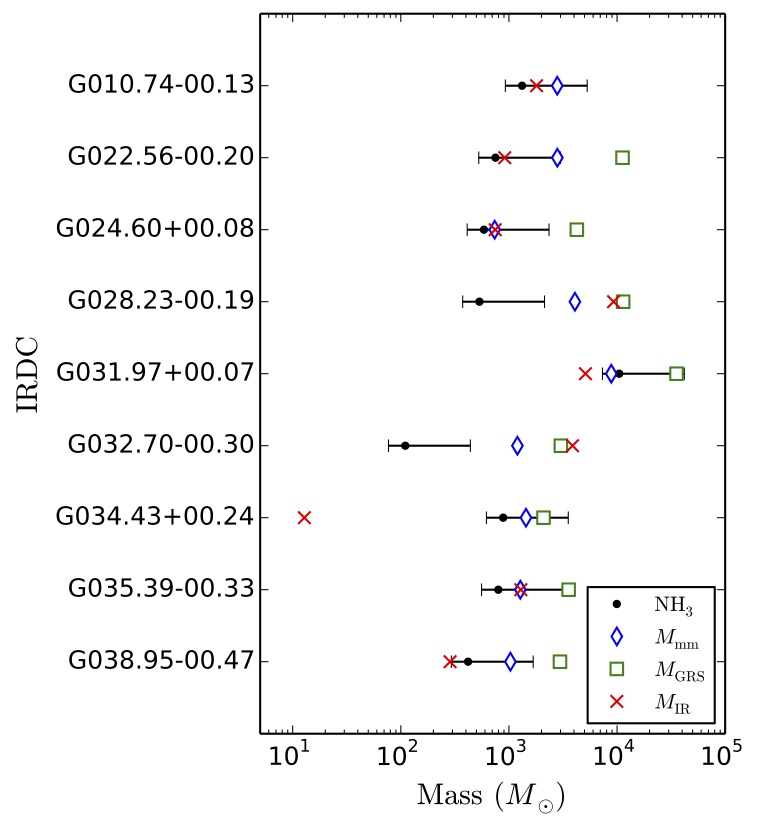}
\caption{Total molecular gas masses for our IRDC sample using three different method of estimation, along with the mass inferred from NH$_{3}$ assuming a relative abundance of $10^{-8}$ compared to molecular hydrogen. The error bars represent a factor of three variation in this abundance. Values of $M_{\mathrm{IR}}$ are the masses estimated from dust extinction, i.e. 24 $\micron$ contrast. Because of the infrared emission from protostellar candidates, IR bubbles, etc., the dust extinction method is not viable everywhere. The mass estimates may therefore be lower limits, particularly for G034.43+00.24. Values of $M_{\mathrm{mm}}$ are the masses estimated from dust emission at 1.12 mm, observed in the BGPS. Values of $M_{\mathrm{13CO}}$ are masses estimated from $^{13}$CO $J$=1-0 in the BU-GRS, except for G010.74-00.13, which was not covered in that survey. The emission is probably optically thick, but also includes the contribution of the lower density molecular envelopes. The masses estimated from the BGPS and the BU-GRS data are typically well-correlated, but the BU-GRS masses are higher by factors of a few. \label{f21}}
\end{center}
\end{figure}

We estimate the H$_{2}$ masses for IRDCs from three independent methods: dust extinction, dust thermal emission, and $^{13}$CO emission, described below. Total IRDC masses for our sample for all three methods are shown in Figure \ref{f21}, with a comparison discussed in \S \ref{sec-compmass}. All masses were measured by integrating over the footprint of our combined GBT+VLA observations.

\subsubsection{IR Extinction Mass Estimates \label{sec-irext}}

For the dust extinction method, we continue to follow the analysis presented by \cite{2009ApJ...696..484B} used to calculate IR extinction maps described in \S \ref{sec-extinction}. The mass surface density is determined from the optical depth by
\begin{equation}
\Sigma_{\lambda} = \dfrac{\tau_{\lambda}}{\kappa_{\lambda}} ,
\end{equation}
where $\kappa_{\lambda}$ is the dust opacity per gas mass at wavelength $\lambda$. Our choice of dust opacity from the literature will affect the mass estimate by as much as factors of a few. For MIPS 24 $\micron$, we adopt $\kappa_{24\mu\mathrm{m}} = 13.3$ cm$^{2}$ g$^{-1}$ from the dust model of \cite{1994A&A...291..943O} for $10^{5}$ year-old, $10^{6}$ cm$^{-3}$ dust with thin ice mantles and assuming a dust-to-gas mass ratio of 1:100. This value has an uncertainty of about a factor of three given the full range of possible $\kappa_{\nu}$ values, depending on the age, thickness of ice mantles, and number density of dust grains, which results in a factor of about three uncertainty in the mass estimates. The angular resolution of the 24 $\micron$ images, and thus our extinction and mass surface density maps, is 6$\arcsec$, compared to the 3$\farcs$2-5$\farcs$4 resolution of the NH$_{3}$ maps. This is the closest resolution match of the three mass estimates, however it lacks any velocity information to disentangle multiple clumps along a single line of sight.

 For the total mass, we simply sum over the IRDC:
\begin{align}
M_{\mathrm{IR}} = D^{2} \Omega \sum_{\mathrm{IRDC}} \Sigma_{\lambda} \notag\\ = 30 \left( \dfrac{D}{\mathrm{kpc}} \right)^{2} \left( \dfrac{\Omega}{\mathrm{arcmin}^{2}} \right) \left( \sum_{\mathrm{IRDC}} \tau_{\lambda} \right) M_{\sun} .
\end{align}
IRDC masses estimated with this method range from about 280 $M_{\sun}$ to 3900 $M_{\sun}$, excluding G034.43+00.24. In this IRDC, $M_{\mathrm{IR}}$ is a very weak lower limit on the mass because there is significant IR emission from IR point sources that prevents us from calculating the extinction across most of the IRDC.

\subsubsection{BU-GRS $^{13}$CO $J$=1-0 Mass Estimates \label{sec-coem}}

Next, we estimate the gas mass from the $^{13}$CO $J$=1-0 BU-GRS data cubes. Because the spectral line data have velocity information, we restrict measurements from this emission to velocity ranges with NH$_{3}$ (1,1) emission greater than 5$\sigma$. While the velocity resolution of 0.2 km s$^{-1}$ is in fact better than our 0.6 km s$^{-1}$ resolution NH$_{3}$ data cubes, the beam FWHM is 46$\arcsec$, so we are not able to resolve individual clumps. G010.74-00.13 is outside the coverage of the BU-GRS, and so is not included in this analysis.

To calculate the $^{13}$CO $J$=1-0 column density, we assume that the gas is in LTE, fills the beam, and is optically thin, and thus use the standard equation:
\begin{align}
N \left(^{13}\mathrm{CO} \right) = \dfrac{3k_{\mathrm{B}}}{8 \pi^{3} \nu_{\mathrm{13CO}} S( J_{u}) \mu_{\mathrm{13CO}}^{2}} \dfrac{ Z_{\mathrm{13CO}}}{g_{u} g_{K} g_{\mathrm{nuclear}}} \notag\\ \times \exp \left( \dfrac{E_{u,\mathrm{13CO}}}{k_{\mathrm{B}} T_{\mathrm{ex,13CO}}} \right) \int T_{\mathrm{B},\mathrm{13CO}}~dv ,
\end{align}
where $\nu_{\mathrm{13CO}}$ = 110.201353 GHz, $S( J_{u}) = J_{u}/(2J_{u} + 1)$ is the statistical weight of the upper level, $J_{u} = 1$ for the upper level, $\mu_{\mathrm{13CO}}$ = 0.112 debye is the electric dipole moment, $g_{u} = 2J_{u} + 1$ is the rotational degeneracy, $g_{K} = 1$ is the $K$ degeneracy for a linear molecule, $g_{\mathrm{nuclear}} = 1$ is the nuclear spin degeneracy, and $E_{u,\mathrm{13CO}}/k_{\mathrm{B}} = 5.29$ K for the energy of the upper level. We assume a partition function of the form $ Z_{\mathrm{13CO}} \approx 0.38(T_{\mathrm{ex,13CO}}/\mathrm{K} + 0.88)$ \citep{2009tra..book.....W} and an excitation temperature $T_{\mathrm{ex,13CO}} = 10$ K. Using a higher excitation temperature of 20 K would raise the mass estimates by approximately 40\% compared to 10 K. We adopt a conversion factor $N(\mathrm{H}_{2})/N(^{13}\mathrm{CO}) = 5 \times 10^{5}$ from \cite{2001ApJ...551..747S} (assuming $R(^{12}\mathrm{CO}/^{13}\mathrm{CO}) = 45$ \citep{1990ApJ...357..477L} and $X(^{12}\mathrm{CO}) = 8 \times 10^{-5}$ \citep{1987ApJ...315..621B}), and thus the column density of the total molecular gas is
\begin{equation}
N_{\mathrm{13CO}}(\mathrm{H}_{2}) = 4.2 \times 10^{20} \left( \dfrac{\int T_{\mathrm{B},\mathrm{13CO}}~dv}{\mathrm{K~km~s}^{-1}} \right) \mathrm{cm}^{-2} .
\end{equation}
To calculate the total mass of the gas, we sum over the IRDC and use
\begin{align}
M_{\mathrm{13CO}} = \mu m_{\mathrm{H}} D^{2} \Omega \sum_{\mathrm{IRDC}} N_{\mathrm{13CO}}(\mathrm{H}_{2}) \notag\\ = 0.7 \left( \dfrac{D}{\mathrm{kpc}} \right)^{2} \left( \dfrac{\Omega}{\mathrm{arcmin}^{2}} \right) \left( \dfrac{I_{\mathrm{13CO}}}{\mathrm{K~km~s}^{-1}} \right) M_{\sun} ,
\end{align}
where $\mu$ is the mean molecular weight in multiples of the proton mass (assumed to be 2.33), D is the distance to the region, $\Omega$ is the solid angle occupied by the gas, and $I_{\mathrm{13CO}}$ is the sum of $T_{\mathrm{B},\mathrm{13CO}}$ over the IRDC in position and velocity.

IRDC masses estimated with this method range from about 2090 $M_{\sun}$ to 36,000 $M_{\sun}$. We note that our mass estimates from this method are systematically lower than those reported by \cite{2006ApJ...653.1325S} by factors of about two to six. However, the \cite{2006ApJ...653.1325S} analysis uses ellipses larger than the IRDCs themselves, sometimes including complex networks of filaments extending beyond our maps, to get total masses, and so they expect their own values to be overestimates. Both our mass estimates with this method and those of \cite{2006ApJ...653.1325S} are systematically higher than those derived from IR extinction. We discuss this further in \S \ref{sec-compmass}.

\subsubsection{BGPS Dust Emission Mass Estimates \label{sec-dustem}}

To calculate the gas mass from thermal dust emission, we follow the method of \cite{2009ApJ...697.1457F} applied to $\lambda$ = 1.12 mm maps from the Bolocam Galactic Plane Survey (BGPS) \citep{2011ApJS..192....4A}. The hydrogen column can be estimated by
\begin{equation}
N_{\mathrm{mm}}(\mathrm{H}_{2}) = \dfrac{S_{1.12\mathrm{mm}}}{\Omega_{\mathrm{mb}} \mu m_{\mathrm{H}} \kappa_{1.12\mathrm{mm}} B_{\lambda}(T_{d})} ,
\end{equation}
where $S_{1.12\mathrm{mm}}$ is the flux density of the Bolocam emission, $\Omega_{\mathrm{mb}}$ is the solid angle of the main beam (33$\arcsec$ FWHM), $\mu$ is the mean molecular weight, $m_{\mathrm{H}}$ is the mass of a hydrogen atom, $\kappa_{1.12\mathrm{mm}}$ is the dust emissivity per unit gas mass, and $B_{\lambda}(T_{d})$ is the Planck function for a dust temperature $T_{d}$. We adopt $\kappa_{1.12\mathrm{mm}} = 0.012$ cm$^{2}$ g$^{-1}$, again assuming a dust-to-gas mass ratio of 1:100 and using the dust model of \cite{1994A&A...291..943O} for $10^{5}$ year-old, $10^{6}$ cm$^{-3}$ dust with thin ice mantles. The total mass summed over the IRDC is then
\begin{align}
M_{\mathrm{mm}} = \mu m_{\mathrm{H}} D^{2} \Omega \sum_{\mathrm{IRDC}} N_{\mathrm{mm}}(\mathrm{H}_{2}) \notag\\ = 41 \left( \dfrac{S_{\nu}}{\mathrm{Jy}} \right) \left( \dfrac{D}{\mathrm{kpc}} \right)^{2} \left( \dfrac{\Omega}{\mathrm{arcmin}^{2}} \right) \notag\\ \times \left( \exp \left[ \dfrac{12.85~\mathrm{K}}{T_{d}} \right] - 1 \right) M_{\sun} .
\end{align}
Our choice of dust temperature is informed by the findings of \cite{2014ApJ...786..116B}. They compared the dust temperature inferred from \emph{Herschel} SEDs to the gas (kinetic) temperature measured from the NH$_{3}$ (1,1), (2,2), and (4,4) transitions with the VLA in G031.97+00.07 (called G32.02+0.05 in their work). The dust temperatures, 11.6$\pm$0.2 K, were systematically lower than the gas temperatures, 15.2$\pm$1.5 K, however the two temperatures were generally correlated regardless of local star formation activity. We therefore adopt the median of the clump-averaged kinetic temperatures in Table \ref{t5} for each IRDC multiplied by the ratio 11.6/15.2 as the dust temperature for each IRDC. We do not use our $T_{K}$ maps directly because of the resolution mismatch and the lack of kinematic information in the Bolocam data to separate multiple velocity components. IRDC masses estimated with this method range from about 740 $M_{\sun}$ to 8900 $M_{\sun}$. It should be noted that the resolution of the BGPS is poorer than that of the ammonia maps by a factor of a few, and also does not contain velocity information.

\subsubsection{Comparison of Mass Estimates \label{sec-compmass}}

The IRDC mass estimates from the three methods are shown in Figure \ref{f21}. For the majority of our sample, the masses derived from dust emission are typically consistent with those estimated from the IR extinction within a factor of a few, well within our estimated uncertainties. We note that in G028.23-00.19 and G032.70-00.30 , $M_{\mathrm{IR}}$ and $M_{\mathrm{13CO}}$ agree well, but $M_{\mathrm{mm}}$ is well below the other estimates. These are both quiescent clouds, so it is possible that the typical dust temperature in these IRDCs is lower than the median clump-averaged kinetic temperature we used, and so the dust emission mass estimate is low. In G034.43+00.24, $M_{\mathrm{IR}}$ is a very weak lower limit on the mass because of IR emission from IR point sources, as discussed above.

In every case, $M_{\mathrm{13CO}}$ is greater than $M_{\mathrm{mm}}$ by at least factors of a few, and greater than an order of magnitude in G024.60+00.08. However, these two estimates are clearly well correlated. The $^{13}$CO $J$=1-0 emission is probably optically thick, and may miss gas mass because of CO depletion, but also includes the contribution of the lower density molecular envelopes surrounding the IRDCs. \cite{2011ApJ...730...44H} compared the mass surface density estimates from 8 $\micron$ extinction to that from the BU-GRS, taking care to subtract the contribution from the molecular envelope surrounding the IRDC itself (their study was performed in G034.43+00.24 and G035.39-00.33, though selected slightly different portions of the IRDCs that we studied here). Their mass estimates were also well-correlated though systematically offset, and they estimated the mass surface density of the envelopes to be about 50\% higher that within the IRDC itself. Considering that the envelopes have a greater physical extent than the IRDCs, this effect can easily explain the difference in masses derived from dust and from $^{13}$CO emission.

\subsection{Ammonia Abundance \label{sec-abundance}}

Total masses of the individual clumps are necessary to investigate their stability against gravitational collapse to form stars. We cannot use any of the three methods used above for the total IRDC masses because none of the datasets have both sufficient angular and velocity resolution to be used on individual clumps. It is straightforward to calculate a total NH$_{3}$ column from the spectral line fit, as shown in Appendix \ref{sec-linefit}. We then must assume a fractional abundance, $X$(NH$_{3})$, to convert from NH$_{3}$ column density to H$_{2}$ column density. Our ability to compute the total mass of molecular gas in individual clumps, and thus their gravitational stability, is limited by our knowledge of the ammonia abundance relative to molecule hydrogen. Previous studies of the fractional abundance of NH$_{3}$ in physically similar environments yield a range of values from $10^{-9}$ to $10^{-7}$ \citep[see for example][]{2011ApJ...736..163R,2013A&A...552A..40C,2014ApJ...786..116B}, and may depend on galactocentric radius, IRDC age, and physical environment \citep{2014ApJ...786..116B}. The uncertainty in this conversion factor will translate linearly to the uncertainty in clump masses, and so will affect our analysis of gravitational stability.

Because of the different resolutions and the fact that our ammonia column and mass measurements are restricted to the dense gas, we cannot probe how the abundance varies with environment. However, as shown in Figure \ref{f21}, adopting a value of $X(\mathrm{NH}_{3}) = 10^{-8}$ for the ammonia abundance relative to hydrogen typically gives good agreement with dust emission and IR extinction estimates within a factor of about three. We therefore assume a nominal value of $10^{-8}$ for calculating the masses of clumps, $M_{\mathrm{cl}}$, however the uncertainty of any individual clump mass is likely at least a factor of three.

\subsection{Gravitational Stability \label{sec-clumpstab}}

When calculating the parameters concerning gravitational stability of clumps, we implicitly assume that all filamentary structures are elongated completely within the plane of the sky. We discuss the effects of three-dimensional geometry in Appendix \ref{sec-projection} assuming a random distribution of orientations in space. We cannot correct values for individual clumps with existing observations, but the typical corrections for these values are less than a factor of 2 and often close to 1. The projection effects will be quite large for clumps elongated nearly along the line of sight, but we expect that only a small fraction of clumps have this property. Values reported in this paper are presented without correction for projection effects.

From the mass and size of a clump, we can calculate the mean mass density $\rho_{0}$, and thus the spherical free-fall time,
\begin{align}
t_{\mathrm{ff,sph}} = \sqrt{\dfrac{3 \pi}{32 G \rho_{0}}} = \sqrt{\dfrac{\pi^{2} R_{\mathrm{eff}}^{3}}{8 G M_{\mathrm{cl}}}} \notag\\ = 16.6\left( \dfrac{R_{\mathrm{eff}}}{\mathrm{pc}} \right)^{3/2} \left( \dfrac{M_{\mathrm{cl}}}{M_{\sun}} \right)^{-1/2} \mathrm{Myr} ,
\end{align}
or the filament free-fall time for a cylinder collapsing along its axis,
\begin{align}
t_{\mathrm{ff,cyl}} = \sqrt{\dfrac{2}{3}} A_{0} t_{\mathrm{ff,sph}} \notag\\ = 13.5 A_{0} \left( \dfrac{R_{\mathrm{eff}}}{\mathrm{pc}} \right)^{3/2} \left( \dfrac{M_{\mathrm{cl}}}{M_{\sun}} \right)^{-1/2} \mathrm{Myr} ,
\end{align}
where $A_{0}$ is the initial aspect ratio of the filament \citep{2011ApJ...740...88P}. We adopt the current aspect ratio of the clumps for $A_{0}$, implicitly assuming that these structures are either young or have not evolved substantially since their formation, so our values of $t_{\mathrm{ff,cyl}}$ will be lower limits. Spherical free-fall times for these clumps range from $2 \times 10^{4}$ to $2 \times 10^{5}$ years, and cylindrical free-fall times range from $2 \times 10^{4}$ to $10^{6}$ years. These timescales are consistent either with the expectation that these IRDCs are still in the earliest phases of star formation or that they are supported against free-fall collapse.

The first models of filamentary structures collapsing along characteristic length scales was presented by \cite{1953ApJ...118..116C}. They determined that a cylindrical cloud of incompressible fluid with an axial magnetic field separated into collapsing fragments whose length was initially 12 times the diameter of the cylinder. Following collapse and redistribution of mass, the length to diameter ratio shrinks to 4.5 if the average density remains unchanged, or 2 if there is no magnetic field. \cite{1979ApJS...41...87S} identified filaments in optically dark nebulae and measured ratios of core separation to filament diameter of 3 $\pm$ 1, consistent with these predictions. The aspect ratios of approximately two-thirds of the clumps in our sample are between 1 and 2, but the most filamentary of our clumps have aspect ratios greater than 3, consistent with these predictions and studies.

It is common to calculate the virial mass and virial parameter for spherical clumps via
\begin{equation}
\alpha_{\mathrm{vir,sph}} = \dfrac{M_{\mathrm{vir,sph}}}{M_{\mathrm{cl}}} = \dfrac{5 \sigma_{\mathrm{line}}^{2} R_{\mathrm{eff}}}{G M_{\mathrm{cl}}} ,
\end{equation}
using the total clump mass determined from NH$_{3}$, $M_{\mathrm{cl}}$, the line of sight velocity dispersion $\sigma_{\mathrm{line}}$, and the effective radius. A virial parameter less than 1 indicates a spherical clump prone to collapse, and greater than 1 is resistant to collapse. This formulation of the parameter assumes a uniform density sphere, which is unlikely to be an accurate description. If the clump is spherical but the mass density has a power law distribution, i.e. $\rho \propto r^{-p}$, then the gravitational potential energy, and thus the virial parameter, will change by a factor of $(15-6p)/(15-5p)$. For example, the virial parameter will be a factor of 9/10 as large if $p=1$ or 3/5 as large if $p=2$. Power law density distributions can then reduce the virial parameters by about a factor of 2, and so clumps will be more prone to collapse.

To calculate a virial parameter for filamentary clumps, we use the equilibrium relation from \cite{1953ApJ...118..116C} for an infinite, isothermal cylinder:
\begin{equation}
2 \left( \gamma - 1 \right) \left( \dfrac{\mathcal{U}}{L} \right) + 2 \left( \dfrac{\mathcal{B}}{L} \right) - G \left( \dfrac{M_{\mathrm{vir,cyl}}}{L} \right) = 0 .
\end{equation}
We take $\gamma = 7/5$ for the adiabatic index of diatomic gas, $\left( \mathcal{U}/L \right)$ is the internal kinetic energy of the clump per unit length, $\left( \mathcal{B}/L \right)$ is the magnetic energy per unit length, and $\left( M_{\mathrm{vir,cyl}}/L \right)$ is the cylindrical virial mass per unit length. We ignore the contribution from magnetic fields for now, and assume the internal kinetic energy is turbulence dominated since the linewidths are largely nonthermal, such that
\begin{equation}
\dfrac{\mathcal{U}}{L} = \dfrac{3}{2} \left( \dfrac{M_{\mathrm{vir,cyl}}}{L} \right) \sigma_{\mathrm{line}}^{2} .
\end{equation}

For cylinders of finite length, we note that the $L \approx 2 R_{\mathrm{eff}} \sqrt{A_{0}}$ by approximating the filament as an ellipse (as in \texttt{cprops} to calculate the aspect ratio) and noting that it has a projected area equal to $\pi R_{\mathrm{eff}}^{2}$. We then recover an equivalent virial parameter for cylindrical clumps:
\begin{align}
\alpha_{\mathrm{vir,cyl}} = \dfrac{M_{\mathrm{vir,cyl}}}{M_{\mathrm{cl}}} \notag\\ = \dfrac{12}{5} \sqrt{A_{0}} \dfrac{\sigma_{\mathrm{line}}^{2} R_{\mathrm{eff}}}{G M_{\mathrm{cl}}} = \dfrac{12}{25} \sqrt{A_{0}} \alpha_{\mathrm{vir,sph}}.
\end{align}
For the most filamentary clumps in our sample $A_{0} \gtrsim 3$ (with one being approximately 6), so $\alpha_{\mathrm{vir,cyl}} \approx \alpha_{\mathrm{vir,sph}}$. For clumps with smaller aspect ratios $\alpha_{\mathrm{vir,cyl}} < \alpha_{\mathrm{vir,sph}}$, however this estimate is less reliable because the initial assumption of a cylinder of infinite length is certainly violated in this case. The median value of $\alpha_{\mathrm{vir,sph}}$ is about 1.5, and is about 1 for $\alpha_{\mathrm{vir,cyl}}$.

Given the factor of three uncertainty in the individual clump masses and the assumption of relatively simple geometry, we cannot confidently say which individual clumps are gravitationally bound; however, it is significant that across our large sample most clumps are approximately in equipartition of gravitational and kinetic energies. For clumps that are interacting and not in isolation, then the simple assumptions of virial equilibrium are not met and it becomes more difficult to determine if these structures are bound. However, if the clumps are completely collapsing, then the kinetic energy should be twice that of the gravitational energy. Since the sample average kinetic and potential energies are approximately equal it is more likely that these clumps are prevented from collapsing by their internal motions, rather than that they are already collapsing. This is consistent with the fact that the star formation is not predominantly in the center of the clumps, but at the interfaces between clumps.

We further investigate the role of turbulent support in these clumps by calculating the sound speed of the gas as given by \cite{1953ApJ...118..116C},
\begin{equation}
c_{s} = \sqrt{\dfrac{\gamma k_{\mathrm{B}} T_{K}}{\mu m_{\mathrm{H}}}} ,
\end{equation}
and then the Mach number,
\begin{equation}
\mathcal{M} = \dfrac{\sigma_{\mathrm{NT}}}{c_{s}} .
\end{equation}
We take $\gamma = 7/5$ as above, and list the sound speeds and Mach numbers for the clumps in Table \ref{t6}. The Mach numbers are typically 3 to 5 and all greater than 1 (with the exception of one clump in G034.43+00.24 for which the linewidth is approximately the thermal linewidth). This implies that the clumps are dominated by nonthermal motions. These nonthermal motions may be dominated by turbulence that provides support against gravitational collapse of the clumps. It is also possible that high nonthermal motions arise from YSOs, for example from infall motions, which do not provide support. However, it is unlikely that gas motions directly associated with YSOs can explain the high nonthermal motions across the entirety of the IRDCs, so turbulence is likely the dominate cause and thus an important source of support for these clumps.

\begin{deluxetable}{cccccccccc}
\tablecolumns{10}
\tablewidth{0pt}
\tablecaption{Clump Stability \label{t6}}
\tablehead{
 \colhead{$M_{\mathrm{cl}}$} &
 \colhead{$t_{\mathrm{ff,sph}}$} &
 \colhead{$t_{\mathrm{ff,cyl}}$} &
 \colhead{$M_{\mathrm{vir,sph}}$} &
 \colhead{$M_{\mathrm{vir,cyl}}$} &
 \colhead{$\alpha_{\mathrm{vir,sph}}$} &
 \colhead{$\alpha_{\mathrm{vir,cyl}}$} &
 \colhead{$c_{s}$} &
 \colhead{$\mathcal{M}$} &
 \colhead{$B_{\mathrm{cr}}$} \\
 \colhead{$\left( M_{\sun} \right)$} &
 \colhead{(Myr)} &
 \colhead{(Myr)} &
 \colhead{$\left( M_{\sun} \right)$} &
 \colhead{$\left( M_{\sun} \right)$} &
 &
 &
 \colhead{(km s$^{-1}$)} &
 &
 \colhead{(mG)}
 }
\startdata
 409 & 0.06 & 0.10 & 217 & 152 & 0.53 & 0.37 & 0.26 & 3.9 & 1.37 \\
 454 & 0.05 & 0.05 & 171 & 91 & 0.38 & 0.20 & 0.27 & 3.5 & 1.88 \\
 112 & 0.06 & 0.10 & 59 & 41 & 0.53 & 0.36 & 0.26 & 2.5 & 0.87 \\
 348 & 0.06 & 0.12 & 195 & 149 & 0.56 & 0.43 & 0.27 & 3.8 & 1.36 \\
 486 & 0.05 & 0.11 & 287 & 238 & 0.59 & 0.49 & 0.29 & 4.4 & 2.09 \\
 162 & 0.04 & 0.05 & 264 & 158 & 1.63 & 0.97 & 0.28 & 5.6 & 1.95 \\
 103 & 0.04 & 0.10 & 257 & 207 & 2.50 & 2.01 & 0.28 & 5.5 & 1.33 \\
 174 & 0.05 & 0.06 & 271 & 155 & 1.56 & 0.89 & 0.31 & 4.5 & 1.28 \\
 243 & 0.07 & 0.09 & 136 & 80 & 0.56 & 0.33 & 0.26 & 3.2 & 0.88 \\
  63 & 0.06 & 0.06 & 40 & 21 & 0.63 & 0.33 & 0.27 & 2.2 & 0.72 \\
  88 & 0.05 & 0.05 & 119 & 63 & 1.36 & 0.72 & 0.30 & 3.5 & 0.99 \\
   6 & 0.04 & 0.10 & 3 & 3 & 0.52 & 0.46 & 0.24 & 1.2 & 0.67 \\
  14 & 0.05 & 0.06 & 10 & 6 & 0.70 & 0.42 & 0.28 & 1.4 & 0.59 \\
 199 & 0.03 & 0.04 & 156 & 91 & 0.78 & 0.46 & 0.26 & 4.5 & 2.34 \\
  80 & 0.04 & 0.04 & 85 & 46 & 1.06 & 0.57 & 0.26 & 3.7 & 1.30 \\
  99 & 0.04 & 0.05 & 199 & 128 & 2.01 & 1.30 & 0.25 & 6.0 & 1.73 \\
  21 & 0.03 & 0.05 & 41 & 26 & 1.93 & 1.24 & 0.26 & 3.3 & 1.05 \\
  24 & 0.05 & 0.11 & 82 & 67 & 3.45 & 2.81 & 0.27 & 4.1 & 0.72 \\
  41 & 0.03 & 0.12 & 74 & 74 & 1.81 & 1.81 & 0.25 & 4.2 & 1.31 \\
  42 & 0.03 & 0.04 & 32 & 20 & 0.77 & 0.48 & 0.26 & 2.9 & 1.84 \\
  28 & 0.02 & 0.03 & 58 & 38 & 2.09 & 1.36 & 0.24 & 5.0 & 2.30 \\
 871 & 0.04 & 0.05 & 212 & 118 & 0.24 & 0.13 & 0.28 & 3.7 & 2.89 \\
 103 & 0.05 & 0.08 & 36 & 24 & 0.35 & 0.24 & 0.25 & 2.2 & 1.18 \\
 116 & 0.05 & 0.05 & 631 & 336 & 5.43 & 2.89 & 0.31 & 7.5 & 1.14 \\
 454 & 0.05 & 0.07 & 301 & 179 & 0.66 & 0.39 & 0.30 & 4.1 & 1.62 \\
 408 & 0.06 & 0.09 & 319 & 210 & 0.78 & 0.51 & 0.30 & 4.2 & 1.45 \\
  59 & 0.04 & 0.05 & 39 & 24 & 0.65 & 0.41 & 0.29 & 2.5 & 1.37 \\
  59 & 0.04 & 0.10 & 87 & 71 & 1.46 & 1.20 & 0.29 & 3.5 & 1.16 \\
 551 & 0.05 & 0.06 & 568 & 335 & 1.03 & 0.61 & 0.29 & 5.8 & 1.95 \\
 104 & 0.04 & 0.04 & 225 & 129 & 2.17 & 1.25 & 0.29 & 5.5 & 1.77 \\
  93 & 0.04 & 0.08 & 300 & 219 & 3.21 & 2.35 & 0.31 & 5.7 & 1.40 \\
  96 & 0.03 & 0.07 & 91 & 80 & 0.94 & 0.83 & 0.25 & 4.5 & 2.63 \\
  26 & 0.10 & 0.11 & 383 & 212 & 14.81 & 8.20 & 0.29 & 6.3 & 0.28 \\
 715 & 0.04 & 0.06 & 895 & 560 & 1.25 & 0.78 & 0.30 & 7.2 & 2.71 \\
 109 & 0.04 & 0.11 & 95 & 84 & 0.87 & 0.77 & 0.26 & 3.7 & 1.54 \\
2759 & 0.05 & 0.04 & 1082 & 536 & 0.39 & 0.19 & 0.31 & 5.9 & 3.47 \\
 654 & 0.05 & 0.08 & 177 & 119 & 0.27 & 0.18 & 0.26 & 3.6 & 2.00 \\
2110 & 0.05 & 0.08 & 817 & 566 & 0.39 & 0.27 & 0.29 & 5.8 & 3.39 \\
 102 & 0.03 & 0.13 & 209 & 218 & 2.04 & 2.13 & 0.25 & 6.2 & 1.78 \\
  37 & 0.04 & 0.06 & 104 & 72 & 2.84 & 1.96 & 0.30 & 4.2 & 1.14 \\
 826 & 0.04 & 0.05 & 786 & 436 & 0.95 & 0.53 & 0.30 & 6.5 & 2.69 \\
 219 & 0.05 & 0.09 & 120 & 91 & 0.55 & 0.41 & 0.30 & 3.1 & 1.59 \\
  17 & 0.03 & 0.03 & 9 & 5 & 0.54 & 0.29 & 0.23 & 2.0 & 1.32 \\
  79 & 0.04 & 0.12 & 54 & 50 & 0.69 & 0.63 & 0.27 & 2.9 & 1.31 \\
  14 & 0.05 & 0.08 & 9 & 6 & 0.65 & 0.44 & 0.25 & 1.5 & 0.54 \\
  24 & 0.10 & 0.17 & 222 & 154 & 9.27 & 6.43 & 0.31 & 4.5 & 0.27 \\
  59 & 0.04 & 0.07 & 216 & 149 & 3.65 & 2.51 & 0.31 & 5.1 & 1.06 \\
  12 & 0.06 & 0.09 & 25 & 16 & 2.12 & 1.36 & 0.25 & 2.4 & 0.39 \\
  96 & 0.04 & 0.04 & 670 & 342 & 6.97 & 3.55 & 0.35 & 7.5 & 1.41 \\
   5 & 0.06 & 0.14 & 6 & 5 & 1.35 & 1.09 & 0.24 & 1.5 & 0.31 \\
   7 & 0.04 & 0.09 & 71 & 54 & 10.43 & 8.04 & 0.26 & 5.0 & 0.53 \\
   3 & 0.04 & 0.06 & 4 & 3 & 1.50 & 1.06 & 0.26 & 1.5 & 0.48 \\
   3 & 0.04 & 0.06 & 43 & 28 & 16.55 & 11.03 & 0.26 & 4.6 & 0.42 \\
   3 & 0.04 & 0.08 & 5 & 4 & 1.98 & 1.50 & 0.24 & 1.8 & 0.44 \\
  59 & 0.04 & 0.05 & 449 & 254 & 7.56 & 4.27 & 0.37 & 6.1 & 1.06 \\
  37 & 0.07 & 0.14 & 103 & 78 & 2.79 & 2.10 & 0.26 & 3.8 & 0.48 \\
 196 & 0.06 & 0.14 & 489 & 385 & 2.49 & 1.96 & 0.32 & 5.4 & 1.01 \\
 116 & 0.06 & 0.06 & 305 & 163 & 2.63 & 1.41 & 0.31 & 4.9 & 0.97 \\
  16 & 0.05 & 0.09 & 59 & 43 & 3.65 & 2.64 & 0.27 & 3.6 & 0.58 \\
  55 & 0.04 & 0.05 & 156 & 96 & 2.86 & 1.75 & 0.33 & 4.5 & 1.41 \\
 116 & 0.07 & 0.16 & 176 & 144 & 1.51 & 1.24 & 0.26 & 4.1 & 0.74 \\
  76 & 0.05 & 0.06 & 146 & 83 & 1.91 & 1.09 & 0.32 & 3.7 & 0.93 \\
   3 & 0.03 & 0.05 & 20 & 13 & 7.53 & 4.87 & 0.26 & 3.5 & 0.62 \\
  58 & 0.06 & 0.10 & 73 & 50 & 1.27 & 0.87 & 0.25 & 3.3 & 0.74 \\
 249 & 0.06 & 0.07 & 117 & 71 & 0.47 & 0.28 & 0.27 & 3.1 & 1.23 \\
  12 & 0.05 & 0.07 & 9 & 6 & 0.78 & 0.51 & 0.24 & 1.7 & 0.59 \\
  55 & 0.06 & 0.06 & 93 & 52 & 1.70 & 0.95 & 0.27 & 3.6 & 0.76 \\
   6 & 0.04 & 0.07 & 19 & 13 & 3.41 & 2.33 & 0.26 & 2.7 & 0.55 \\
 126 & 0.05 & 0.05 & 107 & 57 & 0.84 & 0.45 & 0.26 & 3.6 & 1.21 \\
 106 & 0.06 & 0.08 & 125 & 79 & 1.18 & 0.75 & 0.27 & 3.7 & 0.89 \\
 187 & 0.06 & 0.07 & 74 & 42 & 0.39 & 0.23 & 0.26 & 2.6 & 1.05 \\
 115 & 0.05 & 0.05 & 173 & 94 & 1.51 & 0.82 & 0.30 & 4.2 & 1.22 \\
   1 & 0.10 & 0.16 & 16 & 11 & 12.00 & 8.22 & 0.34 & 1.8 & 0.10 \\
  26 & 0.06 & 0.10 & 155 & 104 & 5.88 & 3.93 & 0.26 & 5.3 & 0.50 \\
  71 & 0.06 & 0.09 & 207 & 129 & 2.94 & 1.82 & 0.27 & 4.8 & 0.70 \\
   7 & 0.06 & 0.05 & 17 & 9 & 2.50 & 1.27 & 0.26 & 2.2 & 0.38 \\
   9 & 0.07 & 0.09 & 75 & 45 & 8.59 & 5.18 & 0.27 & 4.1 & 0.33 \\
  55 & 0.07 & 0.11 & 192 & 130 & 3.48 & 2.36 & 0.28 & 4.7 & 0.60 \\
  61 & 0.08 & 0.07 & 120 & 61 & 1.96 & 0.99 & 0.29 & 3.3 & 0.50 \\
  68 & 0.05 & 0.05 & 112 & 64 & 1.65 & 0.94 & 0.30 & 3.6 & 1.05 \\
   6 & 0.06 & 0.07 & 35 & 19 & 5.95 & 3.29 & 0.27 & 3.0 & 0.30 \\
\enddata
\tablecomments{Clumps appear in the same order as Table \ref{t5}.}
\end{deluxetable}

In the absence of magnetic field strength measurements, we can estimate the critical magnetic field strength required to support uniform spherical clumps against gravitational collapse in the absence of turbulence, $B_{\mathrm{cr}}$, by taking the condition that
\begin{equation}
\dfrac{3GM_{\mathrm{cl}}^{2}}{5 R_{\mathrm{eff}}} \approx \dfrac{B_{\mathrm{cr}}^{2}}{8 \pi} \dfrac{4}{3} \pi R_{\mathrm{eff}}^{3} ,
\end{equation}
so that
\begin{equation}
B_{\mathrm{cr}} \approx 1.02 \times 10^{-4} \left( \dfrac{M_{\mathrm{cl}}}{M_{\sun}} \right) \left( \dfrac{R_{\mathrm{eff}}}{\mathrm{pc}} \right)^{-2} \mathrm{mG} .
\end{equation}
If we postulate that the clouds are all supported entirely by magnetic pressure, the required magnetic field strengths to prevent gravitational collapse would be $13~\mu \mathrm{G} \lesssim B_{\mathrm{cr}} \lesssim 3.5~\mathrm{mG} $, well matched to the typical magnetic field strengths measured in molecular clouds \citep{2012ARA&A..50...29C}. These values of $ B_{\mathrm{cr}} $ are upper limits on the critical field strength since the support from turbulence and thermal motions has been ignored, so even smaller magnetic fields will be sufficient to stabilize the clumps. In fact, clumps with virial parameters greater than 1 require no magnetic field to stay supported against collapse. Typical magnetic fields strengths are therefore probably more than sufficient to support these clumps, though unnecessary for support given the supersonic turbulence.

Comparing the free-fall times, the virial parameters, and the turbulent and magnetic support, the majority of clumps in these IRDCs appear to be supported against collapse. The observed nonthermal motions and typical magnetic field strengths in molecular clouds are sufficient to stabilize these clumps against self-gravity. It is likely that many of these clumps are then at least quasi-stable, and may be long-lived compared to the free-fall times. This may explain why we observe quiescent regions that are devoid of any signs of ongoing star formation, and why protostellar candidates are typically found at the sites where clumps overlap. The clumps that are apparently interacting in position-position-velocity space are not evolving in isolation, and so collapse may be progressing in places where clumps are compressed by collisions with each other.

\subsubsection{Size-linewidth Relation \label{sec-sizelinewidth}}

\begin{figure}
\begin{center}
\includegraphics[width=1\textwidth]{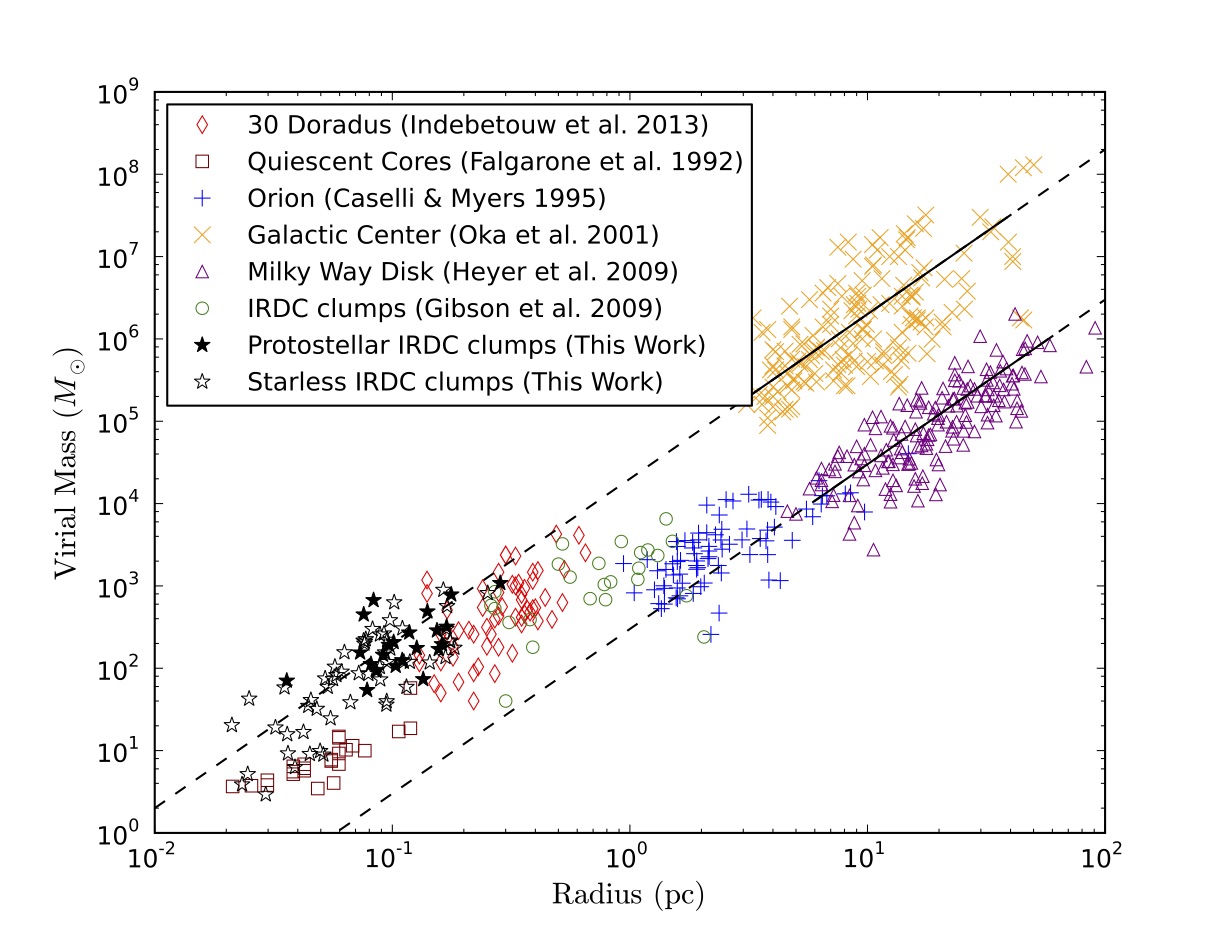}
\caption{A comparison of radius and spherical virial mass (a proxy for linewidth) for the clumps in our sample (using $R_{\mathrm{eff}}$ and $M_{\mathrm{vir,sph}}$) to other studies of molecular gas in different environments. Filled stars indicate clumps from our work that are determined to be protostellar due to the presence of a 70 $\micron$ point source within the clump boundary. Open symbols are starless clumps that lack a 70 $\micron$ point source. The black lines are relations from \cite{2001ApJ...562..348O} and \cite{2009ApJ...699.1092H} for their galactic center clouds and Milky Way disk clouds, respectively, extended to a greater range of cloud radii as dotted lines. See \S \ref{sec-sizelinewidth} for descriptions of the different populations. \label{f22}}
\end{center}
\end{figure}

We can place the clumps identified in this study in the context of other molecular gas clumps in varying environments and physical characteristics by putting them on a plot of size versus linewidth or, equivalently, size and spherical virial mass. In Figure \ref{f22}, red diamonds mark the dense clumps seen in 30 Doradus in the Small Magellanic Cloud in $^{13}$CO $J$=2-1 with ALMA \citep{2013ApJ...774...73I}. Green circles mark dense CS $J$=2-1, 3-2, and 5-4 clumps from a previous study of IRDCs by \cite{2009ApJ...705..123G} with the 14 m FCRAO. \cite{1995ApJ...446..665C} observed massive cores in Orion in $^{13}$CO and C$^{18}$O $J$=1-0, shown in blue crosses. Purple triangles mark galactic clouds listed by \cite{2009ApJ...699.1092H} as observed in $^{12}$CO and $^{13}$CO 1-0 with the FCRAO, with their observed relation plotted in black (extended as a dotted line): $M_{\mathrm{vir,sph}} = 300~M_{\sun} R_{\mathrm{pc}}^{2}$. \cite{2001ApJ...562..348O} observed clouds in the Central Molecular Zone in the galactic center in $^{12}$CO 1-0 with the NRO 45 m telescope (plotted as orange X's), and determined the relation $M_{\mathrm{vir,sph}} = 2 \times 10^{4}~M_{\sun} R_{\mathrm{pc}}^{2}$. Finally, \cite{1992A&A...257..715F} observed quiescent molecular clouds in $^{13}$CO 1-0 and 2-1 with the IRAM 30 m, shown as maroon squares.

We additionally use the presence or absence of a 70 $\micron$ point source within the boundaries of a clump to classify it as ``protostellar'' or ``starless,'' respectively. This classification is listed in Table \ref{t5}. Clumps from this work are shown as black stars in Figure \ref{f22}; protostellar clumps are filled, while starless clumps are empty. It is notable that the trend seen across different studies between radius and linewidth extends over more than 3 orders of magnitude in size, regardless of the molecular line tracers and observations used. We note that both classes of our clumps are along the same trend, though the protostellar clumps are typically larger in size and virial mass (i.e. linewidth). Recall that the thermal linewidths for even $T_{K} = 50$ K gas is only 0.12 km s$^{-1}$, and so the linewidths are largely nonthermal.

\cite{2011ApJS..193...10Z} observed ammonia among other molecular tracers, and also saw that linewidths were elevated around protostellar candidates and were more elevated for more evolved sources. They further found that nonthermal velocity dispersions were supersonic. \cite{2011MNRAS.418.1689U} found that in massive cores the column densities, kinetic temperatures, and linewidths seen in ammonia are all correlated with each other and with bolometric luminosity, indicating that these values are driven largely by the central source. We also observe strong correlations between the column densities, kinetic temperatures, and linewidths in the clumps in our sample.

Molecular clumps may be virialized but still have higher velocity dispersions (and thus virial masses) than Larson's relation for galactic disk clumps if there exists some external pressure. Elevated linewidths may be a result of chaotic gravitational collapse, in which the turbulent motion arises from the collapse. An alternative explanation is pressure from the gravity of the larger IRDC as a whole. In this sample, we have already shown that the clumps have strong supersonic turbulence, which likely accounts for the relationship between the size and velocity dispersions \citep{2011MNRAS.411...65B}. However, the velocity dispersions being elevated above the typical Milky Way relation requires the clumps to not be in equilibrium (as in the case of outflows), or have additional external pressure beyond that of the warm ISM.

\subsection{$N(\mathrm{CCS})/N(\mathrm{NH}_{3})$}

\cite{1992ApJ...392..551S} showed that a relative lack of CCS compared to NH$_3$ indicates a more evolved core. \cite{2009ApJ...699..585H} observed varying abundances of carbon-chain molecules relative to NH$_{3}$ in dark cloud cores and attributed the variation to the different evolution stages of the cores. \cite{2008ApJ...678.1049S} observed 55 millimeter cores in IRDCs with the NRO 45 m telescope and the Atacama Submillimeter Telescope Experiment (ASTE) 10 m telescope. They observed N$_{2}$H$^{+}$, HC$_{3}$N, CCS, NH$_{3}$, and CH$_{3}$OH at 18$\arcsec$-73$\arcsec$ and 0.12 km s$^{-1}$ to 0.5 km s$^{-1}$ resolution. They found that the [CCS]/[N$_{2}$H$^{+}$] ratios of even 24 $\micron$ dark cores were typically less than 1 and concluded that these cores were more chemically evolved than low-mass starless cores where the ratio is 2.6-3.2. \cite{2012A&A...537A...4M}, however, found no variation in the column density ratio, $N(\mathrm{CCS})/N(\mathrm{NH}_{3})$, with evolutionary state of Bok globules, the low-mass dark cloud analogs to IRDCs. All of these studies targeted dense cores but were performed with single-dish observations. \cite{2011ApJ...733...44D} used the VLA and demonstrated that the spatial distributions of the NH$_{3}$ and CCS were highly anticorrelated in a single IRDC. Since these NH$_3$ and CCS transitions have similar excitation conditions (i.e critical densities), these studies show that neither molecule traces all of the gas.

The spatial distributions of NH$_{3}$ and CCS in our sample are shown in Figures \ref{f4}-\ref{f12}. Because of the low signal-to-noise in the CCS line, the GBT observations give a clearer impression the CCS distribution. The CCS generally is not detected over the full spatial extent of the NH$_{3}$ emission. In the scenario where the ratio of these abundances is a viable method for tracing evolutionary stage, we would expect the most evolved (i.e protostellar) cores would have a lower CCS abundance than starless cores.

We investigate the possibility of chemical evolution of different clumps by comparing the column densities of NH$_{3}$ and CCS. \cite{2003AJ....126..311L} calculate the column density of CCS in lower mass star-forming regions from the 22.34403 GHz line using an updated form of the relation from \cite{1992ApJ...392..551S}:
\begin{align}
N(\mathrm{CCS}) = 5.1 \times 10^{11} \tau_{\mathrm{CCS}} \left( \dfrac{\Delta v_{\mathrm{CCS}}}{\mathrm{km~s}^{-1}} \right) Z_{\mathrm{CCS}} \notag\\ \times \dfrac{\exp \left( E_{u,\mathrm{CCS}} / \left[ k_{\mathrm{B}} T_{\mathrm{ex,CCS}} \right] \right)}{\exp \left( E_{u,\mathrm{CCS}} / \left[ k_{\mathrm{B}} T_{\mathrm{ex,CCS}} \right] - 1 \right)} \mathrm{cm}^{-2} ,
\end{align}
where $\Delta v_{\mathrm{CCS}}$ is the FWHM of the CCS line, $ Z_{\mathrm{CCS}}$ is the partition function (24 to 62 for low rotational transitions), $E_{u,\mathrm{CCS}} = 1.12~\mathrm{cm}^{-1}$ is the energy level of the upper state, $T_{\mathrm{ex,CCS}}$ is the excitation temperature of the line, and $\tau_{\mathrm{CCS}}$ is the peak optical depth. The peak optical depth is related to the peak main beam temperature by
\begin{align}
T_{\mathrm{mb,CCS}} = \Phi T_{\mathrm{B},\mathrm{CCS}} = \notag\\ \Phi \left[\mathcal{J}(T_{\mathrm{ex,CCS}}) - \mathcal{J}(T_{\mathrm{bg}}) \right] \left[ 1 - \exp(-\tau_{\mathrm{CCS}}) \right] ,
\end{align}
where $T_{\mathrm{mb,CCS}}$, $T_{\mathrm{B},\mathrm{CCS}}$, and $\mathcal{J}(T)$ are the same as given in Appendix \ref{sec-linefit} but for CCS, $T_{\mathrm{bg}} = 2.73$ K is the background CMB temperature, and $\Phi = 1$ is the assumed beam-filling factor. Linewidths are difficult to measure for the CCS given that it is only detected with a signal-to-noise ratio greater than 5 along few lines of sight (refer to Figures \ref{f4}-\ref{f12}). We therefore take $\Delta v_{\mathrm{CCS}}$ from the ammonia velocity dispersion (second order moment) maps, $U = $ 43, and $T_{\mathrm{ex,CCS}} = 5$ K. With these assumptions, we calculate column densities of approximately $10^{11}$ cm$^{-2}$ and as high as $10^{12}$ cm$^{-2}$ , similar to those observed by \cite{1992ApJ...392..551S} and \cite{2003AJ....126..311L}. Given the scatter in the points, the low signal-to-noise ratio of the CCS emission, and the assumptions made to obtain $N$(CCS) we cannot determine the column densities along individual lines of sight to better than an order of magnitude. We therefore only use the clump averaged values of $N$(CCS) to compare to the column density of the NH$_{3}$.

\begin{figure}[h!]
\begin{center}
\includegraphics[width=1\textwidth]{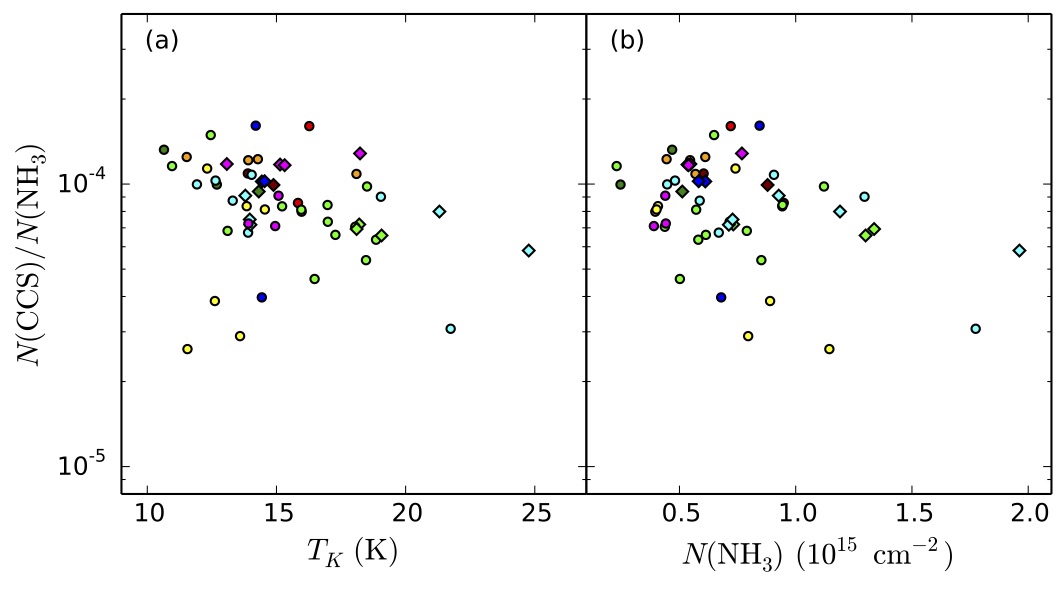}
\caption{A comparison of the ratio of the column densities of CCS and NH$_{3}$ against the kinetic temperature and the ammonia column density, averaged over clumps defined by \texttt{cprops}. Colors are as in Figure \ref{f13}. Diamonds represent clumps that are coincident with a 70 $\micron$ point source (protostellar), and circles represent clumps without such a point source (starless). The distinction between protostellar and starless clumps in $N$(CCS)/$N$(NH$_{3}$) is not as clear as we would expect from previous studies. Given the scatter in the points, the low signal-to-noise ratio of the CCS emission, and the assumptions made to obtain $N$(CCS), more sensitive high resolution studies of CCS, preferably including more than one spectral line, are necessary to say whether or not CCS is relatively depleted in certain environments in IRDCs. \label{f23}}
\end{center}
\end{figure}

We expect from previous studies that CCS will be relatively depleted in the more chemically evolved gas. Without a direct probe of the age of the clumps, we use star formation activity as a proxy. Using the simple classification scheme mentioned in \S \ref{sec-sizelinewidth}, the presence or absence of a 70 $\micron$ point source within the boundaries of a clump are used to classify it as ``protostellar'' or ``starless,'' respectively. We also consider the possibility that temperature and/or density may have an effect on the depletion of CCS onto dust grains. Figure \ref{f23} shows a comparison of the ratio of the column densities of CCS and NH$_{3}$ against the kinetic temperature and the ammonia column density, averaged over clumps defined by \texttt{cprops}. The distinction between protostellar and starless clumps in $N$(CCS)/$N$(NH$_{3}$) is not as clear as we would expect. Figure \ref{f23} shows marginal evidence of a trend for CCS to be relatively depleted compared to NH$_{3}$ in the highest column density gas, but one should be extremely cautious of interpreting this as a true correlation given the uncertainties involved. More sensitive high resolution studies of CCS, preferably including more than one spectral line, are necessary to say whether or not CCS is relatively depleted in certain environments in IRDCs.

\section{Conclusions \label{sec-conclusions}}

In this study, we have mapped the NH$_{3}$ and CCS in nine IRDCs, combining single dish and interferometric data from the GBT and the VLA. From fitting the NH$_{3}$ spectral lines, we probed the physical conditions of these regions. We find typical values of physical parameters in agreement with other studies: kinetic temperatures around 12-25 K, linewidths around 1-2 km s$^{-1}$, clump sizes from $<$0.1 to 0.2 pc, and clump masses of tens to thousands of $M_{\sun}$.

The kinematics seen in NH$_{3}$ reveal a diverse set of configurations and substructure. It appears that no one description of internal structure works for all IRDCs. We observe IRDCs with gradients in the velocity field and velocity dispersion, IRDCs with discrete clumps possibly interacting with each other, and IRDCs that clearly have both of these scenarios occurring simultaneously in different locations. YSOs have a tendency to form at the interface of different clumps. We find that these clumps are typically near virial equilibrium and can easily be supported by typical molecular cloud magnetic field strengths, so these structures may survive for multiple free-fall times. This result is consistent with turbulent core accretion models of massive and intermediate star formation \citep{1997ApJ...476..750M,2003ApJ...585..850M}. The turbulent motions might also arise from the early phases of star formation and thus provide support against gravitational collapse only after some YSOs have been formed.

At least three of these IRDCs are consistent with the picture of hub-filament structure, in which filaments feed molecular gas toward a cental, dense structure harboring star formation. All nine of these IRDCs have lower contrast IR extinction features that extend beyond our molecular maps, and so it is possible that all of these IRDCs are consistent with hub-filament structure. High resolution studies of other physical probes, including shock tracers like SiO, are necessary to further test whether the gas is in fact colliding in these regions.

While this study has one of the largest samples of IRDCs studied at this level of detail, the variability of properties within this sample indicates that it is still not large enough to draw firm conclusions about the entire population of IRDCs. Aside from the dichotomy between globular and filamentary morphologies, there is also variation from more quiescent, slightly supersonic motion to highly supersonic motion. With evidence for at least two different classes of quite distinct morphologies and a possible hybrid classification but only 10 sources we cannot determine with much significance which is most prevalent. A sample of a about 30 would show better if the categories are evenly populated and (about 15 in each category) or if one is more important and whether the hybrid classification is prevalent.

The signal-to-noise ratio of the CCS line is typically low, and so deeper observations and observations of other CCS lines are necessary to better probe the possible chemical differentiation and evolution in IRDCs.

\acknowledgments

The National Radio Astronomy Observatory is a facility of the National Science Foundation operated under cooperative agreement by Associated Universities, Inc. William Dirienzo is a student at the National Radio Astronomy Observatory. This work is based (in part) on observations made with the \emph{Spitzer Space Telescope}, which is operated by the Jet Propulsion Laboratory, California Institute of Technology under a contract with NASA. Support for this work was provided by NASA through an award issued by JPL/Caltech. This research has made use of NASA's Astrophysics Data System. This research has made use of the SIMBAD database, operated at CDS, Strasbourg, France. This research made use of APLpy, an open-source plotting package for Python hosted at http://aplpy.github.com. This research made use of matplotlib, a Python library for publication quality graphics \citep{2007CSE.....9...90H}. This research has made use of SAOImage DS9, developed by Smithsonian Astrophysical Observatory. This work was funded in part by the Jefferson Scholars Foundation at the University of Virginia. This work was funded in part by the Virginia Space Grant Consortium. We thank the referee for detailed and useful comments.

\appendix

\section{IR Extinction Maps \label{sec-irextmaps}}

First, a model of the IR background is generated using a 40$\arcmin$ square median filter sampled every 24$\arcsec$, and then smoothed by a 24$\arcsec$ Gaussian filter. The median filter preferentially skews the model toward pixels lacking strong emission or extinction, provided the size of the filter is larger than the typical size of a dark cloud or other neighboring structures. Increasing the size of the filter also causes the model to be less local, so it is not beneficial to sample the model at the same pixel scale as the original images. The Gaussian smoothing removes unphysically discrete (sharp) features introduced by the discrete sampling of the model.

The fraction of the IR emission that is foreground to the IRDC, $ f_{\mathrm{fore}} $, is estimated assuming a galactic dust distribution, adopting a solar galactocentric radius of 8.4 kpc (consistent with our distance determination in \S \ref{sec-sample}), and using the galactic longitude and distance for each IRDC. Following \cite{2009ApJ...696..484B}, we assume the hot dust follows an azimuthally symmetric distribution matching the surface density of Galactic OB associations,
\begin{equation}
\Sigma_{\mathrm{OB}} \propto \exp \left( - \dfrac{R}{3.5~\mathrm{kpc}} \right) ,
\end{equation}
where $R$ is the galactocentric radius extending to 16 kpc. The values of $f_{\mathrm{fore}}$ are listed in Table \ref{t5}, and range from 0.058 to 0.328. If $ I_{\lambda,0} $ is the intensity of radiation just behind the IRDC and $ I_{\lambda,1} $ is the intensity of radiation just in front of the IRDC, then they are related to the optical depth through the IRDC, $ \tau_{\lambda} $, by
\begin{equation}
I_{\lambda,1} = I_{\lambda,0} \exp \left( - \tau_{\lambda} \right) .
\end{equation}
Accounting for the foreground contribution, the intensity of our background model, $ I_{\lambda,0,\mathrm{obs}} $, is related to the true intensity just behind the IRDC by
\begin{equation}
I_{\lambda,0} = \left( 1 - f_{\mathrm{fore}} \right) I_{\lambda,0,\mathrm{obs}} .
\end{equation}
Similarly, the observed intensity in the images, $ I_{\lambda,1,\mathrm{obs}} $, is related to the true intensity just in front of the IRDC by
\begin{equation}
I_{\lambda,1} = I_{\lambda,1,\mathrm{obs}} - f_{\mathrm{fore}} I_{\lambda,0,\mathrm{obs}} .
\end{equation}
Therefore, the optical depth through the IRDC is obtained by
\begin{equation}
\tau_{\lambda} = - \ln \left( \dfrac{I_{\lambda,1,\mathrm{obs}} - f_{\mathrm{fore}} I_{\lambda,0,\mathrm{obs}}}{ \left( 1 - f_{\mathrm{fore}} \right) I_{\lambda,0,\mathrm{obs}}} \right) .
\end{equation}
\cite{2010ApJ...721..222B} make the point that extended sources in IRAC must be corrected for internal scattering in the instrument, and they include this correction for their analysis of the optical depth at 8 $\micron$. \cite{2007PASP..119..994E}, however, find there is no need to apply such a correction for 24 $\micron$ images from MIPS over a range of a factor of about five in background level, so we do not apply a scattering correction in our calculations.

An optical depth map is then generated on the same pixel scale as the IR images. Point sources and other emission features appear as negative optical depth in the maps, and are thus excluded from analysis. Featureless portions of the IR image correspond to approximately 0 in the optical depth maps as expected.

\section{Ammonia Spectral Line Fitting and Physical Parameters \label{sec-linefit}}

The basic functional form of the fit to a single NH$_{3}$ inversion line is
\begin{equation}
\Delta T^{*}_{A,\nu}(J,K) = \eta_{\mathrm{mb}} \Phi \left[ \mathcal{J} \left( T_{\mathrm{ex}}(J,K) \right) - \mathcal{J} \left( T_{\mathrm{bg}}(J,K) \right) \right] \left[ 1 - \exp \left( -\tau_{\nu}(J,K) \right) \right] ,
\end{equation}
where $\Delta T^{*}_{A,\nu}(J,K)$ is the increase in corrected antenna temperature at the telescope from the line emission, $ \eta_{\mathrm{mb}} $ is the main beam efficiency, $\Phi$ is the beam-filling factor, $ T_{\mathrm{ex}}(J,K) $ is the excitation temperature of the line, $ T_{\mathrm{bg}}(J,K) $ is the background temperature of the line, $ \tau_{\nu}(J,K) $ is the frequency-dependent optical depth, and
\begin{equation}
\mathcal{J}(T) = \dfrac{h \nu}{k_{\mathrm{B}}} \left[ \exp \left( \dfrac{h \nu}{k_{\mathrm{B}} T} \right) - 1 \right]^{-1}
\end{equation}
is the Rayleigh-Jeans temperature for the Planck constant $ h $, the Boltzmann constant $ k_{\mathrm{B}} $, and frequency $ \nu $ \citep{1983ARA&A..21..239H}. We assume $ \Phi = 1 $ (i.e the brightness temperature, $ T_{\mathrm{B}} $, equals the main beam temperature, $ T_{\mathrm{mb}} $). In principle the background temperature includes a contribution from the continuum temperature, $ T_{C}(J,K) $, such that $ T_{\mathrm{bg}}(J,K) = T_{C}(J,K) + T_{\mathrm{CMB}} $, where $ T_{\mathrm{CMB}} = 2.73 $ K is the contribution of the cosmic microwave background. While we cannot exclude the possibility of weak continuum emission, we can set a limit on its contribution. The continuum emission is within the noise of 0 K over the majority of our observed area. The strongest possible continuum emission in our study is about 1 K, but is in a region in G034.43+00.24 with peak emission over 15 K and large linewidths such that the hyperfine components here are blended. This is then an upper limit on the continuum emission taken from the lowest flux density channel between the peaks of the hyperfine components. We set $ T_{C}(J,K) = 0 $ everywhere due to the absence of evidence for continuum emission or a way to measure it reliably.

The line shape as a function of frequency, $\nu$, is primarily determined by the optical depth profile:
\begin{equation}
\tau_{\nu}(J,K) = \tau_{0}(J,K) \sum_{i=1}^{\mathcal{N}(J,K)} a_{i}(J,K) \exp \left( -4 \ln 2 \left( \dfrac{\nu - \nu_{c}(J,K) - \delta \nu_{i}(J,K)}{\Delta \nu(J,K)} \right)^{2} \right) ,
\end{equation}
where $ \tau_{0}(J,K) $ is the opacity in the $ (J,K) $ line, $ \mathcal{N}(J,K) $ is the number of hyperfine components in the $ (J,K) $ line, $ i $ is the index for the hyperfine components, $ a_{i}(J,K) $ is the scale factor for the relative intensity contained in each hyperfine component, $ \nu_{0}(J,K) $ is the rest frequency of the $ (J,K) $ line, $ \nu_{c}(J,K) $ is the Doppler shifted cental frequency of the $ (J,K) $ line, $ \delta \nu_{i}(J,K) $ is the difference in frequency of the $i$th hyperfine component of the $ (J,K) $ line and the main hyperfine component, and $ \Delta \nu (J,K) $ is the FWHM of the $ (J,K) $ line \citep{2009ApJ...697.1457F}.

Treating NH$_{3}$ as a two-state system, the ratio between the optical depths of two different fine structure lines is given by \cite{1983ARA&A..21..239H} as
\begin{align}
\dfrac{\tau_{0}(J',K')}{\tau_{0}(J,K)} = \left( \dfrac{\nu_{0}(J',K')}{\nu_{0}(J,K)} \right)^{2} \left( \dfrac{\Delta \nu(J',K')}{\Delta \nu(J,K)} \right)^{-1} \left( \dfrac{T_{\mathrm{ex}}(J',K')}{T_{\mathrm{ex}}(J,K)} \right)^{-1} \left( \dfrac{\left|\mu(J',K') \right|^{2}}{\left|\mu(J,K) \right|^{2}} \right) \left( \dfrac{g(J',K')}{g(J,K)} \right) \notag\\ \times \exp \left( - \dfrac{\Delta E(J',K';J,K)}{k_{\mathrm{B}} T_{\mathrm{R}}(J',K';J,K)} \right) ,
\end{align}
where $ \Delta E (J',K';J,K) $ is the difference in energy of the lower levels of the transitions, $T_{\mathrm{R}}(J',K';J,K)$ is the rotation temperature of the lines, $ \left|\mu(J,K) \right|^{2} = \mu_{D}^{2} K^{2} / \left( J(J+1)
\right) $ for the electric dipole moment of the molecule $\mu_{D}$, and $ g(J,K) $ is the statistical weight given by \cite{2009ApJ...694...29O} as
\begin{equation}
g(J,K) = \left\{ \begin{array}{ll}
4(2J+1) & \mathrm{for}~K \neq \dot{3}~\mathrm{or}~K = 0 \\
8(2J+1) & \mathrm{for}~K = \dot{3}~\mathrm{and}~K \neq 0 ,
\end{array} \right.
\end{equation}
where $\dot{3}$ denotes being a multiple of 3. This accounts for the rotational degeneracy ($g_{u} = 2J_{u}+1$), the $K$ degeneracy ($g_{K} = 1$ for $K = 0$ or $g_{K} = 2$ for $K \neq 0$ in symmetric top molecules like NH$_{3}$), and the nuclear spin degeneracy ($g_{\mathrm{nuclear}} = 4$ for $K = \dot{3}$ (ortho-NH$_{3}$) or $g_{\mathrm{nuclear}} = 2$ for $K \neq \dot{3}$ (para-NH$_{3}$)). We further assume the excitation temperature, the central velocity, and the velocity FWHM are constant for all $ (J,K) $ (thus $\Delta \nu (J',K') / \Delta \nu (J,K) = \nu_{0}(J',K') / \nu_{0}(J,K) $).

From the fit parameters, it is straightforward to calculate the kinetic temperature of the gas and the total column density of the NH$_{3}$. The kinetic temperature, $T_{K}$, is given by \cite{2004A&A...416..191T} as
\begin{equation}
T_{K} = T_{\mathrm{R}}(2,2;1,1) \left[ 1 - \left( \dfrac{T_{\mathrm{R}}(2,2;1,1)}{42~\mathrm{K}} \right) \ln \left( 1 + 1.1 \exp \left[ - \dfrac{16~\mathrm{K}}{T_{\mathrm{R}}(2,2;1,1)} \right] \right) \right]^{-1} ,
\end{equation}
and the column density of NH$_{3}$ in the (1,1) state is given by \cite{2009ApJ...697.1457F} as
\begin{equation}
N(1,1) = \dfrac{8 \pi \nu_{0}^{2}}{c^{2} A(1,1)} \dfrac{g_{1}}{g_{2}} \dfrac{1 + \exp \left( h \nu_{0} / \left[ k_{\mathrm{B}} T_{\mathrm{ex}} \right] \right)}{1 - \exp \left( h \nu_{0} / \left[ k_{\mathrm{B}} T_{\mathrm{ex}} \right] \right)} \int \tau_{\nu}~d\nu ,
\end{equation}
where the Einstein A coefficient $ A(1,1) = 1.68 \times 10^{-7}~\mathrm{s}^{-1} $ \citep{1998JQSRT..60..883P}, $g_{1} = g_{2}$ for the statistical weights of the upper and lower energy levels of the (1,1) transition, and
\begin{equation}
\int \tau_{\nu}~d\nu = \left( \dfrac{\sqrt{\pi} \tau_{0}(1,1) \Delta \nu (1,1) }{2 \sqrt{\ln 2}} \right) .
\end{equation}
The total NH$_{3}$ column density can then be calculated by considering the partition function for NH$_{3}$ and using the relation $ N(\mathrm{NH}_{3}) = N(1,1) Z_{\mathrm{tot}}/Z(1,1) $ \citep{2009ApJ...697.1457F}. The terms of the partition function are calculated by
\begin{equation}
Z(J,J) = \left(2J+1\right) S(J) \exp \left( \dfrac{-h\left[ B J(J+1)+(C-B) J^{2} \right]}{k_{\mathrm{B}} T_{\mathrm{R}}(2,2;1,1)} \right) ,
\end{equation}
where the rotational constants are $ B = 298.117 $ GHz and $ C = 186.726 $ GHz \citep{1998JQSRT..60..883P}, and $S(J)$ is the statistical weight for ortho- or para-NH$_{3}$. For ortho-NH$_{3}$ ($J$ a multiple of 3), $ S(J) = 2$, and otherwise $ S(J) = 1$ for para-NH$_{3}$, such as (1,1). $ Z_{\mathrm{tot}} $ is then just the sum of the $ Z(J,J) $ terms over all $ J $, which quickly converges. In the rest of this paper, we drop the $(2,2;1,1)$ notation for simplicity.

\section{Projection Effects \label{sec-projection}}

Many of the quantities computed in this paper depend on the inclination of the elongated structures with respect to the line of sight, $\theta$. We cannot make corrections for projection effects for individual clumps without a method to measure $\theta$. We can discuss how various quantities scale with $\theta$ and, assuming a random distribution of inclinations in space, compute typical corrections. Following \cite{2014A&A...561A..83P}, we consider $\theta = 45^{\circ}$ and $\theta = 67^{\circ}$ (the average inclination angle for randomly sampled filaments is $\theta = 67^{\circ}$). The projection effects will be quite large for clumps elongated nearly along the line of sight, but we expect that only a small fraction of clumps have this property.

Masses, column densities, and parameters from extinction mapping and spectral line fitting are all relatively insensitive to projection effects. The viewing angle most directly affects the measured aspect ratio of the clumps. The ``true'' aspect ratio, $A$, is related to the aspect ratio we measure by $A = A_{0} ( \sin \theta )^{-1}$. This will also manifest itself in the calculation of $R_{\mathrm{eff}}$. If we treat elongated clumps as prolate spheroids that are longer by a factor $A_{0}$ along one axis compared to the other two, then the projected area on the sky also scales as $( \sin \theta )^{-1}$. The effective radius then scales as $R_{\mathrm{eff}} \propto ( \sin \theta )^{-1/2}$. We list the parameters that depend on the inclination angle and their scaling in Table \ref{t7}, as well as the correction factor for $\theta = 45^{\circ}$ and $\theta = 67^{\circ}$. One can see the parameter that is most sensitive to viewing angle is $t_{\mathrm{ff,cyl}}$, which is changed by less than a factor of 2 even for $\theta = 45^{\circ}$.

\begin{deluxetable}{rlcc}
\tablecolumns{4}
\tablewidth{0pt}
\tablecaption{Projection Effects \label{t7}}
\tablehead{
 \colhead{Parameter} &
 \colhead{Scaling} &
 \multicolumn{2}{c}{Correction Factor} \\
 \colhead{} &
 \colhead{} &
 \colhead{$\theta = 45^{\circ}$} &
 \colhead{$\theta = 67^{\circ}$}
 }
\startdata
$A_{0}$ & $\propto \left( \sin \theta \right)^{-1}$ & 1.41 & 1.09 \\
$R_{\mathrm{eff}}$ & $\propto \left( \sin \theta \right)^{-1/2}$ & 1.19 & 1.04 \\
$t_{\mathrm{ff,sph}}$ & $\propto \left( \sin \theta \right)^{-3/4}$ & 1.30 & 1.06 \\
$t_{\mathrm{ff,cyl}}$ & $\propto \left( \sin \theta \right)^{-7/4}$ & 1.83 & 1.16 \\
$\alpha_{\mathrm{vir,sph}}$ & $\propto \left( \sin \theta \right)^{-1/2}$ & 1.19 & 1.04 \\
$\alpha_{\mathrm{vir,cyl}}$ & $\propto \left( \sin \theta \right)^{-1}$ & 1.41 & 1.09 \\
$B_{\mathrm{cr}}$ & $\propto \phantom{(} \sin \theta \phantom{)} $ & 0.71 & 0.92 \\
\enddata
\end{deluxetable}

\bibliographystyle{apj}
\bibliography{bibtex}

\begin{thebibliography}{105}
\expandafter\ifx\csname natexlab\endcsname\relax\def\natexlab#1{#1}\fi

\bibitem[{Aguirre {et~al.}(2011)Aguirre, Ginsburg, Dunham, Drosback, Bally,
  Battersby, Bradley, Cyganowski, Dowell, Evans, Glenn, Rosolowsky,
  Stringfellow, Walawender, \& Williams}]{2011ApJS..192....4A}
Aguirre, J.~E., {et~al.} 2011, The Astrophysical Journal Supplement, 192, 4

\bibitem[{Alves {et~al.}(1998)Alves, Lada, Lada, Kenyon, \&
  Phelps}]{1998ApJ...506..292A}
Alves, J., Lada, C.~J., Lada, E.~A., Kenyon, S.~J., \& Phelps, R. 1998, The
  Astrophysical Journal, 506, 292

\bibitem[{Ballesteros-Paredes {et~al.}(2011)Ballesteros-Paredes, Hartmann,
  V{\'a}zquez-Semadeni, Heitsch, \& Zamora-Avil{\'e}s}]{2011MNRAS.411...65B}
Ballesteros-Paredes, J., Hartmann, L.~W., V{\'a}zquez-Semadeni, E., Heitsch,
  F., \& Zamora-Avil{\'e}s, M.~A. 2011, Monthly Notices of the Royal
  Astronomical Society, 411, 65

\bibitem[{Battersby {et~al.}(2014{\natexlab{a}})Battersby, Bally, Dunham,
  Ginsburg, Longmore, \& Darling}]{2014ApJ...786..116B}
Battersby, C., Bally, J., Dunham, M., Ginsburg, A., Longmore, S., \& Darling,
  J. 2014{\natexlab{a}}, The Astrophysical Journal, 786, 116

\bibitem[{Battersby {et~al.}(2010)Battersby, Bally, Jackson, Ginsburg, Shirley,
  Schlingman, \& Glenn}]{2010ApJ...721..222B}
Battersby, C., Bally, J., Jackson, J.~M., Ginsburg, A., Shirley, Y.~L.,
  Schlingman, W., \& Glenn, J. 2010, The Astrophysical Journal, 721, 222

\bibitem[{Battersby {et~al.}(2014{\natexlab{b}})Battersby, Ginsburg, Bally,
  Longmore, Dunham, \& Darling}]{2014ApJ...787..113B}
Battersby, C., Ginsburg, A., Bally, J., Longmore, S., Dunham, M., \& Darling,
  J. 2014{\natexlab{b}}, The Astrophysical Journal, 787, 113

\bibitem[{Benjamin {et~al.}(2003)Benjamin, Churchwell, Babler, Bania, Clemens,
  Cohen, Dickey, Indebetouw, Jackson, Kobulnicky, Lazarian, Marston, Mathis,
  Meade, Seager, Stolovy, Watson, Whitney, Wolff, \&
  Wolfire}]{2003PASP..115..953B}
Benjamin, R.~A., {et~al.} 2003, The Publications of the Astronomical Society of
  the Pacific, 115, 953

\bibitem[{Bergin \& Tafalla(2007)}]{2007ARA&A..45..339B}
Bergin, E.~A., \& Tafalla, M. 2007, Annual Review of Astronomy and
  Astrophysics, 45, 339

\bibitem[{Beuther \& Steinacker(2007)}]{2007ApJ...656L..85B}
Beuther, H., \& Steinacker, J. 2007, The Astrophysical Journal, 656, L85

\bibitem[{Blake {et~al.}(1987)Blake, Sutton, Masson, \&
  Phillips}]{1987ApJ...315..621B}
Blake, G.~A., Sutton, E.~C., Masson, C.~R., \& Phillips, T.~G. 1987,
  Astrophysical Journal, 315, 621

\bibitem[{Blitz \& Shu(1980)}]{1980ApJ...238..148B}
Blitz, L., \& Shu, F.~H. 1980, Astrophysical Journal, 238, 148

\bibitem[{Bowers \& Knapp(1989)}]{1989ApJ...347..325B}
Bowers, P.~F., \& Knapp, G.~R. 1989, Astrophysical Journal, 347, 325

\bibitem[{Butler \& Tan(2009)}]{2009ApJ...696..484B}
Butler, M.~J., \& Tan, J.~C. 2009, The Astrophysical Journal, 696, 484

\bibitem[{Carey {et~al.}(1998)Carey, Clark, Egan, Price, Shipman, \&
  Kuchar}]{1998ApJ...508..721C}
Carey, S.~J., Clark, F.~O., Egan, M.~P., Price, S.~D., Shipman, R.~F., \&
  Kuchar, T.~A. 1998, The Astrophysical Journal, 508, 721

\bibitem[{Carey {et~al.}(2009)Carey, Noriega-Crespo, Mizuno, Shenoy, Paladini,
  Kraemer, Price, Flagey, Ryan, Ingalls, Kuchar, Pinheiro~Gon{\c c}alves,
  Indebetouw, Billot, Marleau, Padgett, Rebull, Bressert, Ali, Molinari,
  Martin, Berriman, Boulanger, Latter, Miville-Deschenes, Shipman, \&
  Testi}]{2009PASP..121...76C}
Carey, S.~J., {et~al.} 2009, Publications of the Astronomical Society of the
  Pacific, 121, 76

\bibitem[{Caselli \& Myers(1995)}]{1995ApJ...446..665C}
Caselli, P., \& Myers, P.~C. 1995, Astrophysical Journal v.446, 446, 665

\bibitem[{Chandrasekhar \& Fermi(1953)}]{1953ApJ...118..116C}
Chandrasekhar, S., \& Fermi, E. 1953, Astrophysical Journal, 118, 116

\bibitem[{Chira {et~al.}(2013)Chira, Beuther, Linz, Schuller, Walmsley, Menten,
  \& Bronfman}]{2013A&A...552A..40C}
Chira, R.~A., Beuther, H., Linz, H., Schuller, F., Walmsley, C.~M., Menten,
  K.~M., \& Bronfman, L. 2013, Astronomy and Astrophysics, 552, 40

\bibitem[{Churchwell {et~al.}(2009)Churchwell, Babler, Meade, Whitney,
  Benjamin, Indebetouw, Cyganowski, Robitaille, Povich, Watson, \&
  Bracker}]{2009PASP..121..213C}
Churchwell, E., {et~al.} 2009, Publications of the Astronomical Society of the
  Pacific, 121, 213

\bibitem[{Contreras {et~al.}(2013)Contreras, Rathborne, \&
  Garay}]{2013MNRAS.433..251C}
Contreras, Y., Rathborne, J., \& Garay, G. 2013, Monthly Notices of the Royal
  Astronomical Society, 433, 251

\bibitem[{Crutcher(2012)}]{2012ARA&A..50...29C}
Crutcher, R.~M. 2012, Annual Review of Astronomy and Astrophysics, 50, 29

\bibitem[{Cyganowski {et~al.}(2008)Cyganowski, Whitney, Holden, Braden, Brogan,
  Churchwell, Indebetouw, Watson, Babler, Benjamin, Gomez, Meade, Povich,
  Robitaille, \& Watson}]{2008AJ....136.2391C}
Cyganowski, C.~J., {et~al.} 2008, The Astronomical Journal, 136, 2391

\bibitem[{Devine {et~al.}(2011)Devine, Chandler, Brogan, Churchwell,
  Indebetouw, Shirley, \& Borg}]{2011ApJ...733...44D}
Devine, K.~E., Chandler, C.~J., Brogan, C., Churchwell, E., Indebetouw, R.,
  Shirley, Y., \& Borg, K.~J. 2011, The Astrophysical Journal, 733, 44

\bibitem[{Dobbs(2008)}]{2008MNRAS.391..844D}
Dobbs, C.~L. 2008, Monthly Notices of the Royal Astronomical Society, 391, 844

\bibitem[{Draine(2003)}]{2003ARA&A..41..241D}
Draine, B.~T. 2003, Annual Review of Astronomy {\&}Astrophysics, 41, 241

\bibitem[{Egan {et~al.}(1998)Egan, Shipman, Price, Carey, Clark, \&
  Cohen}]{1998ApJ...494L.199E}
Egan, M.~P., Shipman, R.~F., Price, S.~D., Carey, S.~J., Clark, F.~O., \&
  Cohen, M. 1998, Astrophysical Journal Letters v.494, 494, L199

\bibitem[{Engelbracht {et~al.}(2007)Engelbracht, Blaylock, Su, Rho, Rieke,
  Muzerolle, Padgett, Hines, Gordon, Fadda, Noriega-Crespo, Kelly, Latter,
  Hinz, Misselt, Morrison, Stansberry, Shupe, Stolovy, Wheaton, Young,
  Neugebauer, Wachter, P{\'e}rez-Gonz{\'a}lez, Frayer, \&
  Marleau}]{2007PASP..119..994E}
Engelbracht, C.~W., {et~al.} 2007, The Publications of the Astronomical Society
  of the Pacific, 119, 994

\bibitem[{Falgarone {et~al.}(1992)Falgarone, Puget, \&
  P{\'e}rault}]{1992A&A...257..715F}
Falgarone, E., Puget, J.~L., \& P{\'e}rault, M. 1992, Astronomy and
  Astrophysics (ISSN 0004-6361), 257, 715

\bibitem[{Fazio {et~al.}(2004)Fazio, Hora, Allen, Ashby, Barmby, Deutsch,
  Huang, Kleiner, Marengo, Megeath, Melnick, Pahre, Patten, Polizotti, Smith,
  Taylor, Wang, Willner, Hoffmann, Pipher, Forrest, McMurty, McCreight,
  McKelvey, McMurray, Koch, Moseley, Arendt, Mentzell, Marx, Losch, Mayman,
  Eichhorn, Krebs, Jhabvala, Gezari, Fixsen, Flores, Shakoorzadeh, Jungo,
  Hakun, Workman, Karpati, Kichak, Whitley, Mann, Tollestrup, Eisenhardt,
  Stern, Gorjian, Bhattacharya, Carey, Nelson, Glaccum, Lacy, Lowrance, Laine,
  Reach, Stauffer, Surace, Wilson, Wright, Hoffman, Domingo, \&
  Cohen}]{2004ApJS..154...10F}
Fazio, G.~G., {et~al.} 2004, The Astrophysical Journal Supplement Series, 154,
  10

\bibitem[{Fiege \& Pudritz(2000)}]{2000MNRAS.311...85F}
Fiege, J.~D., \& Pudritz, R.~E. 2000, Monthly Notices of the Royal Astronomical
  Society, 311, 85

\bibitem[{Foster {et~al.}(2012)Foster, Stead, Benjamin, Hoare, \&
  Jackson}]{2012ApJ...751..157F}
Foster, J.~B., Stead, J.~J., Benjamin, R.~A., Hoare, M.~G., \& Jackson, J.~M.
  2012, The Astrophysical Journal, 751, 157

\bibitem[{Friesen {et~al.}(2009)Friesen, Di~Francesco, Shirley, \&
  Myers}]{2009ApJ...697.1457F}
Friesen, R.~K., Di~Francesco, J., Shirley, Y.~L., \& Myers, P.~C. 2009, The
  Astrophysical Journal, 697, 1457

\bibitem[{Fukui {et~al.}(2014)Fukui, Ohama, Hanaoka, Furukawa, Torii, Dawson,
  Mizuno, Hasegawa, Fukuda, Soga, Moribe, Kuroda, Hayakawa, Kawamura, Kuwahara,
  Yamamoto, Okuda, Onishi, Maezawa, \& Mizuno}]{2014ApJ...780...36F}
Fukui, Y., {et~al.} 2014, The Astrophysical Journal, 780, 36

\bibitem[{Fukui {et~al.}(2015)Fukui, Torii, Ohama, Hasegawa, Hattori, Sano,
  Ohashi, Fujii, Kuwahara, Mizuno, Dawson, Yamamoto, Tachihara, Okuda, Onishi,
  \& Mizuno}]{2015arXiv150405391F}
---. 2015, arXiv.org, 5391

\bibitem[{Furukawa {et~al.}(2009)Furukawa, Dawson, Ohama, Kawamura, Mizuno,
  Onishi, \& Fukui}]{2009ApJ...696L.115F}
Furukawa, N., Dawson, J.~R., Ohama, A., Kawamura, A., Mizuno, N., Onishi, T.,
  \& Fukui, Y. 2009, The Astrophysical Journal Letters, 696, L115

\bibitem[{Gibson {et~al.}(2009)Gibson, Plume, Bergin, Ragan, \&
  Evans}]{2009ApJ...705..123G}
Gibson, D., Plume, R., Bergin, E., Ragan, S., \& Evans, N. 2009, The
  Astrophysical Journal, 705, 123

\bibitem[{Hacar {et~al.}(2013)Hacar, Tafalla, Kauffmann, \&
  Kovacs}]{2013A&A...554A..55H}
Hacar, A., Tafalla, M., Kauffmann, J., \& Kovacs, A. 2013, Astronomy and
  Astrophysics, 554, 55

\bibitem[{Heiderman {et~al.}(2010)Heiderman, Evans, Allen, Huard, \&
  Heyer}]{2010ApJ...723.1019H}
Heiderman, A., Evans, N. J.~I., Allen, L.~E., Huard, T., \& Heyer, M. 2010, The
  Astrophysical Journal, 723, 1019

\bibitem[{Hennebelle {et~al.}(2001)Hennebelle, P{\'e}rault, Teyssier, \&
  Ganesh}]{2001A&A...365..598H}
Hennebelle, P., P{\'e}rault, M., Teyssier, D., \& Ganesh, S. 2001, Astronomy
  and Astrophysics, 365, 598

\bibitem[{Henning {et~al.}(2010)Henning, Linz, Krause, Ragan, Beuther,
  Launhardt, Nielbock, \& Vasyunina}]{2010A&A...518L..95H}
Henning, T., Linz, H., Krause, O., Ragan, S., Beuther, H., Launhardt, R.,
  Nielbock, M., \& Vasyunina, T. 2010, Astronomy and Astrophysics, 518, L95

\bibitem[{Henshaw {et~al.}(2014)Henshaw, Caselli, Fontani, Jim{\'e}nez-Serra,
  \& Tan}]{2014MNRAS.440.2860H}
Henshaw, J.~D., Caselli, P., Fontani, F., Jim{\'e}nez-Serra, I., \& Tan, J.~C.
  2014, Monthly Notices of the Royal Astronomical Society, 440, 2860

\bibitem[{Hernandez \& Tan(2011)}]{2011ApJ...730...44H}
Hernandez, A.~K., \& Tan, J.~C. 2011, The Astrophysical Journal, 730, 44

\bibitem[{Hernandez {et~al.}(2012)Hernandez, Tan, Kainulainen, Caselli, Butler,
  Jim{\'e}nez-Serra, \& Fontani}]{2012ApJ...756L..13H}
Hernandez, A.~K., Tan, J.~C., Kainulainen, J., Caselli, P., Butler, M.~J.,
  Jim{\'e}nez-Serra, I., \& Fontani, F. 2012, The Astrophysical Journal
  Letters, 756, L13

\bibitem[{Heyer {et~al.}(2009)Heyer, Krawczyk, Duval, \&
  Jackson}]{2009ApJ...699.1092H}
Heyer, M., Krawczyk, C., Duval, J., \& Jackson, J.~M. 2009, The Astrophysical
  Journal, 699, 1092

\bibitem[{Hirota {et~al.}(2009)Hirota, Ohishi, \&
  Yamamoto}]{2009ApJ...699..585H}
Hirota, T., Ohishi, M., \& Yamamoto, S. 2009, The Astrophysical Journal, 699,
  585

\bibitem[{Ho \& Townes(1983)}]{1983ARA&A..21..239H}
Ho, P. T.~P., \& Townes, C.~H. 1983, IN: Annual review of astronomy and
  astrophysics. Volume 21 (A84-10851 01-90). Palo Alto, 21, 239

\bibitem[{Indebetouw {et~al.}(2013)Indebetouw, Brogan, Chen, Leroy, Johnson,
  Muller, Madden, Cormier, Galliano, Hughes, Hunter, Kawamura, Kepley,
  Lebouteiller, Meixner, Oliveira, Onishi, \& Vasyunina}]{2013ApJ...774...73I}
Indebetouw, R., {et~al.} 2013, The Astrophysical Journal, 774, 73

\bibitem[{Inoue \& Fukui(2013)}]{2013ApJ...774L..31I}
Inoue, T., \& Fukui, Y. 2013, The Astrophysical Journal Letters, 774, L31

\bibitem[{Jackson {et~al.}(2006)Jackson, Rathborne, Shah, Simon, Bania,
  Clemens, Chambers, Johnson, Dormody, Lavoie, \& Heyer}]{2006ApJS..163..145J}
Jackson, J.~M., {et~al.} 2006, The Astrophysical Journal Supplement Series,
  163, 145

\bibitem[{Jim{\'e}nez-Serra {et~al.}(2014)Jim{\'e}nez-Serra, Caselli, Fontani,
  Tan, Henshaw, Kainulainen, \& Hernandez}]{2014MNRAS.439.1996J}
Jim{\'e}nez-Serra, I., Caselli, P., Fontani, F., Tan, J.~C., Henshaw, J.~D.,
  Kainulainen, J., \& Hernandez, A.~K. 2014, Monthly Notices of the Royal
  Astronomical Society, 439, 1996

\bibitem[{Jim{\'e}nez-Serra {et~al.}(2010)Jim{\'e}nez-Serra, Caselli, Tan,
  Hernandez, Fontani, Butler, \& Van~Loo}]{2010MNRAS.406..187J}
Jim{\'e}nez-Serra, I., Caselli, P., Tan, J.~C., Hernandez, A.~K., Fontani, F.,
  Butler, M.~J., \& Van~Loo, S. 2010, Monthly Notices of the Royal Astronomical
  Society, 406, 187

\bibitem[{Kim {et~al.}(2010)Kim, Lee, Kim, Lee, Ballesteros-Paredes, Myers, \&
  Kurtz}]{2010JKAS...43....9K}
Kim, G., Lee, C.~W., Kim, J., Lee, Y., Ballesteros-Paredes, J., Myers, P.~C.,
  \& Kurtz, S. 2010, Journal of the Korean Astronomical Society, 43, 9

\bibitem[{Kukolich(1967)}]{1967PhRv..156...83K}
Kukolich, S.~G. 1967, Physical Review, 156, 83

\bibitem[{Kurayama {et~al.}(2011)Kurayama, Nakagawa, Sawada-Satoh, Sato, Honma,
  Sunada, Hirota, \& Imai}]{2011PASJ...63..513K}
Kurayama, T., Nakagawa, A., Sawada-Satoh, S., Sato, K., Honma, M., Sunada, K.,
  Hirota, T., \& Imai, H. 2011, Publications of the Astronomical Society of
  Japan, 63, 513

\bibitem[{Lada {et~al.}(1999)Lada, Alves, \& Lada}]{1999ApJ...512..250L}
Lada, C.~J., Alves, J., \& Lada, E.~A. 1999, The Astrophysical Journal, 512,
  250

\bibitem[{Lai {et~al.}(2003)Lai, Velusamy, Langer, \&
  Kuiper}]{2003AJ....126..311L}
Lai, S.-P., Velusamy, T., Langer, W.~D., \& Kuiper, T. B.~H. 2003, The
  Astronomical Journal, 126, 311

\bibitem[{Langer \& Penzias(1990)}]{1990ApJ...357..477L}
Langer, W.~D., \& Penzias, A.~A. 1990, Astrophysical Journal, 357, 477

\bibitem[{Li {et~al.}(2013)Li, Fang, Henning, \&
  Kainulainen}]{2013MNRAS.436.3707L}
Li, H.-b., Fang, M., Henning, T., \& Kainulainen, J. 2013, Monthly Notices of
  the Royal Astronomical Society, 436, 3707

\bibitem[{Lu {et~al.}(2014)Lu, Zhang, Liu, Wang, \& Gu}]{2014ApJ...790...84L}
Lu, X., Zhang, Q., Liu, H.~B., Wang, J., \& Gu, Q. 2014, The Astrophysical
  Journal, 790, 84

\bibitem[{Marka {et~al.}(2012)Marka, Schreyer, Launhardt, Semenov, \&
  Henning}]{2012A&A...537A...4M}
Marka, C., Schreyer, K., Launhardt, R., Semenov, D.~A., \& Henning, T. 2012,
  Astronomy and Astrophysics, 537, 4

\bibitem[{McKee \& Tan(2003)}]{2003ApJ...585..850M}
McKee, C.~F., \& Tan, J.~C. 2003, The Astrophysical Journal, 585, 850

\bibitem[{McLaughlin \& Pudritz(1997)}]{1997ApJ...476..750M}
McLaughlin, D.~E., \& Pudritz, R.~E. 1997, The Astrophysical Journal, 476, 750

\bibitem[{Motte {et~al.}(2014)Motte, Nguy{\^e}n~Luong, Schneider, Heitsch,
  Glover, Carlhoff, Hill, Bontemps, Schilke, Louvet, Hennemann, Didelon, \&
  Beuther}]{2014A&A...571A..32M}
Motte, F., {et~al.} 2014, Astronomy and Astrophysics, 571, A32

\bibitem[{Myers(2009)}]{2009ApJ...700.1609M}
Myers, P.~C. 2009, The Astrophysical Journal, 700, 1609

\bibitem[{Oka {et~al.}(2001)Oka, Hasegawa, Sato, Tsuboi, Miyazaki, \&
  Sugimoto}]{2001ApJ...562..348O}
Oka, T., Hasegawa, T., Sato, F., Tsuboi, M., Miyazaki, A., \& Sugimoto, M.
  2001, The Astrophysical Journal, 562, 348

\bibitem[{Osorio {et~al.}(2009)Osorio, Anglada, Lizano, \&
  D'Alessio}]{2009ApJ...694...29O}
Osorio, M., Anglada, G., Lizano, S., \& D'Alessio, P. 2009, The Astrophysical
  Journal, 694, 29

\bibitem[{Ossenkopf \& Henning(1994)}]{1994A&A...291..943O}
Ossenkopf, V., \& Henning, T. 1994, Astronomy and Astrophysics (ISSN
  0004-6361), 291, 943

\bibitem[{Peretto \& Fuller(2009)}]{2009A&A...505..405P}
Peretto, N., \& Fuller, G.~A. 2009, Astronomy and Astrophysics, 505, 405

\bibitem[{Peretto {et~al.}(2014)Peretto, Fuller, Andr{\'e}, Arzoumanian,
  Rivilla, Bardeau, Duarte~Puertas, Guzman~Fernandez, Lenfestey, Li, Olguin,
  R{\"o}ck, de~Villiers, \& Williams}]{2014A&A...561A..83P}
Peretto, N., {et~al.} 2014, Astronomy and Astrophysics, 561, 83

\bibitem[{Pickett {et~al.}(1998)Pickett, Poynter, Cohen, Delitsky, Pearson, \&
  M{\"u}ller}]{1998JQSRT..60..883P}
Pickett, H.~M., Poynter, R.~L., Cohen, E.~A., Delitsky, M.~L., Pearson, J.~C.,
  \& M{\"u}ller, H. S.~P. 1998, Journal of Quantitative Spectroscopy and
  Radiative Transfer, 60, 883

\bibitem[{Pillai {et~al.}(2006)Pillai, Wyrowski, Menten, \&
  Kr{\"u}gel}]{2006A&A...447..929P}
Pillai, T., Wyrowski, F., Menten, K.~M., \& Kr{\"u}gel, E. 2006, Astronomy and
  Astrophysics, 447, 929

\bibitem[{Pon {et~al.}(2011)Pon, Johnstone, \& Heitsch}]{2011ApJ...740...88P}
Pon, A., Johnstone, D., \& Heitsch, F. 2011, The Astrophysical Journal, 740, 88

\bibitem[{Ragan {et~al.}(2011)Ragan, Bergin, \& Wilner}]{2011ApJ...736..163R}
Ragan, S.~E., Bergin, E.~A., \& Wilner, D. 2011, The Astrophysical Journal,
  736, 163

\bibitem[{Ragan {et~al.}(2012)Ragan, Heitsch, Bergin, \&
  Wilner}]{2012ApJ...746..174R}
Ragan, S.~E., Heitsch, F., Bergin, E.~A., \& Wilner, D. 2012, The Astrophysical
  Journal, 746, 174

\bibitem[{Ragan {et~al.}(2013)Ragan, Henning, \& Beuther}]{2013A&A...559A..79R}
Ragan, S.~E., Henning, T., \& Beuther, H. 2013, Astronomy and Astrophysics,
  559, 79

\bibitem[{Ragan {et~al.}(2014)Ragan, Henning, Tackenberg, Beuther, Johnston,
  Kainulainen, \& Linz}]{2014A&A...568A..73R}
Ragan, S.~E., Henning, T., Tackenberg, J., Beuther, H., Johnston, K.~G.,
  Kainulainen, J., \& Linz, H. 2014, Astronomy and Astrophysics, 568, A73

\bibitem[{Rathborne {et~al.}(2011)Rathborne, Garay, Jackson, Longmore, Zhang,
  \& Simon}]{2011ApJ...741..120R}
Rathborne, J.~M., Garay, G., Jackson, J.~M., Longmore, S., Zhang, Q., \& Simon,
  R. 2011, The Astrophysical Journal, 741, 120

\bibitem[{Rathborne {et~al.}(2005)Rathborne, Jackson, Chambers, Simon, Shipman,
  \& Frieswijk}]{2005ApJ...630L.181R}
Rathborne, J.~M., Jackson, J.~M., Chambers, E.~T., Simon, R., Shipman, R., \&
  Frieswijk, W. 2005, The Astrophysical Journal, 630, L181

\bibitem[{Rathborne {et~al.}(2006)Rathborne, Jackson, \&
  Simon}]{2006ApJ...641..389R}
Rathborne, J.~M., Jackson, J.~M., \& Simon, R. 2006, The Astrophysical Journal,
  641, 389

\bibitem[{Rathborne {et~al.}(2007)Rathborne, Simon, \&
  Jackson}]{2007ApJ...662.1082R}
Rathborne, J.~M., Simon, R., \& Jackson, J.~M. 2007, The Astrophysical Journal,
  662, 1082

\bibitem[{Reid {et~al.}(2009)Reid, Menten, Zheng, Brunthaler, Moscadelli, Xu,
  Zhang, Sato, Honma, Hirota, Hachisuka, Choi, Moellenbrock, \&
  Bartkiewicz}]{2009ApJ...700..137R}
Reid, M.~J., {et~al.} 2009, The Astrophysical Journal, 700, 137

\bibitem[{Rieke {et~al.}(2004)Rieke, Young, Engelbracht, Kelly, Low, Haller,
  Beeman, Gordon, Stansberry, Misselt, Cadien, Morrison, Rivlis, Latter,
  Noriega-Crespo, Padgett, Stapelfeldt, Hines, Egami, Muzerolle,
  Alonso-Herrero, Blaylock, Dole, Hinz, Le~Floc'h, Papovich,
  P{\'e}rez-Gonz{\'a}lez, Smith, Su, Bennett, Frayer, Henderson, Lu, Masci,
  Pesenson, Rebull, Rho, Keene, Stolovy, Wachter, Wheaton, Werner, \&
  Richards}]{2004ApJS..154...25R}
Rieke, G.~H., {et~al.} 2004, The Astrophysical Journal Supplement Series, 154,
  25

\bibitem[{Rosolowsky \& Leroy(2006)}]{2006PASP..118..590R}
Rosolowsky, E., \& Leroy, A. 2006, The Publications of the Astronomical Society
  of the Pacific, 118, 590

\bibitem[{Rosolowsky {et~al.}(2008)Rosolowsky, Pineda, Foster, Borkin,
  Kauffmann, Caselli, Myers, \& Goodman}]{2008ApJS..175..509R}
Rosolowsky, E.~W., Pineda, J.~E., Foster, J.~B., Borkin, M.~A., Kauffmann, J.,
  Caselli, P., Myers, P.~C., \& Goodman, A.~A. 2008, The Astrophysical Journal
  Supplement Series, 175, 509

\bibitem[{Sakai {et~al.}(2008)Sakai, Sakai, Kamegai, Hirota, Yamaguchi, Shiba,
  \& Yamamoto}]{2008ApJ...678.1049S}
Sakai, T., Sakai, N., Kamegai, K., Hirota, T., Yamaguchi, N., Shiba, S., \&
  Yamamoto, S. 2008, The Astrophysical Journal, 678, 1049

\bibitem[{Sanhueza {et~al.}(2010)Sanhueza, Garay, Bronfman, Mardones, May, \&
  Saito}]{2010ApJ...715...18S}
Sanhueza, P., Garay, G., Bronfman, L., Mardones, D., May, J., \& Saito, M.
  2010, The Astrophysical Journal, 715, 18

\bibitem[{Sanhueza {et~al.}(2013)Sanhueza, Jackson, Foster, Jim{\'e}nez-Serra,
  Dirienzo, \& Pillai}]{2013ApJ...773..123S}
Sanhueza, P., Jackson, J.~M., Foster, J.~B., Jim{\'e}nez-Serra, I., Dirienzo,
  W.~J., \& Pillai, T. 2013, The Astrophysical Journal, 773, 123

\bibitem[{Schneider \& Elmegreen(1979)}]{1979ApJS...41...87S}
Schneider, S., \& Elmegreen, B.~G. 1979, Astrophysical Journal Supplement
  Series, 41, 87

\bibitem[{Simon {et~al.}(2001)Simon, Jackson, Clemens, Bania, \&
  Heyer}]{2001ApJ...551..747S}
Simon, R., Jackson, J.~M., Clemens, D.~P., Bania, T.~M., \& Heyer, M.~H. 2001,
  The Astrophysical Journal, 551, 747

\bibitem[{Simon {et~al.}(2006{\natexlab{a}})Simon, Jackson, Rathborne, \&
  Chambers}]{2006ApJ...639..227S}
Simon, R., Jackson, J.~M., Rathborne, J.~M., \& Chambers, E.~T.
  2006{\natexlab{a}}, The Astrophysical Journal, 639, 227

\bibitem[{Simon {et~al.}(2006{\natexlab{b}})Simon, Rathborne, Shah, Jackson, \&
  Chambers}]{2006ApJ...653.1325S}
Simon, R., Rathborne, J.~M., Shah, R.~Y., Jackson, J.~M., \& Chambers, E.~T.
  2006{\natexlab{b}}, The Astrophysical Journal, 653, 1325

\bibitem[{Suzuki {et~al.}(1992)Suzuki, Yamamoto, Kaifu, Ishikawa, Hirahara, \&
  Takano}]{1992ApJ...392..551S}
Suzuki, H., Yamamoto, S., Kaifu, N., Ishikawa, S.-I., Hirahara, Y., \& Takano,
  S. 1992, Astrophysical Journal, 392, 551

\bibitem[{Swift(2009)}]{2009ApJ...705.1456S}
Swift, J.~J. 2009, The Astrophysical Journal, 705, 1456

\bibitem[{Tafalla {et~al.}(2004)Tafalla, Myers, Caselli, \&
  Walmsley}]{2004A&A...416..191T}
Tafalla, M., Myers, P.~C., Caselli, P., \& Walmsley, C.~M. 2004, Astronomy and
  Astrophysics, 416, 191

\bibitem[{Tafalla {et~al.}(2002)Tafalla, Myers, Caselli, Walmsley, \&
  Comito}]{2002ApJ...569..815T}
Tafalla, M., Myers, P.~C., Caselli, P., Walmsley, C.~M., \& Comito, C. 2002,
  The Astrophysical Journal, 569, 815

\bibitem[{Torii {et~al.}(2011)Torii, Enokiya, Sano, Yoshiike, Hanaoka, Ohama,
  Furukawa, Dawson, Moribe, Oishi, Nakashima, Okuda, Yamamoto, Kawamura,
  Mizuno, Maezawa, Onishi, Mizuno, \& Fukui}]{2011ApJ...738...46T}
Torii, K., {et~al.} 2011, The Astrophysical Journal, 738, 46

\bibitem[{Urquhart {et~al.}(2009)Urquhart, Hoare, Purcell, Lumsden, Oudmaijer,
  Moore, Busfield, Mottram, \& Davies}]{2009A&A...501..539U}
Urquhart, J.~S., {et~al.} 2009, Astronomy and Astrophysics, 501, 539

\bibitem[{Urquhart {et~al.}(2011)Urquhart, Morgan, Figura, Moore, Lumsden,
  Hoare, Oudmaijer, Mottram, Davies, \& Dunham}]{2011MNRAS.418.1689U}
---. 2011, Monthly Notices of the Royal Astronomical Society, 418, 1689

\bibitem[{Wang {et~al.}(2008)Wang, Zhang, Pillai, Wyrowski, \&
  Wu}]{2008ApJ...672L..33W}
Wang, Y., Zhang, Q., Pillai, T., Wyrowski, F., \& Wu, Y. 2008, The
  Astrophysical Journal, 672, L33

\bibitem[{Wang {et~al.}(2006)Wang, Zhang, Rathborne, Jackson, \&
  Wu}]{2006ApJ...651L.125W}
Wang, Y., Zhang, Q., Rathborne, J.~M., Jackson, J., \& Wu, Y. 2006, The
  Astrophysical Journal, 651, L125

\bibitem[{Williams {et~al.}(1994)Williams, de~Geus, \&
  Blitz}]{1994ApJ...428..693W}
Williams, J.~P., de~Geus, E.~J., \& Blitz, L. 1994, The Astrophysical Journal,
  428, 693

\bibitem[{Wilson {et~al.}(2009)Wilson, Rohlfs, \&
  H{\"u}ttemeister}]{2009tra..book.....W}
Wilson, T.~L., Rohlfs, K., \& H{\"u}ttemeister, S. 2009, Tools of Radio
  Astronomy

\bibitem[{Wood {et~al.}(1994)Wood, Myers, \& Daugherty}]{1994ApJS...95..457W}
Wood, D. O.~S., Myers, P.~C., \& Daugherty, D.~A. 1994, Astrophysical Journal
  Supplement Series (ISSN 0067-0049), 95, 457

\bibitem[{Xu {et~al.}(2013)Xu, Wang, \& Liu}]{2013A&A...559A.113X}
Xu, J.-L., Wang, J.-J., \& Liu, X.-L. 2013, Astronomy and Astrophysics, 559,
  113

\bibitem[{Zhang {et~al.}(2011)Zhang, Yang, Xu, Pandian, Menten, \&
  Henkel}]{2011ApJS..193...10Z}
Zhang, S.~B., Yang, J., Xu, Y., Pandian, J.~D., Menten, K.~M., \& Henkel, C.
  2011, The Astrophysical Journal Supplement, 193, 10

\end{thebibliography}

\end{document}